 \documentclass[final,5p,times]{elsarticle}

\biboptions{sort&compress}

\usepackage{amssymb}
\usepackage{amsmath}

\begin{document}

\begin{frontmatter}




\title{Uncertainties and understanding of experimental and theoretical results regarding reactions forming heavy and superheavy nuclei}


\author[MIFT]{G.~Giardina}
\ead{ggiardina@unime.it}

\author[CHIM,Catania]{G.~Mandaglio}
\ead{gmandaglio@unime.it}

\author[JINR,Tashkent]{A.K.~Nasirov}
\ead{nasirov@jinr.ru}

\author[MIFT,LNF]{A.~Anastasi}

\author[LNF]{F.~Curciarello}

\author[MIFT]{G.~Fazio}

\address[MIFT]{Dipartimento di Scienze Matematiche e Informatiche, Scienze Fisiche e Scienze della Terra, University of Messina, Messina,  Italy}

\address[CHIM]{Dipartimento di Scienze Chimiche, Biologiche, Farmaceutiche ed Ambientali, University of Messina, Messina,  Italy}

\address[Catania]{INFN Sezione di Catania, Catania, Italy}

\address[JINR]{JINR - Bogoliubov Laboratory of Theoretical Physics, Dubna, Russia}

\address[Tashkent]{Institute of Nuclear Physics Uzbek Academy of Sciences, Tashkent, Uzbekistan}

\address[LNF]{ INFN Laboratori Nazionali di Frascati, Frascati, Italy}
\begin{abstract}
Experimental and theoretical results of the $P_{\mathrm{CN}}$ fusion probability of
reactants in the entrance channel and the $W_{\mathrm{sur}}$ survival probability
against fission at deexcitation of the compound nucleus formed in heavy-ion
collisions are discussed. The theoretical results for a set of nuclear
reactions leading to formation of compound nuclei (CNs) with the charge
number $Z=102\text{--}122$ reveal a strong sensitivity of $P_{\mathrm{CN}}$
to the characteristics of colliding nuclei in the entrance channel,
dynamics of the reaction mechanism, and excitation energy of the system. We
discuss the validity of assumptions and procedures for analysis of
experimental data, and also the limits of validity of theoretical results
obtained by the use of phenomenological models. The comparison of results
obtained in many investigated reactions reveals serious limits of validity
of the data analysis and calculation procedures.
\end{abstract}


\end{frontmatter}

\section{Introduction}

\label{intro}
The study of the nuclear reactions in heavy ion collisions continues to excite great interest in the scientific community to better understanding the
 reaction dynamics  from the stage of capture of the projectile by target-nucleus up to the formation of the reaction products.
 The knowledge about reaction dynamics is important
in planning possible experiments suitable to form heavy and superheavy compound
 nuclei leading to evaporation residues (ERs) and identifiable fragments belonging to
  the fusion-fission process.
 It is clear that the differences between the experimental results measured for
the same quantities in the same nuclear reactions are explained by the specific conditions present in the overall experimental apparatus and data analysis.
The differences between the theoretical results calculated by the different models
 are related to the assumptions made in the procedures of theoretical calculations
 and use of simplified phenomenological models unsuitable to describe the reaction dynamics. In fact, in this last case the use of free parameters can lead to an apparent acceptable agreement between the calculated results and data but actually these results can not prove the effectiveness of the procedures used to obtain the experimental results. Analogously, the procedure of obtaining the best values of free parameters used in the phenomenological model which leads to results of calculation in good agreement with the obtained experimental results can not demonstrate by a clear and unambiguous way the understanding of the reaction dynamics, from the stage of colliding nuclei up to the achievement of the final products.

The reliability of the experimental results can be improved by decreasing the number  of assumptions due to the increase of the measured physical quantities and their correlation.

  The evaporation residues (ERs) are registered enough unambiguously since those  products can be
separated easily from the ones of the other events. Therefore, theoretical results are
 aimed to be close to the experimental data of evaporation residues.
Furthermore, there are difficulties
 in estimating  the incomplete fusion contribution \cite{michel2002,PRC81,Esb2010} in the formation of
 the evaporation residues since the ambiguities of its mechanism are appeared.

The main reason for the differences in the experimental fusion and capture cross sections  is related to the ambiguity of the procedures at the
 separation  of the events corresponding to deep-inelastic collisions (DICs), quasifission (QF)
 and fusion-fission (FIS) processes. The quasifission is
the decay of the DNS into two fragments without
formation of CN.
  There is still no definite understanding
nature of full momentum transfer reactions in the experimental analysis of the
deep-inelastic collisions and quasifission events to estimate capture cross
 sections. The overlap of the mass and/or angular distributions of the
 quasifission and fusion-fission products causes ambiguity in the estimation
 of the experimental fusion cross sections.

 The choice of degrees of freedom and  interaction  forces involved in calculations
 are directed to simplify the complicated or unknown nature of the physical
 processes of the heavy ion collisions.
Therefore, the deviations   between the experimental results and the various theoretical ones are inevitable.

 The $P_{CN}$ fusion probability of reactants in the heavy ion collisions is estimated
 as a ratio of the complete fusion ($\sigma_{\rm fus}$) and capture
 ($\sigma_{\rm cap}$) cross sections:
 \begin{equation}
 P_{\rm CN}=\frac{\sigma_{\rm fus}}{\sigma_{\rm cap}}=
 \frac{\sigma_{\rm fus}}{\sigma_{\rm fus}+\sigma_{\rm qfis}}.
 \label{Pcneq}
\end{equation}
The capture cross section is determined by the estimation of the range of the orbital
angular momentum leading to the full momentum transfer in the entrance
channel of collision.
The evolution of the excited dinuclear system (DNS) formation can lead to complete fusion
in competition with quasifission. The details of this model are
present in many of our papers (see, for example,
\cite{PRC79,FazioMPL2005,JMPE2010,PRC84,NasiEPJ49,FazioJP77,PRC91,Conf205,aglio2012,aglio2015}),
but we give in Appendix A of the paper the main description of the method regarding the reaction in the entrance channel up to the evolution of DNS to complete fusion stage in competition with the decay of DNS into two nuclei by the quasifission process.
 The model takes into account the dependence of the capture cross section on the range of the
 orientation angles of nuclei at the initial stage of nuclear collision,
 the mass asymmetry parameter of reactants in the entrance channel, the considered $E_{c.m}$ energy range
 for the investigated reaction, the orbital angular momentum range that are
 needfull to consider in a refined and sensitive model with the aim
 of studying the evolution of each reaction from the contact of reactants
 to the compound nucleus formation, until to obtaining the final products of deexcitation of CN.

 It is obvious the differences between the values of
 $P_{\rm CN}$  extracted from the measured data of the capture and
 complete fusion events depend on how are correctly estimated
 capture and fusion cross sections from the measured data.
 Therefore, the reliable experimental determinations of $P_{\rm CN}$
 and consequently the understanding the
entrance channel effect on the $P_{\rm CN}$ values are strongly
related to the choice of assumptions for the data analysis.

For example, in refs.\cite{PRC79,GG2011} the ambiguity in the estimation
of the experimental quasifission events for the $^{48}$Ca+ $^{154}$Sm reaction
is discussed.
The strong difference between the experimental data \cite{knya2007} and
theoretical curves of quasifission cross sections in ref. \cite{PRC79}
 is explained by excluding the quasifission
events related to the mass numbers outside the range $60 \leq A \leq 130$
at low energies $E_{\rm c.m.}^* < 140$ MeV.
The authors of ref. \cite{knya2007} considered the reaction products
with mass numbers $A<60$ (or $A>130$) as the ones of the deep-inelastic collisions.
 The yield of products of the full momentum transfer (capture) reaction
 is seen from their total kinetic energy distribution presented in ref.
  \cite{knya2007}.
 So the reason causing the huge difference between theoretical \cite{PRC79}
and experimental \cite{knya2007} results is related to the conditions of determination
of the capture events. The separation of the capture events
  by the restriction of the mass numbers of binary fragments
in the range $60 \leq A \leq 130$ is not completely correct since there are
events of capture related with yield of binary fragments with mass numbers  $A \leq 60$.
 Such a procedure of restriction at the analysis of the measured data
 leads to the loss of an unknown part of the capture cross section, and,
 consequently, the  fusion probability  $P_{\rm CN}$ obtained  by the restriction
 of the capture events is not realistic.
 Since a significant part of the quasifission products with the mass numbers
  $A<60$ (or $A>130$) are excluded from the consideration. Therefore, the reduced
  capture cross section leads to increase the  fusion probability  $P_{\rm CN}$
  (see Eq. (\ref{Pcneq})).
  The presence and overlapping of the quasifission products among DIC products was demonstrated
  in ref. \cite{GG2011} (see  figs. 3 and 4) and ref.\cite{FazioMPL2005} (see Fig. 4).

The competition between the complete fusion of nuclei in DNS and
 quasifission (decay of DNS into two fragments) processes decreases the
value of the fusion cross section \cite{Fazio2004,PRC72,NuclPhys05}:
 \begin{equation}
\sigma_{\rm fus}(E_{\rm c.m.}) = \sum^{\ell_d(E_{\rm c.m.})}_{\ell=0} (2\ell+1)
\sigma_{\rm cap}(E_{\rm c.m.},\ell) P_{\rm CN}(E_{\rm c.m.},\ell).
\label{fus}
\end{equation}
The maximum value of $\ell$ leading to capture $\ell_d(E_{c.m.})$ depends on the beam energy
and it is calculated by the solution of the radial motion equations (see ref. \cite{NuclPhys05}).
Since the  capture cross section is equal to the sum of the complete fusion and quasifission
 cross sections,
$\sigma_{\rm cap}=\sigma_{\rm fus}+\sigma_{\rm qfis}$,
 the quasifission cross section is calculated by the expression
 \begin{equation}
 \sigma_{\rm qfis}(E_{\rm c.m.}) = \sum^{\ell_d}_{\ell=0}(2\ell+1)
 \sigma_{\rm cap}(E_{\rm c.m.},\ell)(1-P_{\rm CN}(E_{\rm c.m.},\ell)).
\label{qfis}
	\end{equation}
It should be stressed that quasifission of dinuclear system can take place
at all angular momentum values from $\ell=0$ to $\ell_d$.
Another binary process which leads to the formation of two fragments
similar to those of fusion-fission and quasifission is the fast fission (FF).
The fast fission occurs only if there is not a fission barrier
for the being formed compound nucleus due to the large values of the angular momentum,
$\ell > \ell_f$.
According to the rotating liquid drop model (see
\cite{SierkPRC33}) a rotating nucleus
with  the angular momentum $\ell_f$ breaks down immediately.
Therefore, FF is determined as the disintegration
  of the fast rotating mononucleus into two fragments, though DNS
survives quasifission to be transformed into CN.
In the case of the superheavy nuclei, the fission barrier providing their
stability against fission appears only due to shell effects in their binding
energy because there is no barrier connected with the liquid-drop model (see ref.\cite{Sobiczewski2007}).
 The damping of the shell effects decreases the
possibility of the deformed mononucleus of reaching the CN equilibrium shape, and
the mononucleus breaks down into two fragments without to reach the
CN stage.
Therefore, the fast fission cross section $\sigma_{\rm ff}$ is calculated
by summing the contributions of the partial fusion cross sections with
$\ell$ values corresponding to the range $\ell_f < \ell < \ell_d$
leading to the formation of the mononucleus,
 \begin{equation}
 \sigma_{\rm ff}(E_{\rm c.m.}) = \sum^{\ell_d}_{\ell=\ell_f} (2\ell+1)
\sigma_{\rm cap}(E_{\rm c.m.},\ell) P_{\rm CN}(E_{\rm c.m.},\ell).
\label{ffis}	
 \end{equation}

The sensitivity of the capture $\sigma_{\rm cap}$ and
fusion cross section $\sigma_{\rm fus}$ to the change of the radius
parameter $r_0$ is discussed in Appendix A of this work.
The low energy part of the
excitation functions of  $\sigma_{\rm cap}$ and $\sigma_{\rm fus}$
is moved to lower energies by the increase of the $r_0$ values.
This means that the variation of  $r_0$ from 1.16 fm to 1.18 fm leads to
 the change of the fusion probability about 2 times at the
 fixed low value of the beam energy. The part of the
 excitation functions of  $\sigma_{\rm cap}$ and $\sigma_{\rm fus}$
 above the Coulomb barrier is less sensitive to the  $r_0$ values.
   It is clear the change of $r_0$ leads to an appreciable modification of the
 interaction barrier of nuclei.
 This property of the excitation function is used in our calculation
 to reach an agreement of the capture cross section at the lowest
 energies with the experimental data. This allows us to use
 the partial fusion cross sections $\sigma_{\rm cap}(E_{\rm c.m.},\ell)$
 to calculate the partial evaporation residues cross sections
 and the sum them is compared with the experimental data.

  As the advantage of our modular system of nuclear reaction codes we stress the possibility to include
  into calculation  explicitly the effect of the orbital angular momentum on the capture, fusion
  and survival probability. This fact allows us to analyze the role of
  the entrance channel effects on the evaporation residue cross section \cite{PRC91}.

\section{About the $P_{\rm CN}$ determination for the CN formation}

\begin{figure}
\resizebox{0.5\textwidth}{!}{\vspace{5cm}\includegraphics{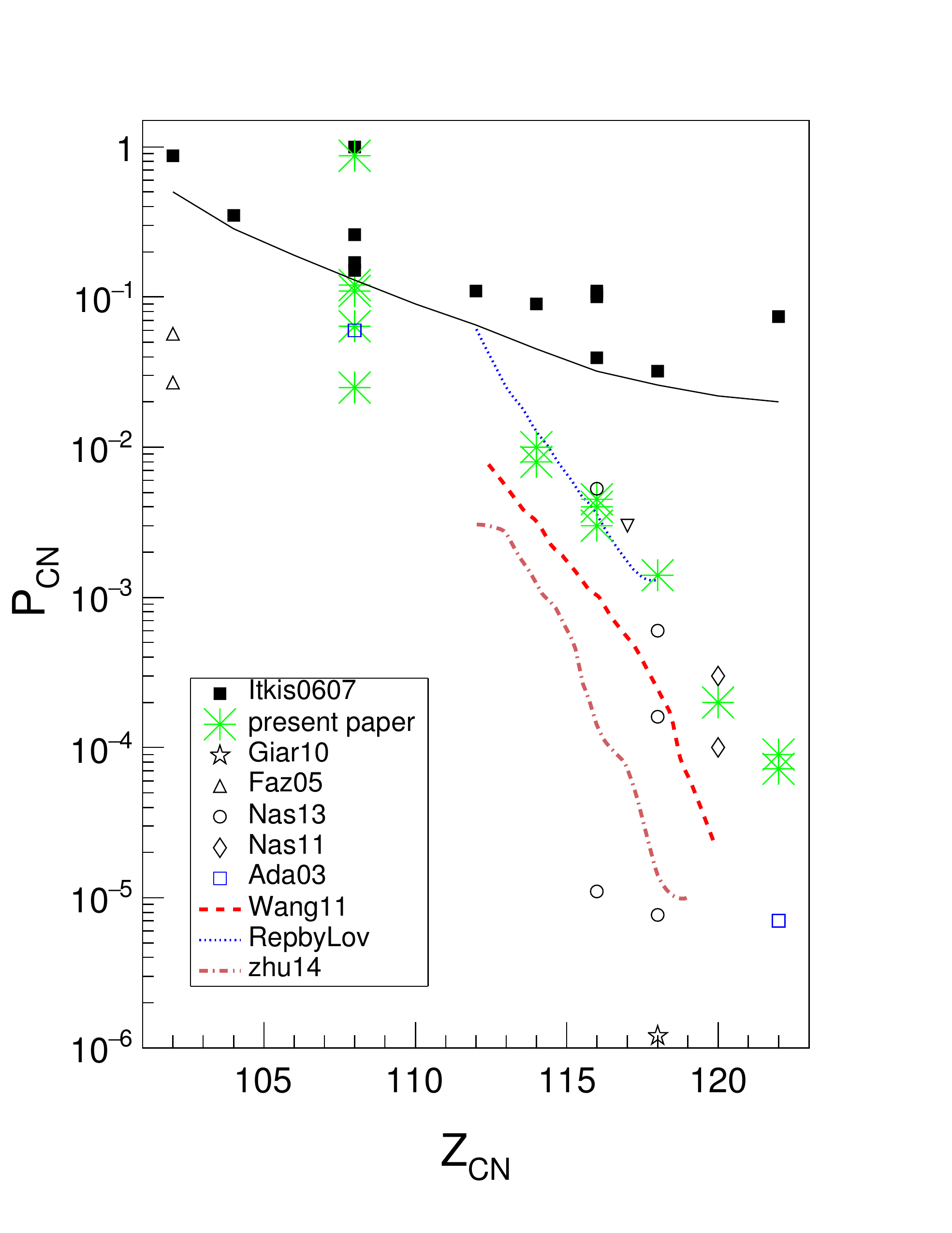}}
\put(-200,190){\large (a)}\vspace{-0.5cm}\\
\resizebox{0.5\textwidth}{!}{\vspace{5cm}\includegraphics{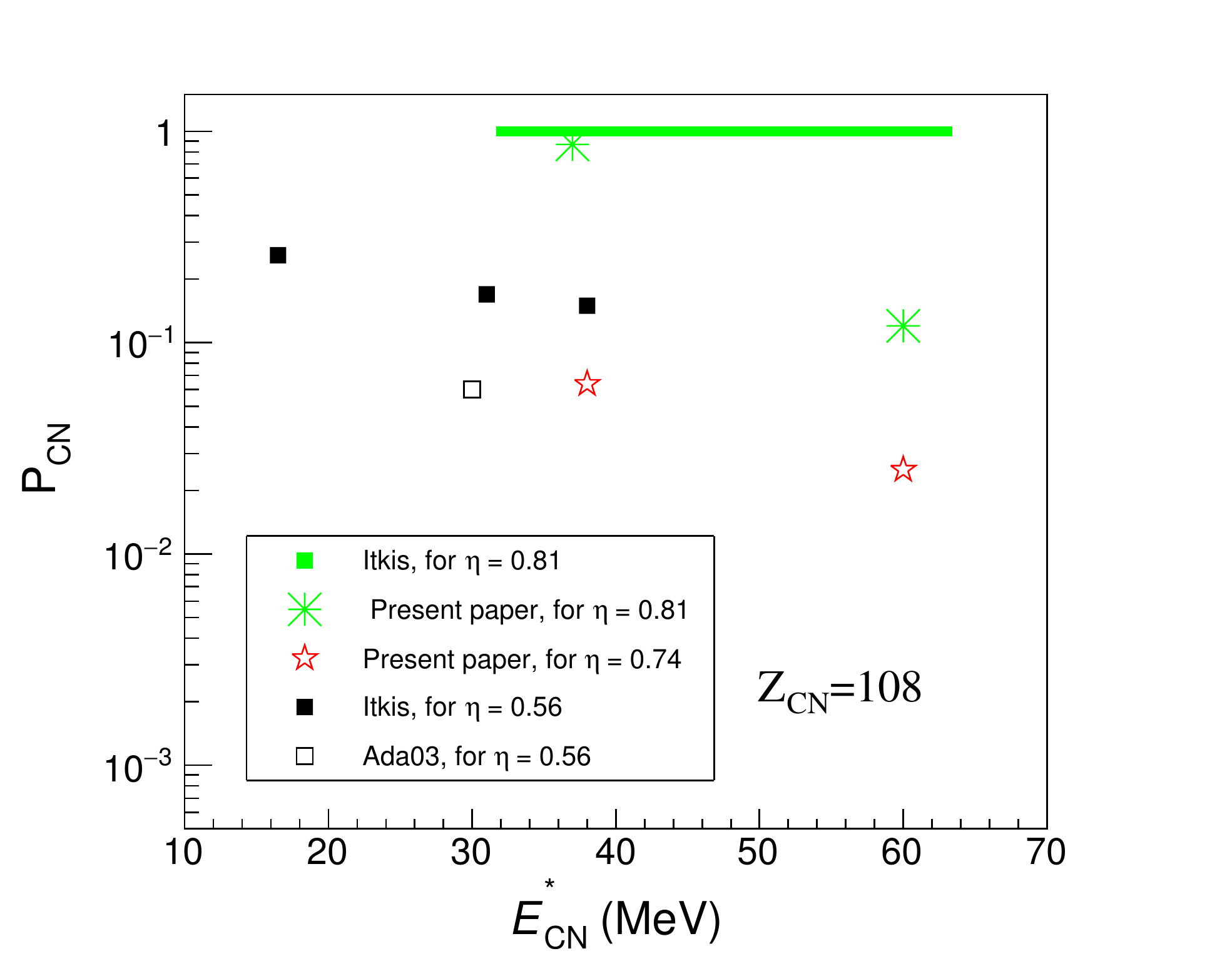}}
\put(-200,165){\large (b)}
\caption{(Color on-line) Comparison of $P_{\mathrm{CN}}$ values for many
asymmetric, less asymmetric, and almost asymmetric reactions leading to
compound nuclei with the atomic numbers $Z_{\mathrm{CN}}$ included in the 102--122
range. (a) The $P_{\mathrm{CN}}$ experimental values from refs.
\cite{AIP2006,ItNPA2007} (full squares), the theoretical values of the
present paper (asterisks); the results presented in refs. \cite{FazioMPL2005}
(open triangles), \cite{JMPE2010} (open star), \cite{PRC84} (open diamonds),
\cite{NasiEPJ49} (open circles), \cite{Adamian2003} (open square),
\cite{PRC79} (open inverse triangle), \cite{Adamian2003,love2015} (dotted
line), \cite{Wang2011,love2015} (dashed line), \cite{Zhu2014,love2015}
(dash--dotted line); the thin full line is a guide for the eye indicating a
clear separation between the experimental $P_{\mathrm{CN}}$ determinations and the
theoretical $P_{\mathrm{CN}}$ values for the investigated reactions. (b) The
$P_{\mathrm{CN}}$ values vs $E^{*}_{\mathrm{CN}}$ for the set of entrance
channel reactions leading to CNs with $Z_{\mathrm{CN}}=108$ and different mass
asymmetry parameters of CN: $^{26}Mg$+$^{248}$Cm
reaction with $\eta =0.81$, $^{32}S$+$^{238}$U with
$\eta=0.74$ and 
$^{58}$Fe+$^{208}$Pb 
with $\eta=0.56$.}\label{pcnvsz}
\end{figure}

 In  Fig. 3 of the ref.~\cite{love2015}, the author  presents
some  values of the $P_{\rm CN}$ fusion probabilitiy extracted from the
experimental data \cite{AIP2006,ItNPA2007}
in  nuclear reactions leading to CNs with $Z_{CN}$=108, 112, 114, and 116, and compares them with the trends of some theoretical results presented in
refs.~\cite{GG2011,Adamian2003,Wang2011,Zhu2014}.
 These theoretical results are different since they are obtained by different
 models and computational procedures. Therefore, the comparison between the experimental results and the various theoretical results seems to be formal
  and does not allow to make analysis of the reasons causing the observed
  behavior of the fusion probability, though in some cases the theoretical
 results are substantially consistent  with  those presented in Fig. 3 of ref. \cite{love2015}.
In order to hold a larger and more general discussion we present in Fig.~\ref{pcnvsz} of the present paper the results of $P_{\rm CN}$ as a function of $Z_{\rm CN} $
for reactions leading to CNs included in the  range 102--122 of atomic number  of superheavy nuclei (SHN). It is  an extension of results presented
 in Fig. 3 of paper \cite{love2015} and covering various kinds of reactions from very asymmetric to almost symmetric reactions characterized by the mass asymmetry parameter $\eta=\frac{|A_2-A_1|}{A_1+A_2}$.
Moreover, in Table \ref{tab:1} there are included all the specific details
 used to estimate $P_{\rm CN}$ for the reactions
 leading to the formation of CNs with the charge number $Z_{\rm CN}$ and excitation energy
  $E^{*}_{\rm CN}$ which was considered by authors of the corresponding papers.
   From a whole view of Fig.\ref{pcnvsz}~(a), it appears that the experimental
    data of $P_{CN}$ are underestimated by the theoretical approaches for
   the same reactions and at the extracting results from the measured data.
   Such systematic difference
  between the experimental and theoretical values of the $P_{\rm CN}$
  is the  evidence of the missing contribution
  of the quasifission process producing the projectile-like and target-like reaction products
 or/and  considering the quasifission products with mass numbers  around
 the symmetrical mass fragments as ones of the fusion-fission process. Unfortunately,
  the mass symmetric contributions cannot be quantified and separated from the one
  resulting from the pure contribution deriving from the fusion-fission process.

  To understand the underestimation of the experimental results for $P_{\rm CN}$
 by the theoretical calculations it is necessary to make the following  remarks and comments
  on the results presented in Fig.~\ref{pcnvsz} and Table \ref{tab:1}, reviewing the various reactions.

\subsection{Comparison of $P_{\rm CN}$ results for reactions forming CNs in the $Z_{CN}$=102-122 range}

A detailed comparison and analysis between $P_{\rm CN}$ experimental determinations and the related theoretical results is needed in order to understand the reasons of some relevant discrepancies.

\begin{table*}
\caption{Measured and calculated $P_{\rm CN}$ fusion probabilities obtained at $E^{*}_{\rm CN}$ excitation energies for the listed reactions with mass asymmetric parameter $\eta=\frac{|A_2-A_1|}{A_1+A_2}$ leading to CNs with atomic numbers $Z_{CN}$ included in the $Z_{CN}$=102-122 range. The presented $P_{\rm CN}$ experimental values range between 1 and $3.2\times10^{-2}$ for the excitation energy values of CN included in the 16.5--63.4~MeV interval, whereas our calculated $P_{\rm CN}$ values range between 0.87 and $7.7\times10^{-6}$ for the corresponding excitation energy values of CN included in 16--60~MeV interval.}
\label{tab:1}       
\begin{tabular}{|l|l|l|l|l|l|l|l|}
\hline\noalign{\smallskip}
Reaction & $\eta$ & $Z_{CN}$ & CN & Measured $P_{\rm CN}$ & Calculated $P_{\rm CN}$ & $E^{*}_{\rm CN}$ (MeV)   & reference  \\
\noalign{\smallskip}\hline\noalign{\smallskip}
$^{48}$Ca+$^{208}$Pb & 0.63 & 102 & 256 & 0.87 & & $\sim$30 & \cite{AIP2006,ItNPA2007}\\
$^{48}$Ca+$^{208}$Pb & 0.63 & 102 & 256 &  & 0.027 & 30 & \cite{FazioMPL2005} \\
$^{48}$Ca+$^{208}$Pb & 0.63 & 102 & 256 &  & 0.057 & 16 & \cite{FazioMPL2005}\\
$^{50}$Ti+$^{208}$Pb & 0.61 & 104 & 258 & 0.35 & & $\sim$30 & \cite{AIP2006,ItNPA2007}\\
$^{58}$Fe+$^{208}$Pb & 0.56 & 108 & 266 & 0.26 & & 16.5 & \cite{AIP2006,ItNPA2007}\\
$^{58}$Fe+$^{208}$Pb & 0.56 & 108 & 266 & 0.17 & & 31 & \cite{AIP2006,ItNPA2007}\\
$^{58}$Fe+$^{208}$Pb & 0.56 & 108 & 266 & 0.15 & & 38 & \cite{AIP2006,ItNPA2007}\\
$^{58}$Fe+$^{208}$Pb & 0.56 & 108 & 266 & & 0.06 & 30 & \cite{Adamian2003}\\
$^{36}$S+$^{238}$U & 0.74 & 108 & 274 & & 0.064 & 38 & present paper \\
$^{36}$S+$^{238}$U & 0.74 & 108 & 274 & & 0.025 & 60 & present paper\\
$^{26}$Mg+$^{248}$Cm & 0.81 & 108 & 274 & 1 & & from 31.7 to 63.4 & \cite{AIP2006,ItNPA2007}\\
$^{26}$Mg+$^{248}$Cm & 0.81 & 108 & 274 & & 0.87& 37 & present paper\\
$^{26}$Mg+$^{248}$Cm & 0.81 & 108 & 274 & & 0.12 & 60  & present paper\\
$^{48}$Ca+$^{238}$U & 0.664 & 112 & 286 & 0.11& & 32 & \cite{AIP2006,ItNPA2007}\\
$^{48}$Ca+$^{244}$Pu & 0.67 & 114 & 292 & 0.08 & & 32 & \cite{AIP2006,ItNPA2007}\\
$^{48}$Ca+$^{244}$Pu & 0.67 & 114 & 292 & & 0.008 & 32 & present paper\\
$^{48}$Ca+$^{244}$Pu & 0.67 & 114 & 292 & & 0.01 & 37 & present paper\\
$^{48}$Ca+$^{246}$Cm & 0.674 & 116 & 294 & 0.11 & & 32.5 & \cite{AIP2006,ItNPA2007}\\
$^{48}$Ca+$^{248}$Cm & 0.676 & 116 & 296 & 0.047 & & 32 & \cite{AIP2006,ItNPA2007}\\
$^{48}$Ca+$^{248}$Cm & 0.676 & 116 & 296 & & 5.3$\times 10^{-3}$ & 33 & \cite{NasiEPJ49}\\
$^{48}$Ca+$^{248}$Cm & 0.676 & 116 & 296 & & 0.004  & 37 & present paper\\
$^{50}$Ti+$^{244}$Pu & 0.66 & 116 & 294 & 0.105 & & 41.5 & \cite{AIP2006,ItNPA2007}\\
$^{50}$Ti+$^{244}$Pu & 0.66 & 116 & 294 & 0.10 & & 51.5 & \cite{AIP2006,ItNPA2007}\\
$^{50}$Ti+$^{244}$Pu & 0.66 & 116 & 294 & & 0.003 & 35 & present paper\\
$^{50}$Ti+$^{244}$Pu & 0.66 & 116 & 294 & & 0.0045 & 41 & present paper\\
$^{58}$Fe+$^{232}$Th & 0.60 & 116 & 290 & & 1.1$\times 10^{-5}$ & 40 & \cite{NasiEPJ49}\\
$^{48}$Ca+$^{249}$Bk & 0.68 & 117 & 297 & & 3$\times 10^{-3}$ & 33 & \cite{PRC79}\\
$^{48}$Ca+$^{249}$Cf & 0.68 & 118 & 297 & & 0.6$\times 10^{-3}$ & 33 & \cite{JMPE2010,NasiEPJ49}\\
$^{48}$Ca+$^{249}$Cf & 0.68 & 118 & 297 & & 0.14$\times 10^{-2}$ & 37 & present paper\\
$^{64}$Ni+$^{232}$Th & 0.57 & 118 & 296 & & 7.7$\times 10^{-6}$ & 35 & \cite{NasiEPJ49}\\
$^{64}$Ni+$^{232}$Th & 0.57 & 118 & 296 & & 1.6$\times 10^{-4}$ & 40 & \cite{NasiEPJ49}\\
$^{86}$Kr+$^{208}$Pb & 0.42 & 118 & 294 & 0.032 & & $\sim$30 & \cite{AIP2006,ItNPA2007}\\
$^{86}$Kr+$^{208}$Pb & 0.42 & 118 & 294 & & 2$\times 10^{-5}$ & 30 & \cite{JMPE2010}\\
$^{50}$Ti+$^{249}$Cf & 0.666 & 120 & 299 & & 3$\times 10^{-4}$ & 33 & \cite{PRC84}\\
$^{54}$Cr+$^{248}$Cm & 0.64 & 120 & 302 & & 1$\times 10^{-5}$ & 30 & \cite{PRC84}\\
$^{54}$Cr+$^{248}$Cm & 0.64 & 120 & 302 & & 2$\times 10^{-4}$ & 37 & present paper\\
$^{58}$Fe+$^{248}$Cm & 0.63 & 122 & 306 & 0.07 & & 33 & \cite{AIP2006,ItNPA2007}\\
$^{58}$Fe+$^{248}$Cm & 0.63 & 122 & 306 & & 7$\times 10^{-6}$ & 33 & \cite{Adamian2003}\\
$^{54}$Cr+$^{249}$Cf & 0.644 & 122 & 303 & & 7.2$\times 10^{-5}$ & 33 & present paper\\
$^{54}$Cr+$^{249}$Cf & 0.644 & 122 & 303 & & 9$\times 10^{-5}$ & 37 & present paper\\
\noalign{\smallskip}\hline
\end{tabular}
\end{table*}

\begin{enumerate}
\item For the $^{48}$Ca+$^{208}$Pb reaction leading to the $^{256}_{102}$No CN the
 value of the fusion probability $P_{\rm CN}$=0.87 has been extracted at
 $E^*_{\rm CN}$ of about 30~MeV from the experimental data of the capture and
 fusion cross sections in refs. \cite{AIP2006,ItNPA2007} while the theoretical value $P_{\rm CN}$=0.027
has been found in ref. \cite{FazioMPL2005} at the  excitation energy
$E^{*}_{\rm CN}$=30 MeV (see Fig. \ref{pcnvsz}(a).
There are two reasons causing this large discrepancy between theoretical and experimental results for $P_{\rm CN}$.
\textbf{a)} At the extraction of $P_{\rm CN}$ values from the
experimental data  the contribution of the quasifission events into capture cross section with the mass numbers around $A$=50
has been neglected although those events are the full momentum transfer events
 \cite{FazioProc2002,NasirovHung2004,AriOhta}. This means that
  the quasifission events with the mass numbers around initial mass numbers  $A$=50 considered as deep-inelastic
 collisions and the quasifission cross section $\sigma_{\rm qfis}$ is
 underestimated of one order of magnitude in calculations of $P_{\rm CN}$ by experimentalists.
As a result the value of $P_{\rm CN}$ increased (see Fig. 2 of ref.~\cite{FazioMPL2005}).
\textbf{  b)} On the other hand, due to the inclusion of the quasifision and fast fission events
  occurring at  large values  of the orbital angular momentum of collision
 into the fusion-fission events, the experimental fusion cross section
 $\sigma_{\rm fus}$ was overestimated. As a result, the experimental
 value of  $P_{\rm CN}$ appears larger.
  It is well known that the fast fission products are mixed
   with the fusion-fission ones and it is impossible to separate
   the contributions between these two processes.
  This circumstance leads to an increase of the fusion cross section $\sigma_{\rm fus}$.
    Obviously these two defects of the procedure at the analysis
   of the experimental data increase the $P_{\rm CN}$ value.
     The similar difference is seen from the comparison of the experimental
     and theoretical  results of the  $P_{\rm CN}$ values  0.35 from
     \cite{AIP2006,ItNPA2007} and 0.027 from  \cite{FazioMPL2005}, respectively,
     for two close mass asymmetric reactions ($\eta=0.63$ and 0.61, respectively)  $^{50}$Ti+$^{208}$Pb and $^{48}$Ca+$^{208}$Pb at the same excitation energy $E^{*}_{\rm CN}$ = 30 MeV.

 \item
  In Fig.~\ref{pcnvsz}~(b) the $P_{\rm CN}$ values vs  $E^{*}_{\rm CN}$
 for the reactions  with different mass asymmetry parameters $\eta=|A_1-A_2|/(A_1+A_2)$ leading to CNs with $Z_{CN}=108$ demonstrate the role of the entrance channel:
  thick line represents the experimental determinations  $P_{\rm CN}=1$ given in \cite{AIP2006,ItNPA2007} for
   the very asymmetric reaction $^{26}$Mg+$^{248}$Cm ($\eta$=0.81) leading to
   $^{274}$108 CN; in this figure, asterisks represent the theoretical values of $P_{\rm CN}$ for the same reaction found in the present paper; open stars represent the $P_{\rm CN}$ theoretical values obtained
  in the present paper for the less asymmetric reaction $^{36}$S+$^{238}$U
  ($\eta=0.74$) leading to the same $^{274}$108 CN. In the same figure full and open squares
  represent the experimental  \cite{AIP2006,ItNPA2007} and  theoretical value \cite{Adamian2003} obtained for
   the $^{58}$Fe+$^{208}$Pb   reaction. It is seen that the experimental data are about 3 times higher than the theoretical ones.
   It can be concluded  that:
 i) the dependence of the experimental values of $P_{\rm CN}$ \cite{AIP2006,ItNPA2007} are less sensitive to $E^{*}_{\rm CN}$
 than  the one of its  theoretical  values: $P_{\rm CN}$ value for the
 $^{26}$Mg+$^{238}$U  reaction decreases by more than 7 times at the increase of the $E^{*}_{\rm CN}$ excitation energy from 37 to 60 MeV;
 ii) similarly, our theoretical  $P_{\rm CN}$ values for the $^{36}$S+$^{238}$U less asymmetric reaction ($\eta=0.74$) decrease with increasing
 the $E^{*}_{\rm CN}$ values;
 iii) the $P_{\rm CN}$ values for a less asymmetric reaction are smaller than the ones for a more asymmetric reaction;
 iv) the experimental $P_{\rm CN}$ result extracted at $E^{*}_{\rm CN}=31$~MeV by using the capture and fusion cross sections reported in refs.~\cite{AIP2006,ItNPA2007}  for the $^{58}$Fe+$^{208}$Pb reaction leading to $^{266}$108 CN  with mass asymmetry parameter $\eta=0.56$ is about three times higher than the value found in ref. \cite{Adamian2003} at the comparable excitation energy of $E^{*}_{\rm CN}=30$~MeV; moreover, we have to observe that in Table \ref{tab:1} the  $P_{\rm CN}=0.064$ value  found in the present paper for the more asymmetric  $^{36}$S+$^{238}$U reaction with mass asymmetry parameter  $\eta=0.74$ leading to the same element with $Z_{CN}$=108 and mass number A=274 at excitation energy $E^{*}_{\rm CN}=38$~MeV is consistent with the $P_{\rm CN}=0.06$ value found in  \cite{Adamian2003}.
   These results clearly demonstrate the great sensitivity of the
 $P_{\rm CN}$ function to the $E^{*}_{\rm CN}$ excitation energy of CN for each
  considered entrance channel reaction.
It is well known that the $P_{\rm CN}$ value is smaller for a less asymmetric reaction
(and a fortiori for the less asymmetric reaction as for example the considered
$^{58}$Fe+$^{208}$Pb) than the  one obtained for a very asymmetric reaction
at a given  $E^{*}_{\rm CN}$ excitation energy of CN.
 This phenomenon is  related to the landscape of driving potential
 where the two different entrance channels has different initial conditions
to reach the same CN.
  The different properties of the CN formation are caused by the different values of
  the intrinsic fusion  $B^{*}_{fus}$ and quasifission $B_{qf}$ barriers
  for the two mass asymmetry parameters characterizing the entrance channels
  (see for example \cite{PRC84,NasiEPJ49,PRC91}).

  The increased sensitivity of the change of $P_{\rm CN}$ obtained in theoretical
   estimations to the mass asymmetry in the entrance channel allows us
   to separate the quasifission products
  from the ones of the fusion-fission process  contributing to the total fragment formation, while this cannot be unambiguously experimentally verified  \cite{PRC79,FazioProc2002,NasirovHung2004,AriOhta,PLB2010}. Since the products of the quasifission process are strongly prominent with respect to the ones of the fusion-fission process when comparing fragments produced by  symmetric (or almost symmetric) reactions  with those produced by asymmetric reactions, the correct analysis of the mass, energy and angular distribution of the reaction fragments allows us to establish process producing them to a reliable description of the reaction dynamics.
\item The fusion probabilities determined from  the  experimental capture and fusion excitation functions of the $^{48}$Ca+$^{238}$U and $^{48}$Ca+$^{244}$Pu  reactions leading to the different $^{286}$112 and $^{292}$114 CNs, respectively, presented in refs. \cite{AIP2006,ItNPA2007} at $E^{*}_{\rm CN}$=32 MeV are very close like $P_{\rm CN}$=0.11 and 0.08 (see Table \ref{tab:1}), respectively, whereas in our theoretical study on the $^{48}$Ca+$^{244}$Pu reaction  in the present work we find the value $P_{\rm CN}$=0.008 for the fusion probability at $E^{*}_{\rm CN}$=32 MeV (see Table \ref{tab:1}) that is 1 order of magnitude smaller than the one found in \cite{AIP2006,ItNPA2007}.
 The reason of this relevant difference is related with procedures at extraction of the experimental values of the $P_{\rm CN}$ by restriction of the mass and angular distributions of the binary reaction products to determine ones related with quasifission process.

\item By considering the group of reactions leading to isotopes of the $Z_{\rm CN}=116$ element, one needs to make the following comments: the results of the
$P_{\rm CN}$ extracted from the measured data \cite{AIP2006,ItNPA2007} show that the values for the  $^{48}$Ca+ $^{246}$Cm  and  $^{48}$Ca+ $^{248}$Cm reactions
  are 0.11 (at $E^{*}_{\rm CN}$=32.5 MeV)  and   0.05 (at $E^{*}_{\rm CN}$=32 MeV), respectively, whereas our calculations \cite{NasiEPJ49} give  values   $P_{\rm CN}$= 5.3$\times$10$^{-3}$
  at $E^{*}_{\rm CN}$=33 MeV) for the $^{48}$Ca+$^{248}$Cm reaction (1 order of magnitude
 smaller than the experimental determination at about the same $E^{*}_{\rm CN}$), and
 $P_{\rm CN}=1.1\times 10^{-5}$ (at $E^{*}_{\rm CN}$=40 MeV) for the $^{58}$Fe+$^{232}$Th  reaction (see Table \ref{tab:1}). Therefore, our theoretical results of $P_{\rm CN}$ are strongly sensitive to the entrance channel reaction and excitation energy (see Table \ref{tab:1}), whereas, the experimental determinations of $P_{\rm CN}$ appear essentially insensitive to the above-mentioned reactions and excitation energy  (see Table \ref{tab:1}). In fact, in all these cases the experimental values appear almost insentive to the various entrance channels and values of the $E^{*}_{\rm CN}$ excitation energy; therefore, also these experimental results are very different from our calculated values. Moreover, the measurements \cite{AIP2006,ItNPA2007} on the $^{50}$Ti+$^{244}$Pu reaction give the same values as $P_{\rm CN}$=0.1 at both $E^{*}_{\rm CN}$=41.5 and 51.5 MeV excitation energies, instead we find the value 4.5$\times$ 10$^{-3}$ at $E^{*}_{\rm CN}$=41 MeV that is over 20 times smaller than the measured one.
\item For the $^{48}$Ca+$^{249}$Bk reaction (CN=$^{297}$117) we obtained \cite{PRC79} $P_{\rm CN}$=$3\times 10^{-3}$ at $E^{*}_{\rm CN}$=33 MeV. The values of $P_{\rm CN}$ found for the close
    $^{48}$Ca+$^{248}$Cm and $^{48}$Ca+$^{249}$Cf reactions leading to the  $^{296}$116
    and $^{297}$118 CNs, respectively,  are equal to 5.3$\times 10^{-3}$  and
    6$\times 10^{-4}$  (see Table \ref{tab:1}), respectively, at $E^{*}_{\rm CN}$=33 MeV.
    These results demonstrate the sensitivity of the $P_{\rm CN}$ value with the entrance channel and at the same time the consistency with reliable results. In the reaction induced by the $^{48}$Ca beam and different targets like $^{248}$Cm, $^{249}$Bk, and $^{249}$Cf with  $Z_{CN}$=96, 97, 98, respectively, the $P_{\rm CN}$ values appreciably decrease (see Table \ref{tab:1}) due to  the increase of the Coulomb potential.
\item For the reaction leading to compound nucleus with $Z_{\rm CN}=118$ we can  compare
 the experimental and theoretical $P_{\rm CN}$ values. The $P_{\rm CN}$=0.032 value (at $E^{*}_{\rm CN}$=30 MeV) experimentally found in \cite{AIP2006,ItNPA2007}
 for the $^{86}$Kr+$^{208}$Pb reaction is three orders of magnitude greater than the theoretical estimation $P_{\rm CN}$=2$\times 10^{-5}$ (at $E^{*}_{\rm CN}$=30 MeV)
 presented by us in \cite{JMPE2010}.
Analogously, the measured value $P_{\rm CN}$=0.07 at $E^{*}_{\rm CN}$=33 MeV obtained in the experiment \cite{AIP2006,ItNPA2007} with the $^{58}$Fe+$^{248}$Cm reaction leading to the $^{306}$122 CN is four orders of magnitude greater than the theoretical value $P_{\rm CN}=7\times 10^{-6}$ at $E^{*}_{\rm CN}$=33 MeV obtained in \cite{Adamian2003}. Therefore, even in these cases of the less mass asymmetric reactions leading to superheavy CNs it is possible to observe  unreliable results of $P_{\rm CN}$ deduced by the analysis of the  experimental data.
\end{enumerate}

In conclusion, a large part of our $P_{\rm CN}$ theoretical results reported in Fig.~\ref{pcnvsz} is in agreement with the results obtained in \cite{Adamian2003} and for some reactions our results are consistent with the ones obtained in \cite{Wang2011,Zhu2014}. Apart from some differences in $P_{\rm CN}$ values obtained with different theoretical models,  it is relevant the common sensitivity of the theoretical results as a function of the mass asymmetry parameter, excitation energy $E^{*}_{\rm CN}$, and the
$Z_{\rm CN}$ atomic number of the CN reached. Conversely, the  $P_{\rm CN}$ results  as a function of $Z_{\rm CN}$  obtained from the analysis of the experimental data given in \cite{AIP2006,ItNPA2007} are larger in comparison with our theoretical results and those obtained by other models (see Fig. 1), where the experimental $P_{\rm CN}$ values range between 1 and $3.2\times10^{-2}$ in the $\Delta E^*_{\rm CN} = 16.5-63.4$~MeV interval, whereas our calculated $P_{\rm CN}$ values range between 0.87 and $7.7\times10^{-6}$  for the comparable excitation energy range of CNs included in the 16--60~MeV interval. The results presented in Fig~\ref{pcnvsz} and Table~\ref{tab:1} for all reactions leading to compound nuclei included in $Z_{\rm CN}$=102--122 interval  of atomic number clearly demonstrate that the $P_{\rm CN}$ experimental determinations appear to be poorly sensitive to the mass asymmetry parameter $\eta$ of the entrance channel and also to the excitation energy $E^{*}_{\rm CN}$ value. Such a large general difference between the $P_{\rm CN}$ experimental determinations and the corresponding theoretical values, for a wide set of investigated reactions leading to heavy and superheavy compound nuclei,   is the clear evidence of the unreliable experimental estimation of the contribute due to the quasifission process  during the evolution of the capture events into compound nucleus formation. In fact, as already explained, there is in experimental analysis some ambiguity in the separation of  the capture events from the huge contribution coming from the deep-inelastic collisions; therefore, the adopted assumptions in the data analysis lead to a relevant uncertainty in the capture cross section determination. Moreover, for the experimental determination of the compound nucleus cross section it is necessary to made some assumption for the mass of the detected fragments which contribute to the true fusion-fission process. Even for this experimental determination, the constraint used in the analysis  to select the events with symmetric mass only do not overcome the problem of the correct determination of the fusion cross section because in the reactions considered in
Fig~\ref{pcnvsz} and Table~\ref{tab:1}, the mass symmetric distribution contributed by the quasifission and fast-fission processes are very relevant.
Therefore, the experimental determination of the capture and fusion cross section are affected by strong uncertainties, and consequently the experimental
 $P_{\rm CN}$ ratio determinations between the capture and fusion cross sections are unreliable.

\subsection{Discussion on the $P_{\rm CN}$ results}

The reason for  this discrepancy between the measured data and theoretical
values of $P_{\rm CN}$ is connected with the experimental difficulties in the identification
 of the fusion-fission fragments produced by fission of the compound nucleus to
 determine the fusion cross section. The mass distribution of the fast fission
 and sometimes quasifission  processes overlaps with the mass symmetrical
 fusion-fission product distributions. The ability of the
  correct extraction of the fusion-fission cross section from the mixed data
 of the reaction products decreases due to intensive population
 of the mass-symmetric region by the fast fission or/and the quasifission fragments
 by the increase of $E^{*}_{\rm CN}$.
 Therefore, the extraction of the fusion cross section from the experimental data is strongly affected by the
  underestimation of contribution of the fast fission and quasifission fragments.
 This problem is inherent  to all kinds of reactions, leading to the formation of superheavy nuclei, when experimentalists selecting only mass symmetric fragments with the mass
 numbers in the range $A_{\rm CN}/2\pm20$,  assume such products belong only to the fusion-fission process. In fact, this assumption is completely doubtful because the yield of mass symmetric fragments produced by the quasifission and fast fission processes are competitive and often some orders of magnitude higher than the ones produced by the fusion-fission process (see for example, refs.~\cite{Fazio2004,mandaconf38}.

 The estimate of the capture cross section is affected by relevant uncertainty in the  separation of capture events of projectile by the target nucleus  from the  deep inelastic collisions with high yield.  The missing the quasifission events at the restriction  of the mass distribution of the binary products
 also leads to increase the experimental $P_{\rm CN}$ values.
 Therefore, the experimental estimate of the $P_{\rm CN}$ fusion probability by the $\sigma_{fus} / \sigma_{cap}$ ratio (where the $\sigma_{fus}$ and $\sigma_{cap}$ values are determined in experiments with large uncertainty) is also affected by great uncertainty by a factor that changes with $E^{*}_{\rm CN}$ excitation energy of CN, and with the asymmetry/symmetry of the entrance channel. Of course the $P_{\rm CN}$ value also strongly changes with the mass number $A$ of CN  at the same atomic number $Z_{CN}$, and even at different $Z_{CN}$ (see Fig. 3 of paper \cite{love2015} and with more details in Fig.~\ref{pcnvsz} of the present paper, especially the set of the reactions leading to CNs with $Z_{CN}$=108, 116, 118, and 122).

\begin{figure}[h]
\hspace*{-0.5cm}
\resizebox{0.5\textwidth}{!}{\includegraphics{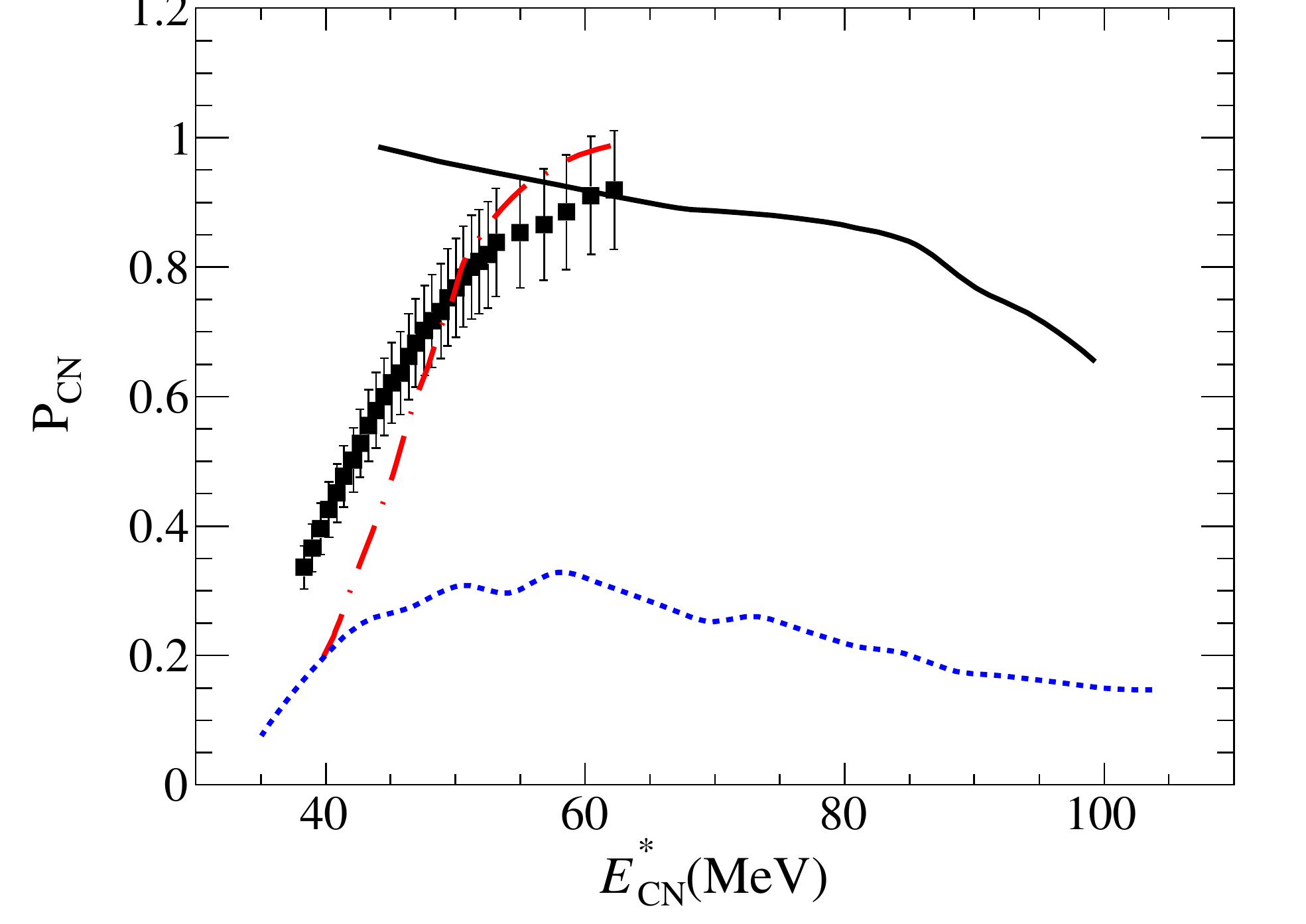}}
\caption{(Color on-line) The $P_{\rm CN}$ fusion probability as a function of the $E^{*}_{\rm CN}$ excitation energy for the $^{16}$O+$^{186}$W very asymmetric reaction and for the $^{48}$Ca+$^{154}$Sm less asymmetric reaction presented in \cite{GG2011} (full and dotted lines, respectively). Full squares are the experimental determinations by ref. \cite{knya2007} for the $^{48}$Ca+$^{154}$Sm reaction and dash-dotted line represents the predicted $P_{\rm CN}$ values for the same reaction by authors of ref. \cite{zagre2008}. \label{pcnvse}}
\end{figure}

By regarding   Fig. 4 of paper \cite{love2015} where the measured $P_{\rm CN}$ values of ref. \cite{knya2007} are compared with the predicted $P_{\rm CN}$ values of ref. \cite{zagre2008} against the $E^{*}_{\rm CN}$ excitation energy of CN. The author used there $P_{\rm CN}$ values  from the paper \cite{GG2011} presented
 against the collision energy relative to the interaction barriers $E_{\rm c.m.} - E_{B}$. Unfortunately the author \cite{love2015} has not performed appropriately transformation
of the $E^{*}_{\rm CN}$ excitation energy from the $E_{\rm c.m.} - E_{B}$ values: the position of the dashed curve from \cite{GG2011} is moved on 15 MeV to higher energy. Why?
We add in Fig.~\ref{pcnvse} of the present paper our calculated $P_{\rm CN}$ values  for the $^{16}$O+$^{186}$W very asymmetric reaction and for the less asymmetric $^{48}$Ca+$^{154}$Sm reaction, also including for a comparison the sets of $P_{\rm CN}$ values cited in paper \cite{love2015} as the measured values of ref.\cite{knya2007} and the predicted values of ref.\cite{zagre2008}.
As Fig.~\ref{pcnvse} shows the experimental $P_{\rm CN}$ values for the $^{48}$Ca+$^{154}$Sm strongly increase with the increase of the $E^{*}_{\rm CN}$ excitation energy of the $^{202}$Pb CN from 0.33 at $E^{*}_{\rm CN}$ = 38 MeV to 0.93 at $E^{*}_{\rm CN}$ = 62 MeV in about 24 MeV of the $\Delta E^{*}_{\rm CN}$ energy interval.  Instead, the theoretical results of $P_{\rm CN}$ presented in Fig.~\ref{pcnvse}, obtained by us for the $^{16}$O+$^{186}$W and $^{48}$Ca+$^{154}$Sm very asymmetric and less asymmetric reactions, respectively, show complete different trends for shape and values. At $E^{*}_{\rm CN}$ = 35 MeV $P_{\rm CN}$ is 0.075 for the reaction induced by $^{48}$Ca, while at $E^{*}_{\rm CN}$ = 44 MeV the $P_{\rm CN}$ is 1 for the asymmetric reaction induced by $^{16}$O, both leading to the same $^{202}$Pb CN.  Moreover, at $E^{*}_{\rm CN}$ = 62 MeV the $P_{\rm CN}$ value calculated by us for the asymmetric $^{16}$O induced reaction is about 0.91 and decreases with the increase of $E^{*}_{\rm CN}$, while the $P_{\rm CN}$ calculated by authors \cite{zagre2008} is 0.99 for the less asymmetric reaction induced by $^{48}$Ca (see dash-dotted line in Fig.~\ref{pcnvse}).

\begin{figure}[h]
\resizebox{0.5\textwidth}{!}{\includegraphics{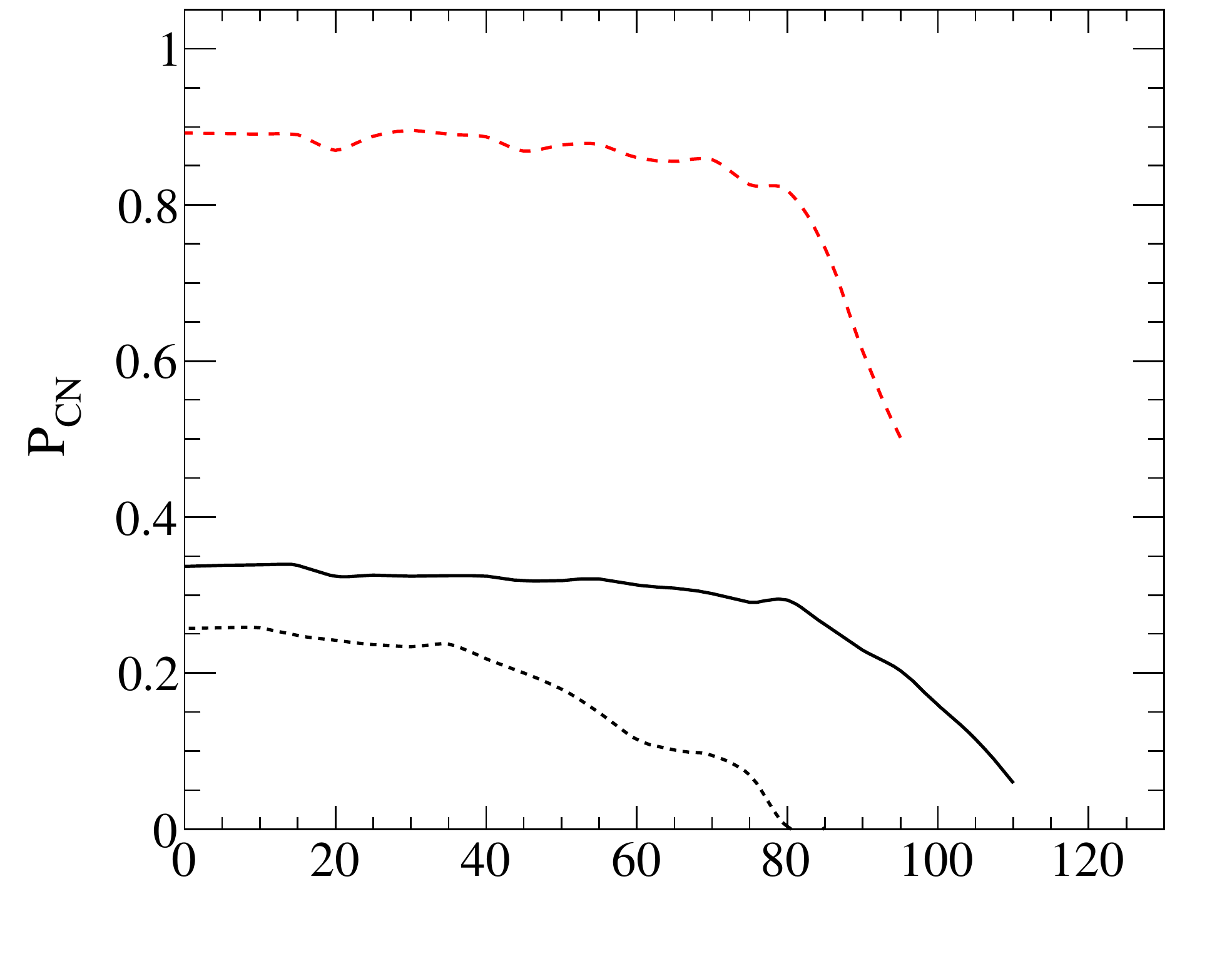}}
\put(-120,5){\large $\ell (\hbar)$}
\caption{(Color on-line) The $P_{\rm CN}$ fusion probability as a function of the angular momentum $\ell$ for the $^{16}$O+$^{186}$W very asymmetric reaction at excitation energy $E^{*}_{\rm CN}$=75 MeV (dashed line) of the $^{202}$Pb CN is shown; in the same figure,  the $P_{\rm CN}$ values vs $\ell$ for the $^{48}$Ca+$^{154}$Sm less asymmetric reaction at excitation energies $E^{*}_{\rm CN}$= 49 and 63 MeV, respectively (dotted and full lines), are also reported. \label{pcnvsl} }
\end{figure}

The trend of the $P_{\rm CN}$ values presented by us for the two above-mentioned
 reactions (full and short-dashed lines) shows the specific sensitivity of the
reaction mechanism for the two different entrance channels; instead, the trend of
results (full squares and dash-dotted line) presented by authors in papers
\cite{knya2007} and \cite{zagre2008} for the $^{48}$Ca+$^{154}$Sm reaction appears
fully inconsistent with our sensitive results and therefore they are very
questionable. Moreover, in order to show the sensitivity of the $P_{\rm CN}$ with
the angular momentum $\ell$ at two different excitation energies of 49 and 63 MeV
(dotted and full lines, respectively) of the formed $^{202}$Pb compound nucleus, we
present in Fig.~\ref{pcnvsl} our results obtained for the $^{48}$Ca+$^{154}$Sm less
 asymmetric reaction.
We also present in the same figure the $P_{\rm CN}$ values vs $\ell$ obtained for the $^{16}$O+$^{186}$W very asymmetric reaction, leading to the same $^{202}$Pb CN 
at $E^{*}_{\rm CN}$=63 MeV (dashed line). 
This figure clearly shows the strong dependence of the $P_{\rm CN}$ results on the excitation energy $E^{*}_{\rm CN}$ and/or of the beam-target combination in the entrance channel.

Therefore, the comparison of the $P_{\rm CN}$ values for different conditions of reactions can be made and understood only if somebody is able to explain 
the reasons for different results and trend due to the entrance channel effects and/or characteristics of the reaction mechanism.

\begin{figure}[h]
\resizebox{0.6\textwidth}{!}{\includegraphics{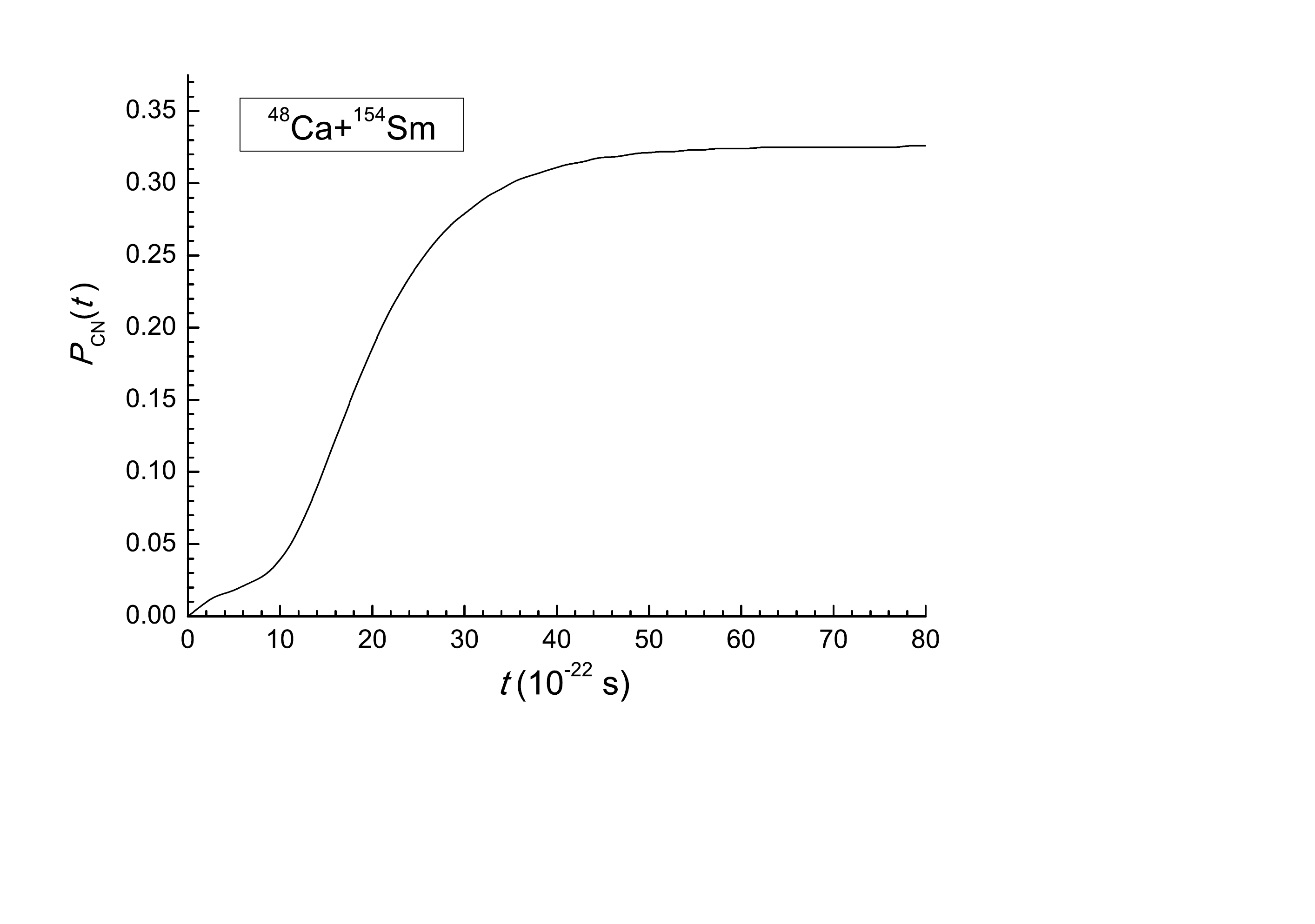}}
\vspace{-2.5cm}
\caption{ The $P_{\rm CN}(t)$ fusion probability calculated by Eq.(\ref{Pcnt})
as a function of time for the $^{48}$Ca+$^{154}$Sm  at excitation energy $E^{*}_{\rm CN}$=49 MeV and $\ell=$20 $\hbar$.
\label{Pcntfig}}
\end{figure}

 There is an alternative way of the estimation of the fusion probability
 by the use of the solution of the master equation (A.9).
$Y_Z$ characterizes the population of the DNS configuration with the charge asymmetry $Z=Z_1$
($Z_2=Z_P+Z_T-Z$). The initial conditions are  $Y_Z(0)=1$ for $Z=Z_P$ and $Z_2=Z_T$, where
 $Z_P$ and $Z_T$ are the charge numbers of the colliding nuclei.
 The fusion probability is found by calculation of
 the total quasifission probability $P_{qf}$  from all charge configuration of DNS.
 The last quantity  is calculated
 by summation of the all decays from the DNS configuration $Z$:
 \begin{eqnarray}
 \label{Pcnt}
P_{\rm CN}(t)&=&1-P_{qf}(t), \\
 P_{qf}(t)&=&\sum_Z\int_0^t \Lambda^{qf}_Z Y_Z(t')dt'.
   \end{eqnarray}
  Its dependence on time
  calculated for the $^{48}$Ca+$^{154}$Sm reaction at the
  excitation energy $E^*_{CN}=49$ MeV is presented in  Fig. \ref{Pcntfig}.
  Its asymptotic value is close to the values of $P_{\rm CN}$ calculated
   by the branching ratio (\ref{Pcn}) of the level densities.
   Our experience shows that the method of calculation presented
    in Section Appendix A allows us to include the peculiarities
    of the driving potential and dependence on the angular momentum
    of the DNS more evidently. As a result the effects of the
    entrance channel on the fusion probability appear more precisely.

%
%

\section{About the $W_{sur}$ survival probability and ER residual nuclei formation}

Another important problem in the analysis of  the experimental data
is related with the determination of the reliable $B_{\rm fis}$ fission barrier values of excited nuclei reached along the deexcitation cascade of CN in order to correctly estimate the overall $W_{\rm sur}(E^*_{\rm CN})$ survival probability against fission which is used to calculate the total ER cross section vs $E^*_{\rm CN}$.
Indeed, the experimental determinations of the ERs values may be  uncertain
whether any of $\alpha$-decay lifetime of reached nuclei along the deexcitation cascade is less than few $\mu$s. The uncertainty in determination of the ERs values appears even more strongly
 when the process of the deexcitation of CN by the evaporation of the charged particles
is neglected in comparison with the evaporation of neutral particles. In fact, in many heavy-ion reactions leading to formation of the
 heavy and superheavy compound nuclei, the total evaporation residue cross sections (when the charged particles are taken into account too) are much higher than the ones obtained for the evaporation of neutrons only.
 For example, the $\sigma_{\rm ER tot}/\sigma_{\rm ER-xn}$ ratio is  5--8 times for the  $^{26}$Mg+$^{248}$Cm and $^{36}$S+$^{238}$U reactions (leading to the same $^{274}$108 CN called $^{274}$Hs$^*$) and it can reach even 1 or 2 orders of magnitude for  some other reactions (see for example \cite{PRC91,APPB2015}). Therefore, neglecting the contribution of the charged particles in the determination of evaporation residue nuclei without the possibility of knowing the effect on the final results is doubtful. We can conclude here that the complete evaporation residue cross section $\sigma_{\rm ER tot}$ can be determined correctly if
 there is a possibility of full detection of all the total evaporation residue nuclei in the reaction.

  But, practically, it is impossible to determine experimentally at each step (or even at the first step only) the probability of the deexcitation cascade
  from a nucleus with excitation  energy  $E^{*}$  emitting only $\nu$
   neutrons.
   In ref. \cite{love2015} the author attracts attention on the deduced value
  $\Gamma_{\rm n} / \Gamma_{\rm total}=0.89\pm0.13$ \cite{yanez2014} (in the  $^{26}$Mg+$^{248}$Cm reaction) for the first step of the $^{274}$Hs CN decay at $E^{*}_{\rm CN}$=63 MeV of excitation energy
 by measurement of the angular distribution of the neutrons
  associated with the fission fragments in the $^{26}$Mg+$^{248}$Cm reaction, while in our investigation \cite{aglio2015}  we obtain the value
    $\Gamma_{\rm n} / \Gamma_{\rm total}$ = 0.17    for the same reaction and conditions.
The author   \cite{love2015} concludes that a highly excited nucleus decays
 with the vanishingly small fission probability and emits more faster
a neutron rather than fission. But this statement in \cite{love2015}    is  clearly based on the information about the $B_{\rm fis}$ fission barrier of about 11 MeV at $E^{*}_{\rm CN}$=63 MeV as reported in Fig.~4 of \cite{yanez2014} for the investigated reaction $^{26}$Mg+$^{248}$Cm.
  In fact, the value $B_{\rm fis}$ = 11 MeV is unjustifiable since the  macroscopic
   component of the fission barrier  is zero for the $^{274}$Hs CN and the
  microscopic  component  is 4.37 MeV (shell correction at $\ell$=0 and
  at ground state)   (see ref.\cite{Kowal2010});
   moreover, with the increase of the excitation energy $E^{*}_{\rm CN}$  and angular
  momentum $\ell$  decreases  the fission barrier of the $^{274}$Hs CN. Therefore,
 at  $E^{*}_{\rm CN}$=60 MeV the effective fission barrier of the $^{274}$Hs
  is lower than 1 MeV. Only assuming the dissipation coefficient $\gamma \simeq$18  \cite{krames1940} it is possible in principle to justify the needed delay of the fission process, but  nobody knows what mechanism, excited nuclear structure or
  large amplitude collective motion could produce  such a high viscosity for this $^{274}$Hs compound nucleus.

 Moreover, it is easy to prove the non physical consequence of the
 result $\Gamma_{\rm n} / \Gamma_{\rm tot}$ = 0.89 found \cite{yanez2014} at the first
 step of deexcitation cascade of the $^{274}$Hs
 CN at $E^{*}_{\rm CN}$ = 63 MeV.
 In fact, if this is true then the following
 $^{273}$Hs$^{*}$, $^{272}$Hs$^{*}$...$^{267}$Hs$^{*}$ excited hassium isotopes
 are reached after neutron emission along the deexcitation cascade of
 the $^{274}$Hs$^{*}$ CN and their shell corrections will be higher than the one
  of $^{274}$Hs$^{*}$ (see for example the tables in
 \cite{MolNix95,MuntSob66,Kowal2010,Mol2009}) due to the decrease
 of damping at the cooling the excited hassium isotopes after neutron emission.
 This circumstance  leads to the conclusion that at each step of the deexcitation
 cascade of the CN the fission barrier $B_{\rm fis}$ increases and the
 $\Gamma_{\rm n} / \Gamma_{\rm tot}$ ratio for each reached intermediate excited hassium
  isotopes ($^{273}$Hs$^{*}$, $^{272}$Hs$^{*}$...$^{267}$Hs$^{*}$..) must be larger
  than the one determined by the authors \cite{yanez2014} at the first neutron
  evaporation of the $^{274}$Hs CN.
  Therefore, since the $\sigma_{\rm fus}$ fusion
  cross section at $E^{*}_{\rm CN}$=63 MeV is about 0.12$\times \sigma_{capture}$
 (where the $\sigma_{\rm capture}$ capture cross section is about $10^{3}$ mb),
 starting from the $\Gamma_{\rm n} / \Gamma_{\rm tot}$ value of 0.9  determined \cite{yanez2014} at first step of neutron
 evaporation of CN, one should find a value of the total ER$_{\rm xn}$ cross section of
 about $10^{-3}$ mb, instead  the experimental value found
 for the evaporation residue cross sections in the $^{26}$Mg+$^{248}$Cm reaction  is some picobarn \cite{dvorak2006}. In fact, we find for the $\Gamma_{\rm n} / \Gamma_{\rm tot}$ ratio the value of 0.17
 \cite{aglio2015} at first step of neutron evaporation from the $^{274}$Hs CN at
  $E^{*}$=63 MeV; this value is consistent with the other following $\Gamma_{n} / \Gamma_{tot}$
 ratios along the deexcitation cascade of the compound nucleus (see Fig.~10 of paper \cite{aglio2015}). Therefore, our present value of  $W_{\rm sur}=6\times 10^{-14}$
  for the complete survival probability is consistent with the measured \cite{dvorak2006} total ER cross section of some  pb after neutron emission only.
Consequently, the $P_{\rm CN}$ fusion probability vs $E^{*}_{\rm CN}$ found in
 \cite{AIP2006,ItNPA2007} for the $^{26}$Mg+$^{248}$Cm reaction, and the $W_{\rm sur}$  survival probability found by \cite{yanez2014} at the first  neutron emission from the $^{274}$Hs CN  with $E^{*}$=63 MeV of excitation energy are clearly inconsistent
 because the combination of these results at first step of deexcitation cascade  with the following steps od the deexcitation  can not be in agreement with the experimental determination  \cite{dvorak2006} of the total ERxn cross section of about 1 pb.

The evaporation residue cross sections at the given values of
the CN excitation energy $E_{x}^*$ at each step $x$ of the
deexcitation cascade by the advanced statistical model \cite{aglio2012}
\begin{equation}
\sigma_{\rm ER}^{(x)}(E^*_{x})=\Sigma^{\ell_d}_{\ell=0}
(2\ell+1)\sigma_{\rm ER}^{(x)}(E^*_{x},\ell),
\end{equation}
where $\sigma_{\rm ER}^{(x)}(E^*_{x},\ell)$ is the partial cross section of ER
formation obtained  after the emission
of particles $\nu(x)$n + $y(x)$p + $k(x)\alpha$ + $s(x)$
(where $\nu(x)$, $y$, $k$, and $s$ are numbers of neutrons, protons,
$\alpha$-particles, and $\gamma$-quanta)  from the  intermediate
nucleus with excitation energy $E^*_{x}$   at each step $x$
of the deexcitation cascade by the formula
(for more details, see papers \cite{FazioMPL2005,aglio2012,PRC72}):
\begin{equation}
\label{SigmaEN}
\sigma_{\rm ER}^{(x)}(E^*_{x},\ell)=
\sigma^{(x-1)}_{\rm ER}(E^*_{x-1},\ell)W^{(x)}_{\rm sur}(E^*_{x-1},\ell).
\end{equation}
At $x=1$ we deal with the partial fusion cross section:
$\sigma^{(0)}_{\rm ER}({  E^*_{0},\ell)}=\sigma_{\rm fus}(E^*_{\rm CN},\ell)$.
In Eq. (\ref{SigmaEN}), $\sigma_{\rm ER}^{(x-1)}(E^*_{x-1},\ell)$
 is the partial cross section
of the intermediate excited nucleus formation at the $(x-1)$th step, and
$W^{(x)}_{\rm sur}$ is the survival probability of the
$x$th intermediate nucleus against fission along all steps of the
deexcitation cascade of the CN.

 In calculation of the $W_{\rm sur}^{(x)}(E^*_{x-1},\ell)$ the fission barrier is used
 as a sum of the parametrized macroscopic fission
barrier $B_{fis}^{m}(\ell)$  depending on the angular momentum $\ell$ \cite{SierkPRC33}
 and  the microscopic (shell) correction $\delta W = \delta W_{sad} -\delta W_{gs}$ due to shell effects;
 by considering the large deformation of a fissioning nucleus at the saddle point, $\delta W_{sad}$ is much smaller than the $\delta W_{gs}$ value and the microscopic shell correction $\delta W$ to the fission barrier
 can be expressed by the relation
 $\delta W \cong -\delta W_{\rm gs} $.
 Therefore, an effective fission barrier, as a function of $\ell$ and $T$ for each excited nucleus formed at various steps along the deexcitation cascade of CN,
 is calculated by the expression
\begin{equation}
\label{fissb} B_{\rm fis}(\ell,T)=c \ B_{fis}^{m}(\ell)-h(T) \ q(\ell) \ \delta
W,
\end{equation}
where the factor $c$ was set to 1 in all our calculations and
$h(T)$ and $q(\ell$) represent the damping functions of the nuclear
shell correction (usually it is $\delta W < 0$) with the increase of the
excitation energy $E^{*}$  and $\ell$ angular momentum, respectively \cite{aglio2012}:
\begin{equation}
h(T) = \{ 1 + \exp [(T-T_{0})/ d]\}^{-1}
\label{hoft}
\end{equation}
and
\begin{equation}
q(\ell) = \{ 1 + \exp [(\ell-\ell_{1/2})/\Delta \ell]\}^{-1},
\label{hofl}
\end{equation}
where, in Eq. (\ref{hoft}), $T=\sqrt{E^*/a}$ represents the nuclear
temperature depending on the excitation energy
$E^*$ and the level density parameter $a$, $d= 0.3$~MeV is the rate of
washing out the shell corrections with
the temperature, and $T_0=1.16$~MeV is the value at which the damping
factor $h(T)$
is reduced by 1/2. Analogously, in Eq. (\ref{hofl}), $\Delta \ell
=3\hbar$ is the rate of washing out the shell corrections with the angular momentum, and $\ell_{1/2}
=20\hbar$ is the value
at which the damping factor $q(\ell)$ is reduced by 1/2.

By regarding the determination of intrinsic level density $\rho_{\rm int}$
which takes into account the density of the intrinsic excitations,
the nucleus is considered as a system made up to non-interacting Fermi gases (proton gas and neutron gas)
at the same thermodynamic temperature $T$.
It is supposed that each of the two gases is in thermodynamic equilibrium and that
the excitation energy $E^{*}$ is distributed in a statistical way between two gases.
In this context for the intrinsic level density parameter $a$ we use the general expression \cite{Ignatyuk}
especially tailored to account for the shell effects in the level density

\begin{equation}
a(E^*)= \tilde{a}\left\{ 1+\delta W \left[\frac{1-exp(-\gamma E^*)}{E^*}\right]     \right\}
\label{level_density_a}
\end{equation}
where $\tilde a = 0.094 \times A$ is the asymptotic value that takes into account the dependence on the mass number $A$,
and $\gamma=$0.0064 MeV$^{-1}$ is the parameter which accounts for the rate at which shell effects wash out with
excitation energy for neutron or other light particle emission.
The general expression (\ref{level_density_a}) works well also for deformed prolate or oblate nuclei.
Physically, the disappearance of the shell effects with $E^{*}$ excitation energy may be seen as a rearrangement of the shell-model orbitals
in such a way that the shell gap between orbitals close to the Fermi energy vanishes.
The value of the $\gamma$ parameter was obtained \cite{Ignatyuk} by fitting the observed density of neutron resonances.

\begin{figure}[h]
\center
\includegraphics[scale=0.4]{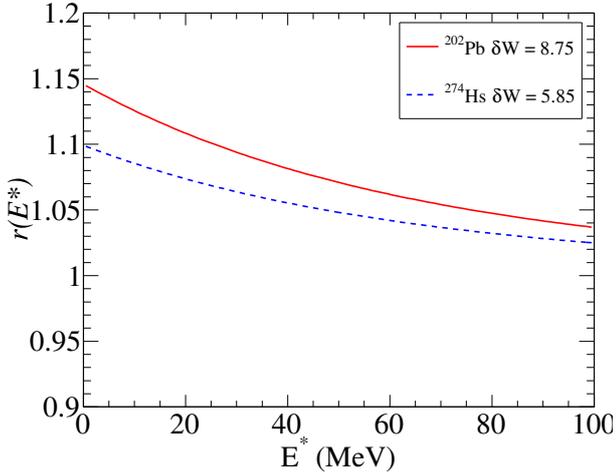}
\caption{(Color on-line) The $a_{fis}/a_{n}$ ratio vs $E^{*}_{\rm CN}$ for the two investigated reactions. Full line for the $^{48}$Ca+$^{154}$Sm reaction leading to the heavy $^{202}$Pb CN and dashed line for the $^{26}$Mg+$^{248}$Cm reaction leading to the superheavy $^{274}$Hs CN. \label{afanratio}}
\end{figure}

In order to determine the $a_{\rm fis}$ level density parameter in the
fission channel we use the relation $a_{fis}(E^*) =a_n(E^*)\times r(E^*)$ found in \cite{darrigo94} where $r(E^{*})$ is given by the relation

\begin{equation}
r(E^*)=\frac{\left[exp(-\gamma_{fis} E^*) - \left(1+\frac{E^*}{\delta W}\right) \right]}{\left[exp(-\gamma E^*) - \left(1+\frac{E^*}{\delta W}\right) \right]}
\label{ratio_a_parameters}
\end{equation}
with $\gamma_{fis}=0.024$~MeV$^{-1}$.
In Fig. \ref{afanratio} are reported, as an example,  the values of the $a_{fis}/a_{n}$ ratio versus $E^{*}$ for two investigated reactions: (a) for the  $^{48}$Ca+$^{154}$Sm reaction leading to the heavy $^{202}$Pb CN$^*$ (red full line); (b) for the $^{26}$Mg+$^{248}$Cm reaction leading to the superheavy  $^{274}$Hs
CN$^*$ (blue dashed line). At any excitation energy $E^{*}$ the $a_{fis}/a_{n}$ ratio is always greater than 1 and asymptotically  tends to unity with increasing the excitation energy $E^{*}$ at very high values.

\begin{figure}[ht!]
\resizebox{0.5\textwidth}{!}{\includegraphics{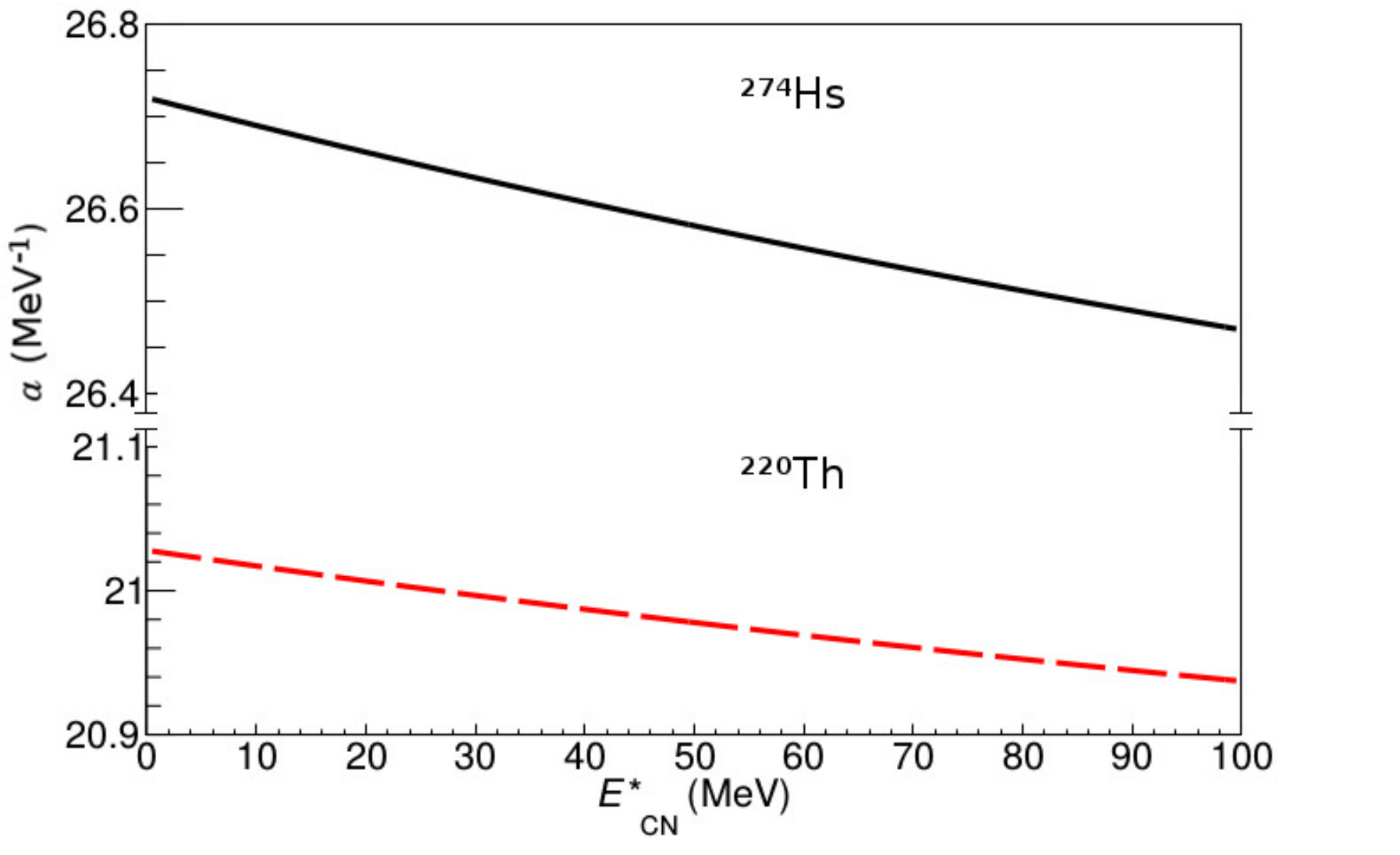}}
 \caption{Level density parameter \emph{a} vs $E^{*}_{\rm CN}$ for the $^{274}$Hs CN obtained by the $^{26}$Mg+$^{248}$Cm reaction (full line)
 and the $^{220}$Th CN obtained by the $^{16}$O+$^{204}$Pb reaction (dashed line).\label{lev_dens_par}}
\end{figure}

We stress that relation (\ref{ratio_a_parameters}) allows one to describe in a consistent approach including
collective effects the relevant functional form of the $a_{fis}(E^{*})/a_{n}(E^{*})$ ratio given by a general expression $r(E^{*})$,
rather than adjust by a phenomenological way the value of the cited $a_{fis}/a_{n}$ ratio for each excited nucleus.
This procedure allows the shell corrections to become sensitive to the excitation energy, as shown in Fig. \ref{lev_dens_par}.

To calculate the intrinsic level density $\rho_{int}(E^{*},\ell)$ we use the general expression

  \begin{equation}
  \begin{aligned}
  \rho_{\rm int}(E,J)= & \frac{1}{16\sqrt{6\pi}}{\left[\frac{\hbar^{2}}{\mathcal{J}_{\parallel}}\right]}^{1/2} a^{-1/4} \\
  & \times\sum_{k=-J}^J [E-E_{\rm rot}(k)]^{-5/4}e^{2\{a[E-E_{\rm rot}(k)]\}^{1/2}}
  \label{lev_dens}
  \end{aligned}
  \end{equation}

  where is

  \begin{equation}
  E_{\rm rot}(k) =\frac{\hbar^2}{2\mathcal{J}_{\perp}}J(J+1)+\frac{\hbar^2K^2}{2}\left[\frac{1}{\mathcal{J}_{\parallel}} -\frac{1}{\mathcal{J}_{\perp}}  \right]
  \label{ene_rot}
  \end{equation}

and where $\mathcal{J}_{\perp}$ and $\mathcal{J}_{\parallel}$  are  moments of inertia perpendicular and parallel to the symmetry axis
and $K$ is the projection of the total spin $J$ on the quantization axis.
Application of the general expression \cite{Ignatyuk} depends on the particular case. Specific cases take into account: the nucleus at the saddle point,
the case of yrast state, and prolate or oblate or triaxial shape. This expression of $\rho_{\rm int}$ works well for both deformed and spherical nuclei as for example nuclei very close to the shell closure.
\textit{The collective level density $\rho_{coll}$ calculated in the adiabatic approach, valid at low excitation energies}, takes into account in addition to the intrinsic excitations
also the rotational and vibrational excitation states by the collective enhancement factor $K_{\rm coll}(E^{*})$:

\begin{equation}
 \rho_{\rm coll}^{\rm adiab}(E^{*},J)=\rho_{\rm int}(E^{*},J)\times K_{\rm coll}^{\rm adiab}(E^{*}) \notag
\end{equation}

where $K_{\rm coll}^{\rm adiab}(E^{*})$ is given by the simple multiplication of the two $K_{\rm rot}^{\rm adiab}(E^{*})$ and $K_{\rm vibr}^{\rm adiab}(E^{*})$ enhancement factors;
therefore, $\rho_{\rm coll}^{\rm adiab}(E^{*},J)$ is determined (for more details see Appendix B) as

\begin{equation}
 \rho_{\rm coll}^{\rm adiab}(E^{*},J)=\rho_{\rm int}(E^{*},J)\times K_{\rm rot}^{\rm adiab}(E^{*})\times K_{\rm vibr}^{\rm adiab}(E^{*}).
 \label{rho_adiabatic}
\end{equation}

In Appendix B we give many other details regarding the intrinsic $\rho_{int}$ and collective $\rho_{coll}$ level density determinations,
 the fission $\Gamma_{fis}$ and particle-x $\Gamma_{x}$ decay widths, and we show the sensitivity of the model on final reaction products
by using mass asymmetric and almost symmetric reactants in the entrance channel.

\begin{figure}
\resizebox{0.4\textwidth}{!}{\includegraphics{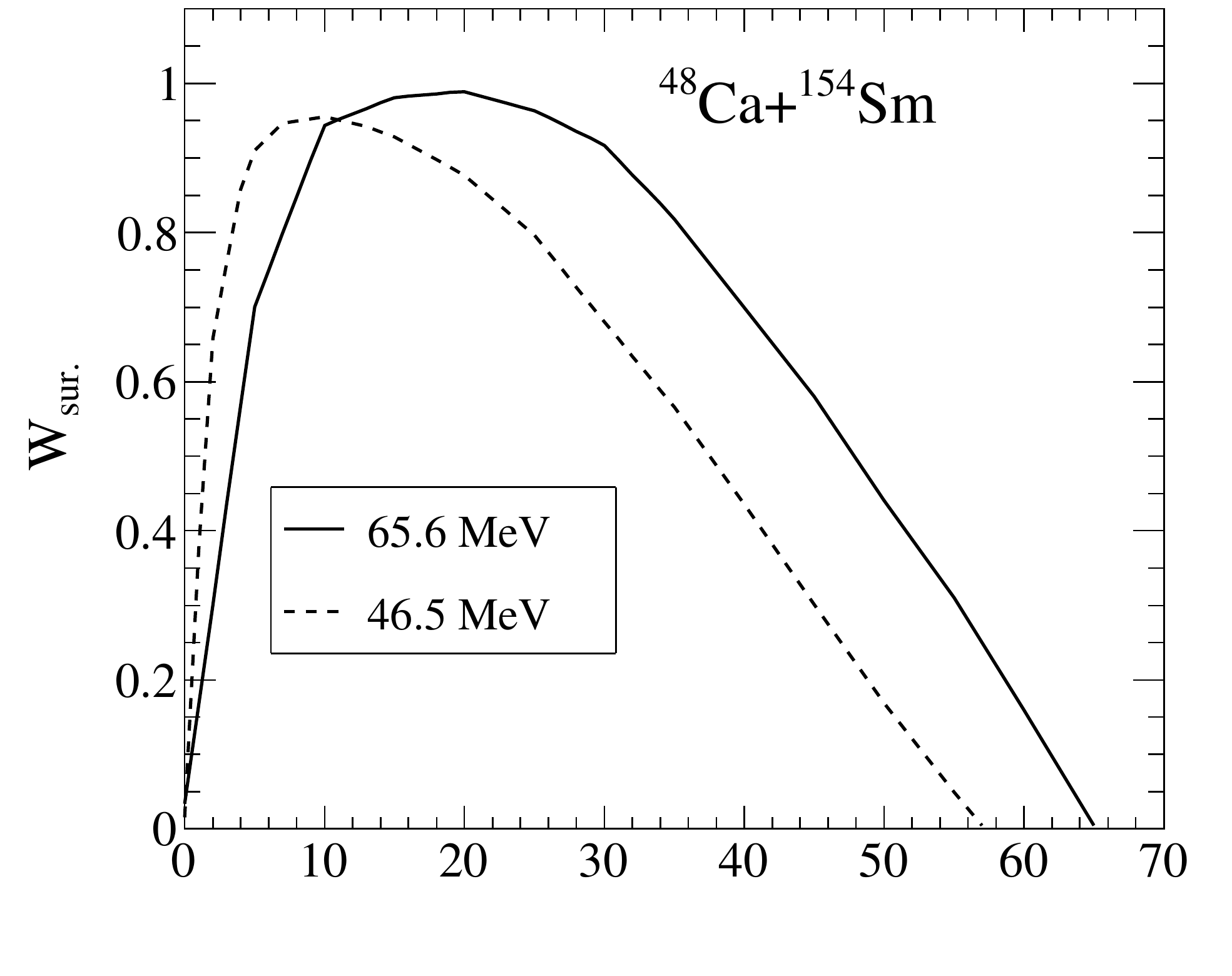}}
\put(-60,120){\large (a)}
\put(-100,5){\large $\ell (\hbar)$}\\
\resizebox{0.4\textwidth}{!}{\includegraphics{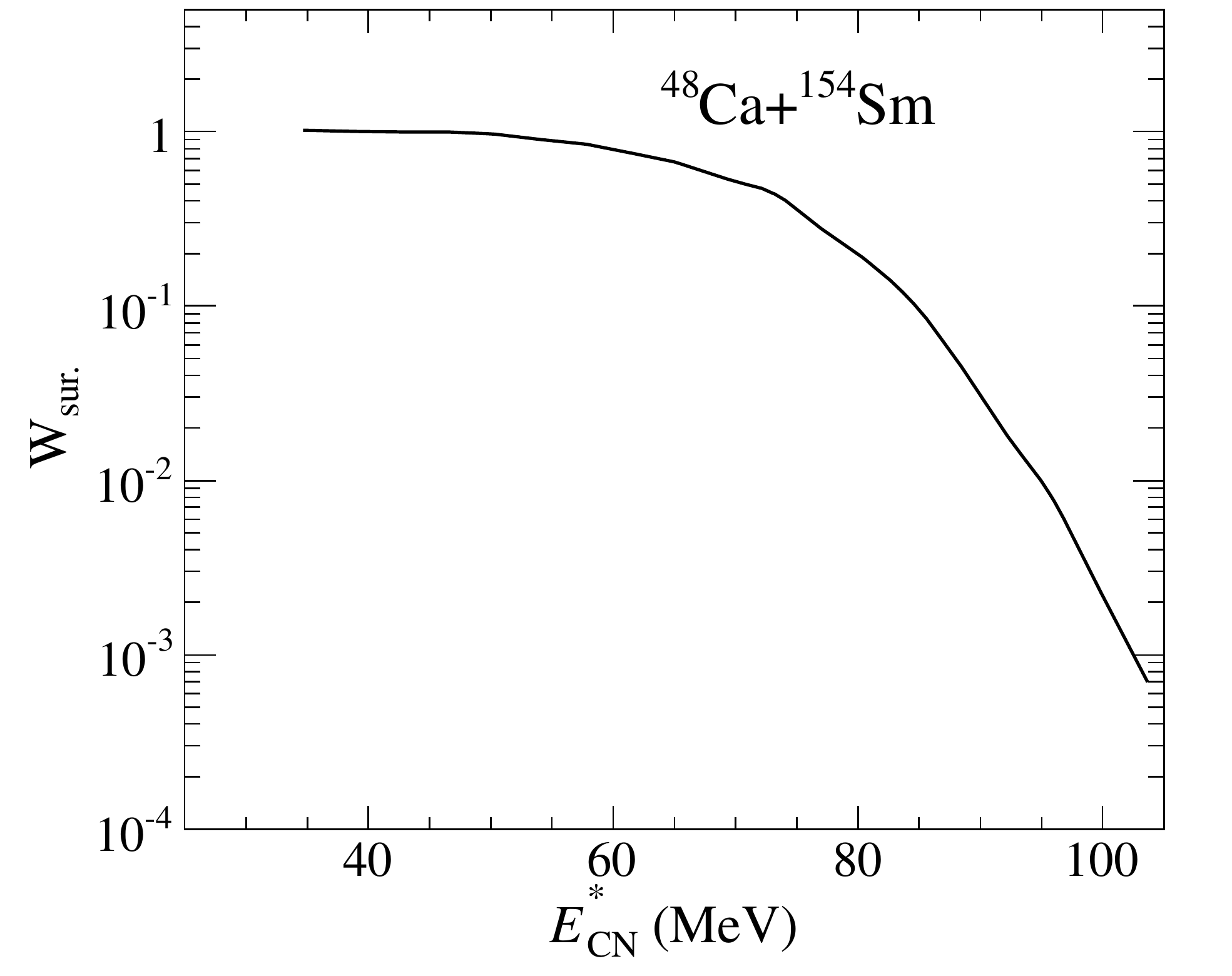}}
\put(-60,120){\large (b)}
\caption{(a) The $W_{sur}$  survival probability against fission of the deexcitation cascade of $^{202}$Pb CN as a function of the $\ell$  angular momentum for the $^{48}$Ca+$^{154}$Sm reaction at $E^{*}_{\rm CN}$=46.5 (dashed line) and 65.6 MeV (full line) excitation energy of CN; (b) the $W_{sur}$ survival probability against fission along the deexcitation cascade of $^{202}$Pb CN as a function of the $E^{*}_{\rm CN}$ excitation energy for the same reaction. \label{Wvsl-WvsE}}
\end{figure}

Moreover, if the capture of projectile by target takes place and complete fusion stage is reached,
for the rotating mononucleus the fission barrier disappears
at $\ell > \ell_{cr}$ (where $\ell_{\rm cr}$ is a critical value characteristic for each nucleus) due to the damping of shell correction with angular momentum $\ell$ by the $q(\ell)$ function.
To demonstrate the result of this effect, as an example, we present in Fig.~\ref{Wvsl-WvsE}(a) the $W_{\rm sur}$ survival probability vs $\ell$ regarding the deexcitation at first step of the $^{202}$Pb CN
formed in the $^{48}$Ca+$^{154}$Sm reaction, at two different values of $E^{*}_{\rm CN}$ excitation energy of 46.5 and 65.6 MeV.
This figure shows the sensitivity and importance of the $\ell$ angular momentum range on the $W_{\rm sur}$ surviving probability at the first step of deexcitation of the $^{202}$Pb CN,
and how its influence changes with increasing excitation energy.
Therefore, the approximation often used in calculations of the fission barrier $B_{\rm fis}$ and $W_{\rm sur}$ surviving probability to fission for $\ell$=0 only leads to an insufficient ERs determination.
Moreover, we present in Fig.~\ref{Wvsl-WvsE} (b) $W_{sur}$ vs $E^{*}_{\rm CN}$ for the same $^{48}$Ca+$^{154}$Sm reaction leading to the $^{202}$Pb CN.
Even in this case, it is easy to observe the sensitivity of the $W_{sur}$ function with the $E^{*}_{\rm CN}$ excitation energy:
at high excitation energies ($E^{*}_{\rm CN}>$70 MeV) the $W_{sur}$ value changes more than one order of magnitude with the change of the $E^{*}_{\rm CN}$ value of about 14 MeV
and more than two orders of magnitude with the change of $E^{*}_{\rm CN}$ of about 23 MeV.

\begin{figure}[h!]
\resizebox{0.4\textwidth}{!}{\includegraphics{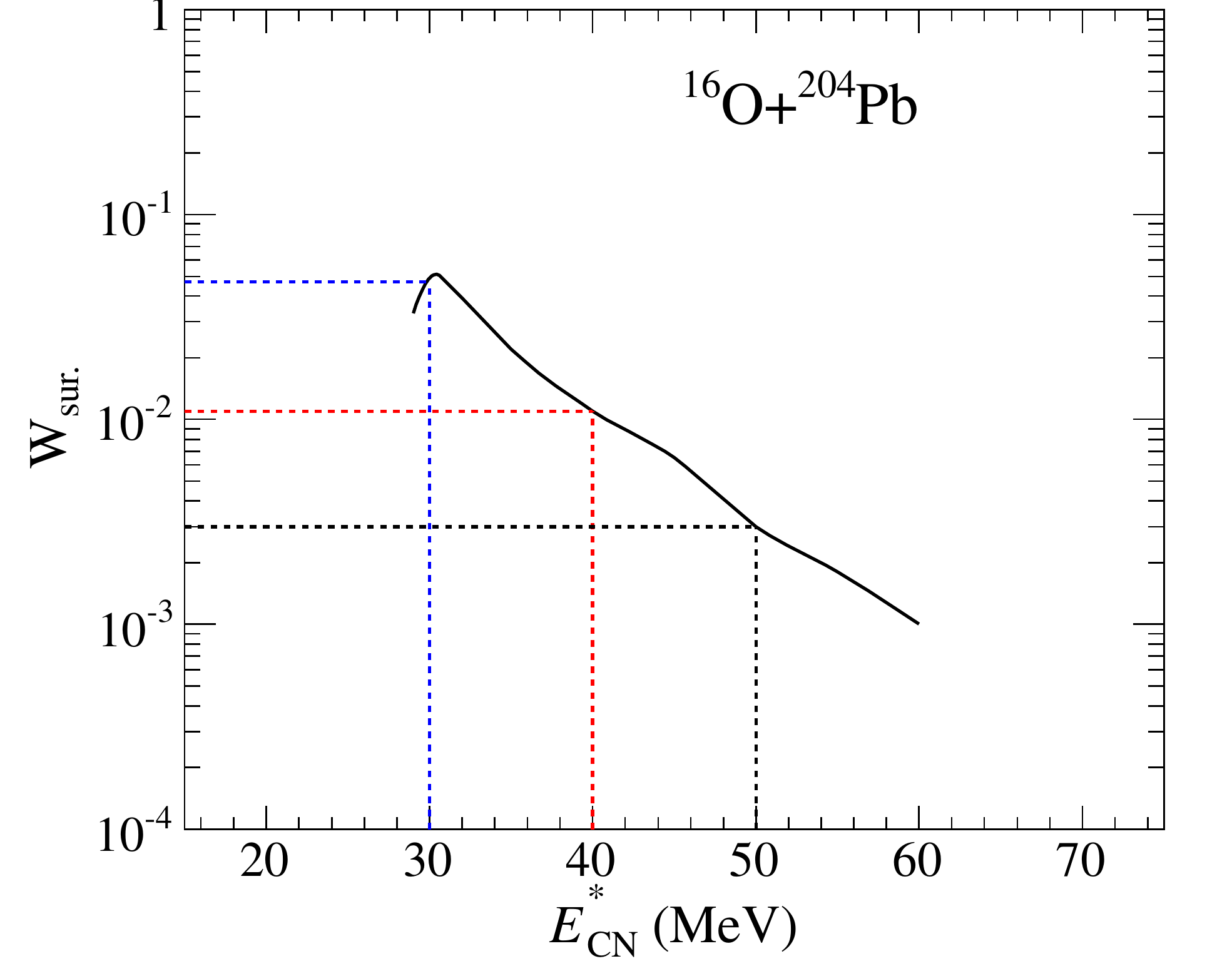}}
\put(-60,120){\large (a)}\\
\resizebox{0.4\textwidth}{!}{\includegraphics{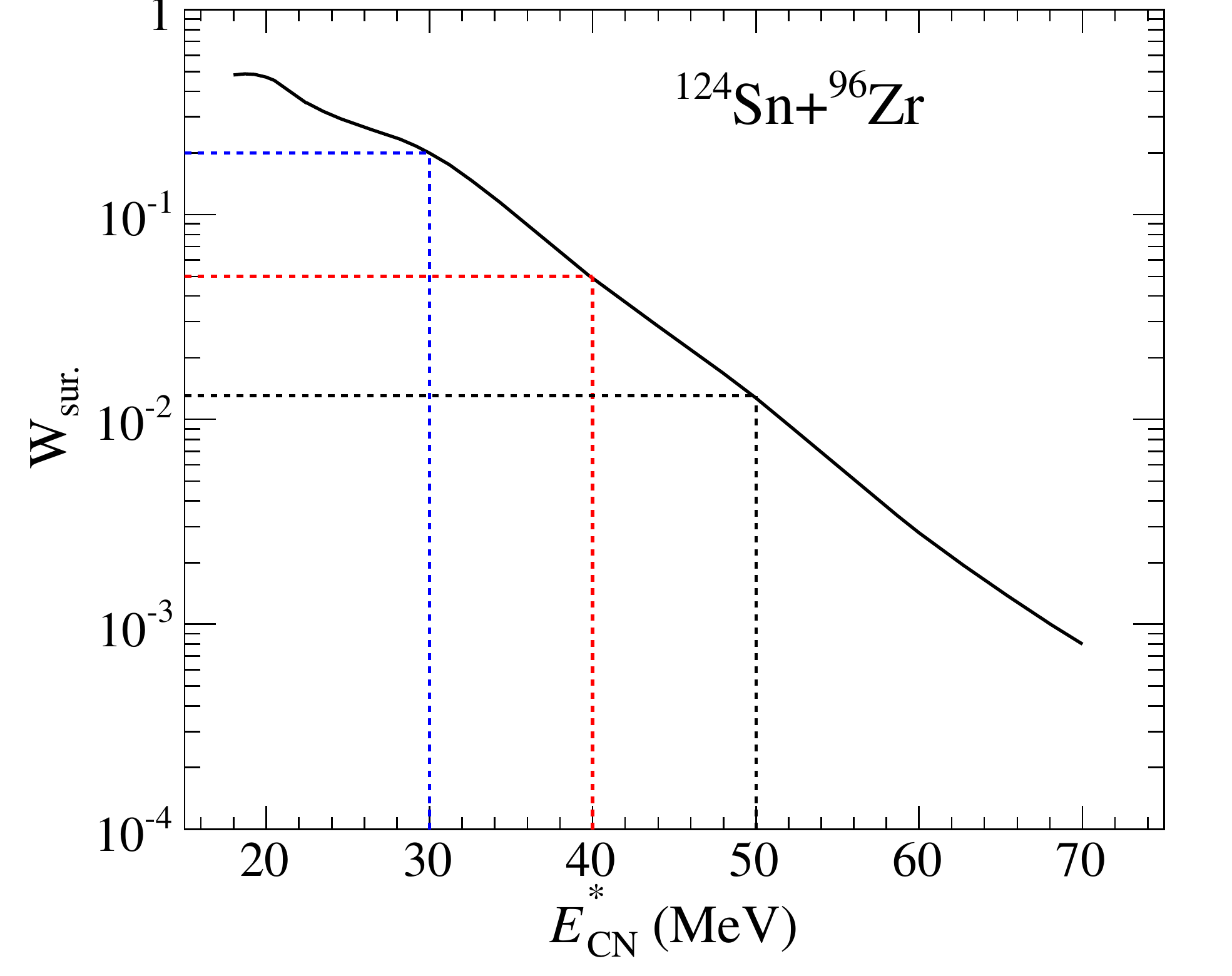}}
\put(-60,120){\large (b)}
\caption{(Color on-line) (a) As Fig.~\ref{Wvsl-WvsE} (b) but for the $^{16}$O+$^{204}$Pb reaction leading to the $^{220}$Th CN; (b) As Fig.~\ref{Wvsl-WvsE} (b) but for the $^{124}$Sn+$^{96}$Zr reaction leading to the same $^{220}$Th CN. \label{WvsETh220}}
\end{figure}

In addition, the panels (a) and (b) of Fig.~\ref{WvsETh220} show the $W_{\rm sur}$ survival probability excitation functions for the $^{220}$Th CN
formed by the $^{16}$O+$^{204}$Pb very asymmetric reaction and $^{124}$Sn+$^{96}$Zr almost symmetric reaction, respectively.

By comparing the $W_{\rm sur}$ values obtained for two very different entrance channels leading to the formation of the $^{220}$Th CN  at excitation energies $E^{*}_{\rm CN}$=30, 40 and 50 MeV,
it is possible to observe the different effects of the angular momentum distribution on the surviving probability excitation functions. The $W_{\rm sur}$ values for the CN formed in
the almost symmetric $^{124}$Sn+$^{96}$Zr reaction are greater than the $W_{\rm sur}$
 values corresponding to the one in the very asymmetric $^{16}$O+$^{204}$Pb
 reaction by factors of 3.5, 4.2, and 4.5 times at the
  $E^{*}_{\rm CN}$=30, 40 and 50  MeV, respectively. Certainly the increase of
  the beam energy $E^{*}_{\rm c.m.}$  leads to an increase of the excitation energy
  $E^{*}_{\rm CN}$ of CN and to an extension of the angular momentum distribution
  of the CN formed in these reactions. But the behavior of the extension
  of the angular momentum distribution is different for the two reactions.
The $\ell$ angular momentum range is larger for the $^{220}$Th CN obtained in the very asymmetric
$^{16}$O+$^{204}$Pb  reaction than for the CN obtained in the almost symmetric
$^{124}$Sn+$^{96}$Zr  reaction at the same considered $E^{*}_{\rm CN}$ excitation energy. This difference in the angular momentum $\ell$ interval increases
with the increase of $E^{*}_{\rm CN}$ since the size of the potential well in the
nucleus-nucleus interaction for the almost symmetric $^{124}$Sn+$^{96}$Zr
 reaction  is smaller than the one for the very asymmetric reaction like $^{16}$O+$^{204}$Pb. Therefore, the number of the angular momentum $\ell$
  contributing to fusion in the $^{124}$Sn+$^{96}$Zr reaction is smaller.
  Moreover, the fusion probability $P_{\rm CN}$ strongly decreases by increasing
 the angular momentum   due to the increase of the $B^{*}_{fus}$ intrinsic fusion
  barrier and due to a decrease of the $B_{qf}$ quasifission barrier for the symmetric
 reaction (see for example ref. \cite{FazioJPSJ72}). The quasifission barrier is
 the depth of the potential well in the nucleus-nucleus interaction (see Fig. \ref{VintQgg}).

  Obviously, the fusion cross section at a considered $E^{*}_{\rm CN}$
 value of CN for the $^{124}$Sn+$^{96}$Zr symmetric reaction is smaller than the one
  of the $^{16}$O+$^{204}$Pb very asymmetric reaction, but with a smaller $\ell$
  interval of the formed CN by the symmetric reaction has a greater $W_{sur}$
   survival probability to fission in comparison with the very asymmetric reaction.

   Moreover, as discussed in our paper \cite{APPB2015}, it is difficult in experiment
    to estimate the $\sigma_{\rm ERtot}$ cross section when the emission of charged
    particles present also, because not all of residue nuclei can be identified.   Therefore, apart from the uncertainties which are inherent to the theoretical
     predictions, there are ambiguities in the estimation of the fusion cross
     sections by the analysis of  experimental data.

\section{Conclusion}

The reasons leading to uncertainties of the experimental and theoretical  values
of the fusion probability $P_{\rm CN}$ in heavy ion collision at lower energies
are discussed.
 It should be stressed that there are two important reasons causing the uncertainties of the experimental values $P_{\rm CN}$.
 The first reason is related to the
 ambiguity in identification of the reaction products formed by the true capture and
 fusion events.
In the analysis of experimental data with full momentum transfer, events with masses around the values of light initial nucleus and conjugate nucleus are not usually taken into consideration.
  Those events
 are considered as originated by the deep-inelastic collisions and this procedure
 of the analysis leads to a decrease the experimental value of the capture
 cross section $\sigma_{\rm cap}^{(exp)}$ and, consequently, to increase fusion probability $P_{\rm CN}$ since it inverse proportional to $\sigma_{\rm cap}^{(exp)}$.
 The authors of ref. \cite{knya2007} considered the reaction
 products with mass numbers $A<60$ as the ones of the deep-inelastic collisions
 and the capture events (characterized by the large energy dissipation
  and with a full momentum transfer) are missed. Therefore, the restriction
  of the mass range $60 \leq A \leq 130$ for the capture products is not completely
  correct because this assumption in the procedure of analysis of selection of experimental capture events leads to decrease the estimated true experimental capture cross sections.
  Therefore, the $P_{CN}$ fusion probability determined by the analysis of experimental events as the ratio between the fusion and capture cross sections is bigger than the true experimental value.
  As a result the experimental fusion probability $P_{\rm CN}$  reported  in Fig.~4 of ref. \cite{love2015} unreasonably appear to be much higher with respect to the various theoretical determinations obtained by different theoretical models.

 The second reason is the ambiguity in the separation of the fusion-fission events
 in the analysis of fission-like products containing quasifission or/and fast
 fission products. Therefore, the number of events seem to be larger than
  true fusion events due to consideration of the part of quasifission and fast
 fission events as fission events of compound nucleus which has not formed in the
 reaction. Certainly the extracted fusion probability $P_{\rm CN}$ will be larger
 than its correct value.

 In fact, in reactions leading to superheavy compound nuclei, the yields of the quasifission and fast fission products are much more than the ones
 due to the fusion-fission products and, besides, the mass and the angular distributions of the reaction fragments can be strongly overlapped.

 The authors of ref.\cite{knya2007} overestimated the fusion cross section by
  including quasifission events producing fragments with mass numbers in the
  range  $60 \leq A \leq 130$. So the second reason of ambiguity in the
  identification of the reaction products also leads to an increase of
  the fusion probability $P_{\rm CN}$ in Fig.~4 of ref. \cite{love2015}.
  The conclusion is that the good agreement between the experimental data
  and their theoretical description by the calculations of ref. \cite{zagre2008}
  does not mean the success in study of the fusion-fission mechanism in the
  $^{48}$Ca+$^{154}$Sm reaction. This question is still open and it must
  be studied by both the experimental and theoretical methods.

  In order to check the reliability of an experimental result, it would be good to be
  able to directly compare the deviations of the final results when they are made vary
  in a controlled manner the assumption made at the beginning of the analysis procedure.
  This methodology is widely used in other field of research, it is not however well practised in the complex study of reaction dynamics between heavy ions.
 Analogously, we demonstrate by figs.~\ref{Wvsl-WvsE} and \ref{WvsETh220} the strong sensitivity of the $W_{\rm sur}$ surviving probability
   excitation function with the $E^{*}_{\rm CN}$ excitation energy and with the
  angular momentum $\ell$ values. In addition, due to difference in the spin distribution probability of
  the heated and rotating nucleus the survival probability $W_{\rm sur}$ is strongly sensitive to  the kind of reactions in the entrance channel even if these reactions lead to the same CN formation
  with the same $E^{*}_{\rm CN}$ excitation energy.

 We have explained the complexity of the fusion-fission and evaporation
 residue formation starting from the DNS formation  in the entrance channel.
 It is important to take into account the role of the intrinsic fusion barrier
 $B^{*}_{\rm fus}$ and quasifission barrier  $B_{\rm qf}$ in the complete fusion/
 quasifission competition that  are sensitive to the DNS lifetime and angular
  momentum range.
 Moreover, the fast fission products caused by the decay of the complete fusion
  deformed mononucleus with high angular momentum values $\ell$
  (because $B_{\rm fis}=0$ for $\ell > \ell_{\rm cr}$) before  reaching the statistically equilibrated shape
 of CN, intensively  populate the symmetric mass distribution at higher
 $E^{*}_{\rm  CN}$ excitation energies.  Thus, these experimental determinations and
 extracted $P_{\rm CN}$ fusion probabilities appear strongly overestimated
 (see the experimental results in Fig.~1 taken from refs.\cite{AIP2006,ItNPA2007}).

On the other hand, it is not realistic to admit as a reliable result that the ratio between the $P_{\rm CN}$ values deduced from experimental observations
of the very asymmetric reaction $^{26}$Mg+$^{248}$Cm leading to CN with $Z_{\rm CN}=108$ ($P_{\rm CN}(Z=108)$) and the one deduced from the less
asymmetrical reaction $^{86}$Kr+$^{208}$Pb leading to CN with $Z_{\rm CN}=118$  ($P_{\rm CN}(Z=118)$), respectively, is  $P_{\rm CN}(Z=108)/P_{\rm CN}(Z=118) = 0.31\times 10^{-1}$,
while the analogous ratio $P_{\rm CN}(Z=108)/P_{\rm CN}(Z=118)$  between $P_{\rm CN}$ theoretical values is 0.23$\times 10^{-4}$ indicating a ratio between experimental and theoretical   values that is at least  3 orders of magnitude higher.
Similarly, the variation of $P_{\rm CN}$ deduced from the experiments for the  $^{48}$Ca+$^{208}$Pb ( $\eta=0.63$ and  $Z_{\rm CN}=102$)
and $^{58}$Fe+$^{248}$Cm ( $\eta=0.63$ and  $Z_{\rm CN}=122$) reactions, having the same asymmetry parameters $\eta$, is approximately one order of magnitude,
while the theoretical results indicate a change of about four orders of magnitude.
If these theoretical predictions were completely unreliable, then it remains incomprehensible why with an experimental  change of  $P_{\rm CN}$ of about one order of magnitude,
the cross sections of evaporation residues pass from values of $\mu$b to that of pb and also much less (about 10 fb) such as it is impossible to detect events of evaporation residues.

The characteristics of our model and procedure used for calculation are based on the possibility to
analyze the different evolution of various nuclear reactions
in the entrance channel with the use of one radius parameter $r_0$
to study  different reactions. The sensitivity of fusion probability is discussed in Appendix A of
this work.
Moreover, in the model all properties of reaction which are responsible for the evolution of reactants with formation of intermediate states and final products
(potentials, barriers, excitation function, reaction mechanisms, competition of processes, cross sections, etc.) are considered as dependent on the energy and angular momentum.
We have shown that the consistent description of the fission cross section reached by consideration of the fade-out of the shell correction to the fission barrier with increasing temperature
and angular momentum. This result is of crucial importance for the synthesis of the superheavy elements, since it extends the stabilizing effects of the shell structure to
higher temperatures, but this stabilizing effect, however, will be partially removed by the decrease of the shell correction with the increasing angular momentum.
Moreover, we note that the investigation of the temperature dependence of the shell correction might be extended to the analysis of the photofission reactions,
in which the angular momentum effects are practically absent. We also have given in the paper a general expression allowing for the determination of the $a_{\rm fis}(E^{*})/a_{\rm n}(E^{*})$
ratio, valid both for spherical and deformed nuclei, and we discussed about the role and modalities of the enhancement factors in the level density at lower and higher excitation energies.
Therefore, the sensitivity and reliability of our modular system of nuclear reaction codes, starting from the contact of reactants in the entrance channel to the formation of final products, have been shown in detail
in order to take into account various reaction mechanisms present at different steps of reaction characterized by different entrance channels.
This means that all properties of reacting nuclei are considered, the orientation angle between the symmetry axes of deformed reactants are taken into account
in order to estimate the real Coulomb barrier between reacting nuclei as a function of the beam energy at the stage of contact.
Moreover, the role of the driving potential as a function of energy and angular momentum is considered, the dependence of the intrinsic fusion barrier $B^{*}_{\rm fus}$(E$^{*},\ell$)
and the quasifission $B_{\rm qf}$(E$^{*},\ell$) barrier are detailed in order to calculate the fusion probability $P_{\rm CN}$, the fusion and quasifission cross sections;
moreover, the complete deexcitation cascade of CN is analyzed in order to calculate at each step the fission and ER cross sections where the fission barrier and the shell effects
are determined by using the damping functions h(T) and q($\ell$) in the competition between light-particle emission and fission processes when the light charged particles are also considered.
Therefore, such our modular system of codes for the study nuclear reactions also represents a powerful predictive theoretical way for new investigations also giving the limits of the reliable expectation.

 Instead, in our conclusion, we affirm that the desire to find a phenomenological model or a simple theoretical model in order to have a detailed knowledge about the reaction dynamics in heavy ion collisions
  and the clear characteristics of the reaction products is a vain hope.
It was believed that it would be enough to make the so-called ``reasonable'' assumptions in treatment of data or in application of models with the aim of simplifying the problem, but in reality the obtained results were strongly affected by large uncertainties as we have explained in ref.~\cite{APPB2015}. Therefore, a simplified and unsuitable model leads to an unhelpful information as it does not provide a realistic understanding of the phenomenon which one wants to study.

\appendix
\section{Procedures for determination of $P_{\rm CN}$}

  The ratio of the sum of the evaporation residue and fusion-fission cross sections
 to the capture cross section is used
  to extract the fusion probability from the experimental data of the reaction products
  (see expression (\ref{Pcneq}) in Introduction). It is clear
 that the experimental results are a sum of contributions from the reactions
 taking place in collisions with different values of the orbital angular momentum.
 Therefore, the fusion probability depends on the
 orbital angular momentum since the intrinsic fusion barrier $B^*_{\rm fus}$ and
 quasifission $B^*_{qf}$ barriers are its function.
 The partial capture cross section is calculated by the estimation of the range of the orbital
angular momentum leading to the full momentum transfer in the entrance
channel of collision.
 This procedure is realized by solution of the equations of motion
 for the relative distance between the centres-of-mass of colliding nuclei and
 orbital angular momentum with the radial and tangential friction coefficients.
 \begin{eqnarray}
\label{maineq} &&\mu(R)\frac{d\dot{R}}{dt} +
\gamma_{R}(R)\dot{R}(t)= F(R),\hspace{0.0cm}\\
\label{maineq2}
 F(R)&=&-\frac {(\partial V(R)+\delta V(R))}{\partial R}-
\dot{R}\hspace{-0.3mm}^2 \frac {\partial \mu(R)}{\partial R}\,,\hspace{0.5cm}\\
\label{maineq3}\frac{dL}{dt}&=&\gamma_{\theta}(R)R(t)\left(\dot{\theta}
R(t) -\dot{\theta_1} R_{1eff} -\dot{\theta_2} R_{2eff}\right), \\
L_0&=&J_R \dot{\theta}+J_1 \dot{\theta_1}+J_2 \dot{\theta_2}\,,\hspace{1.35cm} \\
E_{rot}&=&\frac{J_R \dot{\theta_{}}{}^2}2+\frac{J_1
\dot{\theta_1}^2}2+\frac{J_2 \dot{\theta_2}^2}2\,,\hspace{0.45cm}
\end{eqnarray}
where $R\equiv R(t)$ is the relative motion coordinate;
$\dot{R}(t)$ is the corresponding velocity; $L_0$ and $E_{rot}$ are
defined by initial conditions; $J_R$ and $\dot\theta$, $J_1$ and
$\dot\theta_1$, $J_2$ and $\dot\theta_2$ are  moment of inertia
and  angular velocities of the DNS and its fragments, respectively
($J_R, J_1$ and $J_2$ are defined in Ref. \cite{EPJA19}); $\gamma_{R}$ and
$ \gamma_{\theta}$ are the friction coefficients for the relative
motion along $R$ and the tangential motion when two nuclei roll on
each other's surfaces, respectively;  $V(R)$ is the
nucleus-nucleus potential which includes Coulomb, nuclear and
rotational potentials (see Eq.(A.1) Ref. \cite{NuclPhys05,EPJA19});
$\mu(R,t))$ is the reduced mass of the
system:
\begin{equation}
\label{massredu} \mu(R,t)=\tilde\mu(R,t)+\delta \mu(R,t)\, ,
\end{equation}
where
$$\tilde\mu(R,t)=m_0 A_T(R,t)\cdot A_P(R,t)/(A_T(R,t) +
A_P(R,t)),$$ at $t=0$ $A_T(R)$ and $A_P(R)$ are equal to mass
numbers of the target- and projectile-nucleus, respectively; $m_0$
is the nucleon mass. The time dependencies of
$A_P(t)=Z_P(t)+N_P(t)$ and $A_T(t)=Z_T(t)+N_T(t)$ are found  by
solution of master equation for the evolution of occupation
numbers of single-particle states in nuclei as in \cite{NuclPhys05};
$\delta V(R)$ and $\delta \mu(R,t)$  are changes of the interaction potential
$V(R)$ and reduced mass $\mu$, respectively, during interaction due to
nucleon exchange and overlap of nucleon densities of interacting
nuclei (see Ref. \cite{EPJA19});
$$ R_{1(2)eff}=\frac{R_{01(02)}}{R_{01}+R_{02}}R\,\,,$$
where  $R_{01(02)}$ is the nucleus equilibrium radius: $R_{0i}=r_0
A_i^{1/3}, r_0=1.18$ fm.

\begin{figure}[h!]
 \centering
 \resizebox{0.45\textwidth}{!}{\includegraphics{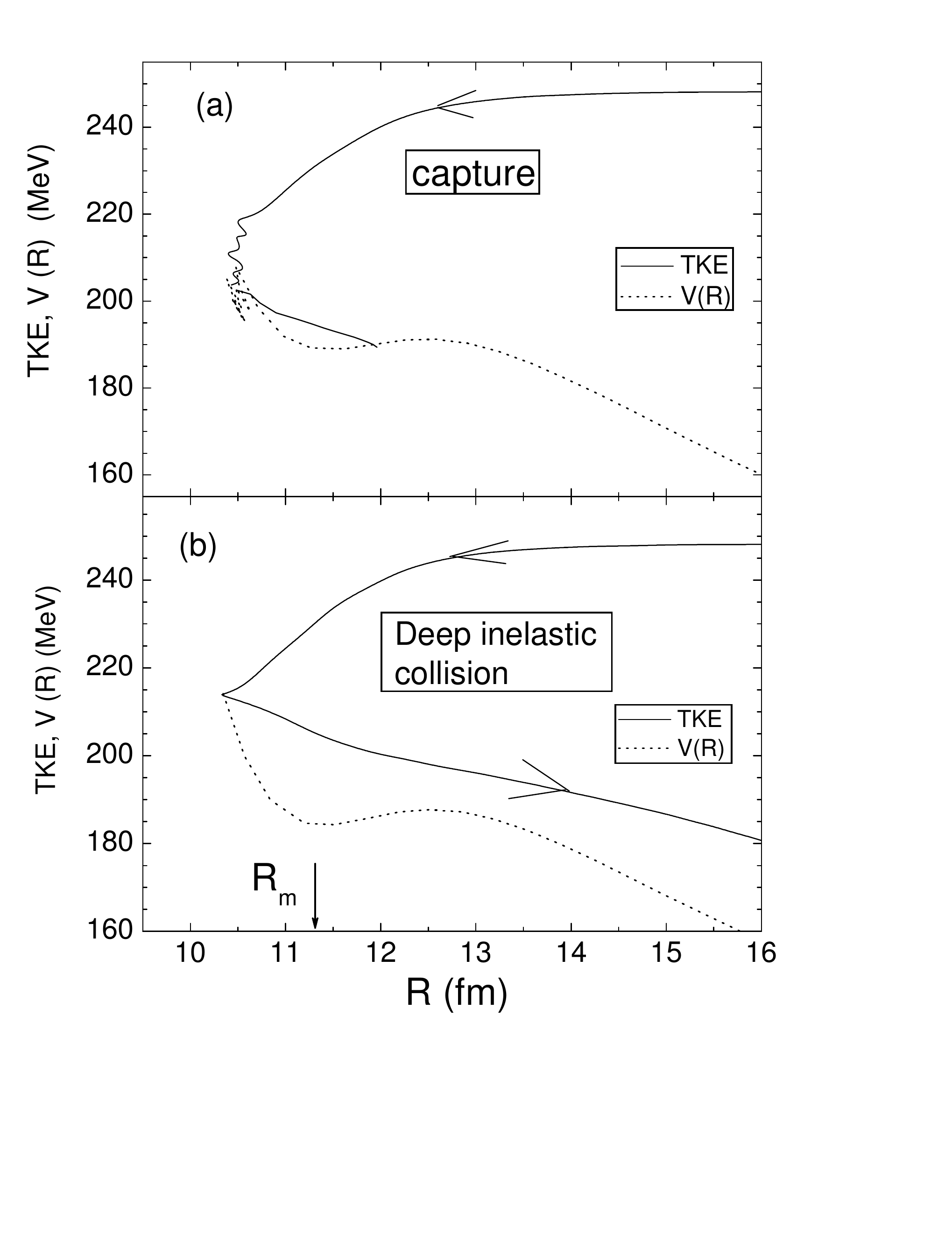}}
\vspace{-2.0cm}
 \caption{The difference between capture (a) and deep inelastic collision (b)
caused by the dependence of the dissipation of the total kinetic energy
 of the relative motion and the nucleus-nucleus potential on the
 radial distance and  orbital angular momentum $L$
 for the $^{48}$Ca+$^{208}$Pb reaction at heavy ion collisions with
 $L_0=40\hbar$ (a) and  20$\hbar$ (b) at $E_{\rm c.m.}$=248 MeV.
\label{TrapWell}}
 \end{figure}

 The use of the friction coefficients
 related with the excitation of intrinsic degrees of freedom allows us
 to separate trajectories of the deep-inelastic collisions and full momentum transfer
 reactions (see Fig. \ref{TrapWell}).

 The partial capture cross section is determined by the capture
probability ${\cal P}^{(\ell)}_{cap}(E)$ which means that the
colliding nuclei are trapped into the well of the nucleus-nucleus
potential after dissipation of a  part of the initial kinetic
energy and orbital angular momentum:
 \begin{equation}
 \label{parcap}
 \sigma^{(\ell)}_{cap}(E,\alpha_1,\alpha_2)=\pi{\lambda\hspace*{-0.23cm}-}^2
{\cal P}_{cap}^{(\ell)}(E,\alpha_1,\alpha_2)
 \end{equation}

Here ${\lambda\hspace*{-0.23cm}-}$ is the de Broglie wavelength of the entrance
 channel.  The capture probability ${\cal P}_{cap}^{(\ell)}(E,\alpha_1,\alpha_2)$  is
  equal to 1 or 0 for the given beam energy and orbital angular momentum.
  Our calculations showed that in dependence on the beam energy, $E=E_{c.m.}$,
  there is a window for capture as a function of orbital angular momentum ($\alpha_1$ and $\alpha_2$
  are omitted here for the simplicity of the formula):
  \[{\cal P}^{\ell}_{cap}(E) = \left\{ \begin{array}{ll} 1,  \hspace*{0.2 cm}  \rm{if}
\ \  \ell_{min}<\ell<\ell_d \ \
\rm{and} \ \ {\it E>V}_{Coul}
  \\ 0, \hspace*{0.2cm} \rm {if}\ \ \ell>\ell_d  \ \
  or \ \   \ell<\ell_{min} \  \rm {and} \ \
  {\it E>V}_{Coul}
  \\ 0,  \hspace*{0.2cm} \rm{for \  all} \ \ell \ \
     \hspace*{0.2cm} \rm{if} \ \
  {\it E\leq V}_{Coul}\:,
 \end{array}
 \right.
 \]
  where $\ell_{min}\ne0$ can be observed  when the beam energy
  is  large  than the Coulomb barrier ($V_{Coul}$).

While exists the DNS formed at capture, we have an ensemble \{$Z$\} of the DNS
configurations which contributes to the competition between
complete fusion and quasifission with probabilities \{$Y_Z$\}.
 The dependence of
  barrier $B^*_{\rm fus}$ and excitation energy of DNS  $E^*_{\rm Z}$
  for given charge $Z$ and mass $A$ on angular momentum and orientation angles
  $\alpha_i$ ($i=$1,2) of the symmetry axis of interacting nuclei
  is connected with the method of calculation  of the interacting potential between nuclei of DNS which is sensitive to those
  variables.
  Consequently, the fusion factor $P_{\rm CN}$  for the given reaction depends
  on the same variables through the charge distribution $Y_Z(E^*_Z)$ and fusion factor $P^{(Z)}_{\rm CN}$
   from charge asymmetry configuration $Z$:
\begin{equation}
\label{pcne} P_{\rm CN}(E^*_Z,\ell; \alpha_1,\alpha_2)=\sum_{Z_{sym}}^{Z_{max}}
Y_Z(E^*_Z,\ell)P^{(Z)}_{\rm CN}(E^*_Z,\ell; \alpha_1,\alpha_2),
\end{equation}
where $P^{(Z)}_{\rm CN}(E^*_Z,\ell; \alpha_1,\alpha_2)$ is the fusion probability
for DNS having excitation energy $E^*_Z$ at charge asymmetry
$Z$ and orientation angles of symmetry axis of its fragments
are equal to $\alpha_1$ and $\alpha_2$.
The evolution of $Y_Z$  is calculated by solving the
transport master equation:
\begin{eqnarray}
\label{massdec}
&&\frac{\partial}{dt}Y_{Z}(E^*_Z(\ell),t)=\Delta^{(-)}_{Z+1}
Y_{Z+1}(E^*_Z(\ell),t)+\nonumber\\
&&\Delta^{(+)}_{Z-1}  Y_{Z-1}(E^*_Z(\ell),t)
-(\Delta^{(-)}_{Z}+\Delta^{(+)}_{Z}+\Lambda^{\rm qf}_{Z})
Y_{Z}(E^*_Z(\ell),t), \nonumber\\
\\
&&\hspace*{0.1cm}\mbox{\rm for} \ Z=2,3,...,
Z_{tot}-2. \nonumber
\end{eqnarray}
Here, the transition coefficients of multinucleon transfer  are
calculated as in Ref.~\cite{Jolos86}
\begin{eqnarray}
\label{delt} \Delta^{(\pm)}_{Z}&=&\frac{1}{\Delta t}
\sum\limits_{P,T}|g^{(Z)}_{PT}|^2 \ n^{(Z)}_{T,P}(E^*_Z(\ell),t) \
(1-n^{(Z)}_{P,T}(E^*_Z(\ell),t)) \nonumber\\
&& \frac{\sin^2(\Delta t(\varepsilon_{P_Z}-
 \varepsilon_{T_Z})/2\hbar)}{(\varepsilon_{P_Z}-
 \varepsilon_{T_Z})^2/4},
\end{eqnarray}
where $\varepsilon_{i_Z}$ and  $n^{(Z)}_{i}(E^*_Z(\ell),t)$ are the single-particle
energies and  occupation numbers of nucleons in the DNS fragments;
 the matrix elements $g_{PT}$  describe one-nucleon exchange
between the nuclei of DNS, and their values are calculated
microscopically using the expression obtained  in
Ref.~\cite{Adam92}. The decay probability of DNS $\Lambda^{\rm qf}_{Z}$ from the
charge asymmetry configuration $Z$  is calculated by the formula used in Ref. \cite{anisEPJA34}:
\begin{eqnarray}
\label{Lambda}
\Lambda^{\rm qf}_Z&=&K_{\rm rot}(E^*_Z) \,\omega_m
\left(\sqrt{\gamma^2/(2\mu_{\rm qf})^2+\omega^2_{\rm qf}}-\gamma/(2\mu_{\rm qf})
\right)\nonumber\\
&\times&\exp\left(-B_{\rm qf}/T_Z(\ell))\right)/(2\pi\omega_{\rm qf}),
\end{eqnarray}
where, $T_Z(\ell)$ is the effective temperature of DNS
  and it is estimated by formula:
\begin{equation}
 T_Z=\sqrt{\frac{E^{(Z)}_{\rm DNS}}{a_{\rm DNS}}},
\end{equation}
 which is determined by the DNS excitation energy  $E^{(Z)}_{\rm DNS}=E_{c.m.}-V_{\rm min}(R)+Q^{(Z)}_{gg}$
 and by the corresponding level density parameter: $a_{\rm DNS}=A/12 MeV^{-1}$.
 The dependence the DNS excitation energy on the charge asymmetry is
 related with the change of its intrinsic energy by the change of
 mass and charge numbers of its constituents from the ones of
 projectile and target nuclei:
\begin{eqnarray}
 Q^{(Z)}_{gg}&=&B_1(Z_P,A_P)+B_2(Z_T,A_T)-\nonumber\\
 &&B_1(Z,A)-B_2(Z_{\rm CN}-Z,A_{\rm CN}-A),
\end{eqnarray}
\begin{figure}[h!]
\centering
 \resizebox{0.5\textwidth}{!}{\includegraphics{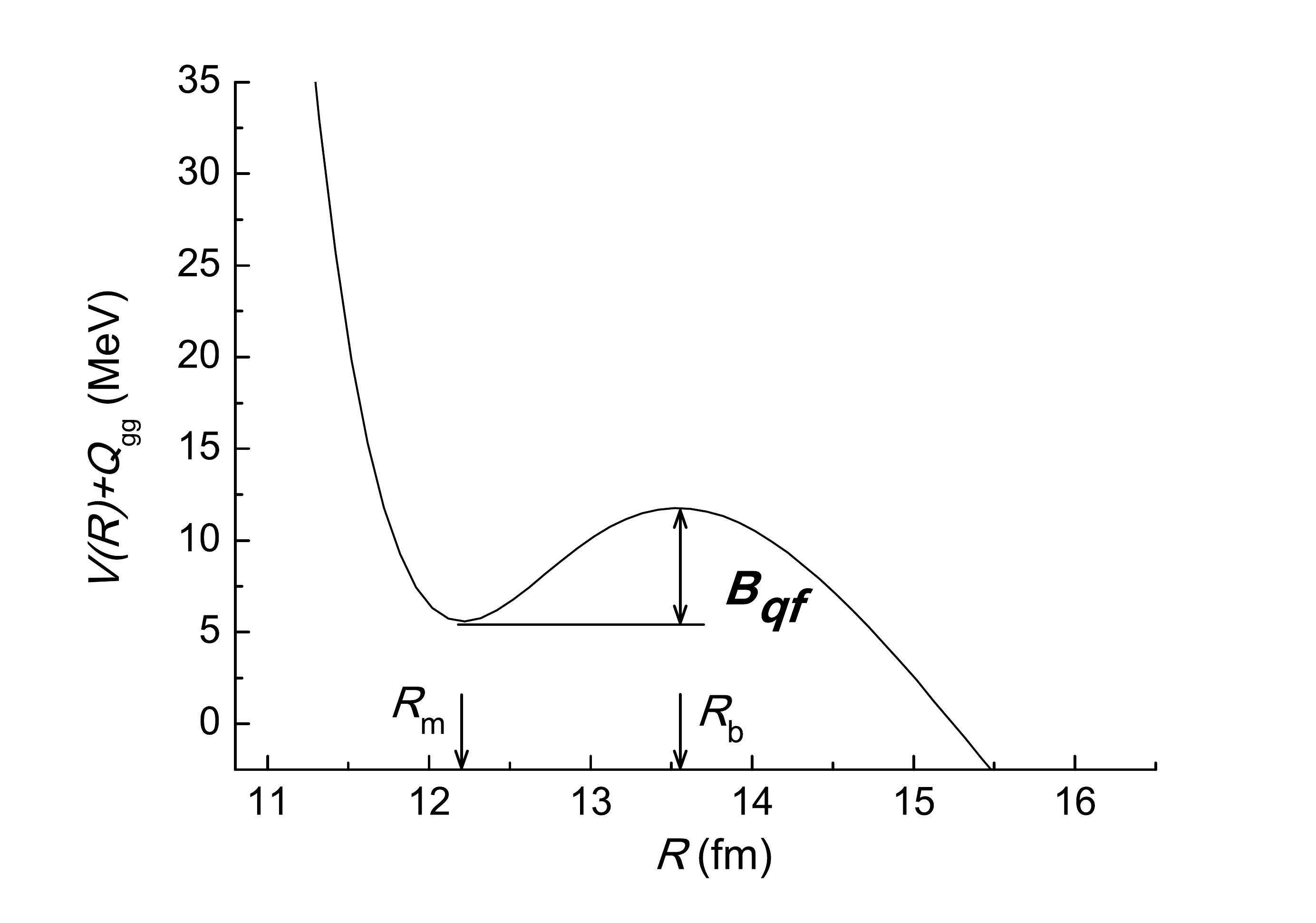}}
\vspace{-1.0cm}
 \caption{The nucleus-nucleus
interaction potential $V(R)$ for the $^{19}$F+$^{208}$Pb system:
the quasifission barrier $B_{qf}$ as the a depth of the potential
well.
\label{VintQgg}}
 \end{figure}

Here the frequency $\omega_m$ and $\omega_{qf}$ are found by the
harmonic oscillator approximation to the nucleus-nucleus potential
$V(R)$ shape for the given DNS configuration $(Z,Z_{tot}-Z)$ on
the bottom of its pocket placed at $R_m$ and  on the top
(quasifission barrier) placed at $R_{qf}$ (see Fig.
\ref{VintQgg}), respectively:
\begin{eqnarray}
 \omega_m^2&=&\mu_{qf}^{-1}\left|\frac{\partial^2 V(R)}{\partial
R^2}\right|_{R=R_m}\,,\\
\omega_{qf}^2&=&\mu_{qf}^{-1}\left|\frac{\partial^2 V(R)}{\partial
R^2}\right|_{R=R_{qf}}.
\end{eqnarray}

 The collective enhancement factor of
the rotational motion $K_{\rm rot}$ to the level density  should be
included because the dinuclear system is a good rotator. It is
calculated by the well known expression \cite{Junghans}:
 \[K_{\rm rot}(E^*_Z) =
 \left\{\begin{array}{ll}(\sigma_{\bot}^2-1)f(E^*_Z)+1,  \hspace*{0.2 cm}  \rm{if}
\ \  \sigma_{\bot}>1 \ \  \
  \\ 1,  \hspace*{0.2cm}  \rm{if} \ \
  \it \sigma_{\bot}\le 1\:,
 \end{array}
 \right.
 \]
where  $\sigma_{\bot}=J_{DNS} T/\hbar^2$;
$f(E)=(1+\exp[(E-E_{cr})/d_{cr}])$;  $E_{cr}=120
\widetilde{\beta}_2^2 A^{1/3}$ MeV; $d_{cr}=1400
\widetilde{\beta}_2^2 A^{2/3}$. $\widetilde{\beta}$ is the
effective quadrupole deformation for the dinuclear
 system. We find it from the results of  $\mathcal{J}^{DNS}_{\bot}$
 calculated as in Ref. \cite{anisEPJA34}.

The fusion probability $P^{(Z)}_{CN}(E^*_{Z}(\ell);\{\alpha_i\})$
 is calculated by  the expression (\ref{Pcn}) presented in our
work \cite{NasiEPJ49}:
\begin{equation}
 \label{Pcn} P^{(Z)}_{CN}(E^*_{Z}(\ell))=\frac{\rho_{\rm fus}(E^*_{Z}(\ell))
}{\rho_{\rm fus}(E^*_{Z}(\ell)) +
\rho_{\rm qf}(E^*_{Z}(\ell))+\rho_{\rm sym}(E^*_{Z}(\ell))}.
 \end{equation}
 The level density of DNS was calculated by formula from Ref. \cite{VolPRC1995}
\begin{eqnarray}
\rho_i(E^*_Z)&=&\left[\frac{g^2}{g_1g_2}\right]^{1/2}\exp\left[2\left(a(E^*_Z-B_i)^{1/2}\right)\right]\nonumber\\
&\cdot&\frac{g}{6^{3/4}\left(2a(E^*_Z-B_i)^{5/4}\right)},
\end{eqnarray}
where $i$=fus, qf, sym; $g_1$ and $g_2$ are densities of single-particle states near the Fermi
surface for the DNS nuclei;  $2g=g_1+g_2$, and $a=\pi^2/6 g$.
 We used the following  set of parameters:  $g=g_1=g_2$ and $a=A/12$ MeV$^{-1}$.

In the DNS model the hindrance to complete fusion is determined by the
peculiarities of the driving potential which is calculated as a sum of the reaction
energy balance $Q_{gg}$ and interaction potential between the DNS nuclei:

\begin{equation}
 \label{Udr} U_{\rm dr}(A,Z,\ell)=Q_{gg}+V(Z_1,A_1,Z_2,A_2,\ell,\{\alpha_i\};R),
 \end{equation}
where $Q_{gg}=B_1+B_2-B_{\rm CN}$,  $B_1$, $B_2$ and $B_{\rm CN}$ are the binding energies of the interacting
nuclei and CN, respectively, which are obtained from the nuclear mass tables in
Refs. \cite{AudiNPA95,MolNix95}.

\begin{figure}[h!]
 \resizebox{0.45\textwidth}{!}{\includegraphics{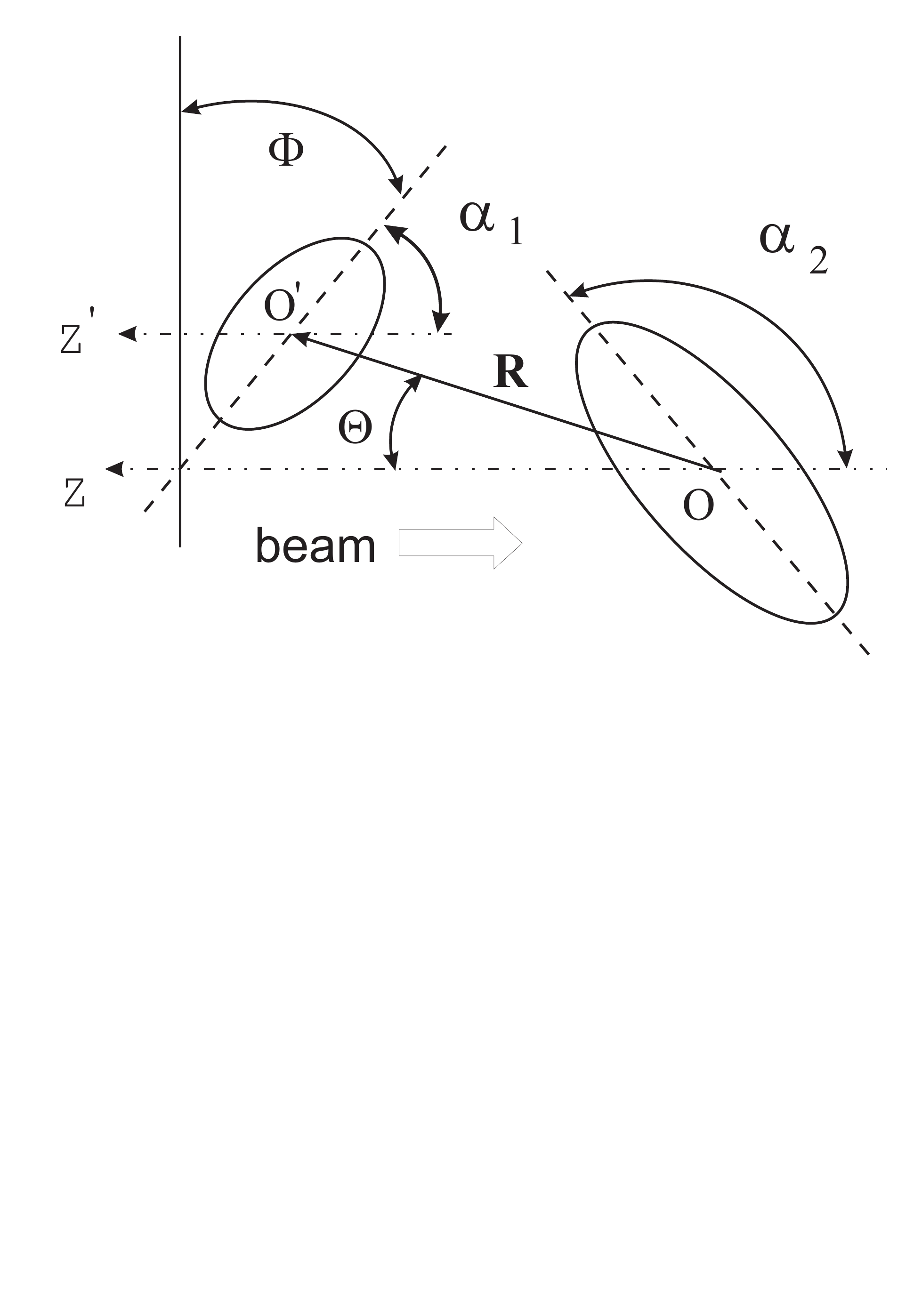}}
 \vspace*{-6.0cm}
 \caption{The coordinate systems and angles which were used for the
description of the initial orientations of projectile and target
nuclei. The beam direction is opposite to $OZ$.
\label{twonuc}}
 \end{figure}

The nucleus-nucleus potential $V$ consists of
the three parts:
\begin{eqnarray}
&&V(Z_1,A_1,Z_2,A_2,\ell,\{\alpha_i\};R)=\nonumber\\
&&V_{\rm Coul}(Z_1,A_1,Z_2,A_2,\{\alpha_i\};R)\nonumber\\
&&+V_{\rm nucl}(Z_1,A_1,Z_2,A_2,\{\alpha_i\};R)
\nonumber\\
&&+V_{\rm rot}(Z_1,A_1,Z_2,A_2,\ell,\{\alpha_i\};R),
\label{nnV}
\end{eqnarray}
where $V_{\rm Coul}$, $V_{\rm nucl}$, and $V_{\rm rot}$ are the Coulomb,
nuclear, and rotational potentials, respectively.
The Coulomb potential $V_{\rm Coul}(R)$ is calculated
 by Wong's formula \cite{Wong}:
\begin{eqnarray}
\label{Coul}
V_C(R,\alpha_1,\alpha_2)&=&\frac{Z_1 Z_2}R e^2 \nonumber\\
&+&\frac{Z_1 Z_2}{R^3}e^2 \left\{\left(\frac9{20\pi}\right)^{1/2}
\sum_{i=1}^2 R_{0i}^2\beta_{2}^{(i)}{ P}_2(\cos\alpha'_i) \right.\nonumber\\
&+&\left. \frac 3{7\pi}\sum_{i=1}^2 R_{0i}^2
\left[\beta_{2}^{(i)}{ P}_2(\cos\alpha'_i)\right]^2\right\}\,,\hspace*{-0.3cm}
\end{eqnarray}
where $\alpha'_1=\alpha_1+\Theta$,
$\alpha'_2=\pi-(\alpha_2+\Theta)$, $\sin\Theta=|{\bf L}|/(\mu
\dot{R} R)$; $Z_i$, $\beta_{2}^{(i)}$,   and $\alpha'_i$ are
the atomic number (for each fragment), the quadrupole deformation
parameter, and the angle (see Fig.\ref{twonuc}) between the line
connecting the centers of masses of the nuclei and the symmetry
axis of the fragment $i (i=1,2)$, respectively.  Here,
 ${ P}_2 (\cos\alpha'_i)$ is the second term of the second type of Legendre polynomial.
The radius parameter $r_0$ used to find the nuclear radius $R_{0i}=r_0 A_i^{1/3}$
 is changed in the range $r_0$=1.16---1.17 fm to reach an agreement with
 the experimental data of the capture cross section at lowest energies.

The nuclear part of the nucleus-nucleus potential is calculated
using the folding procedure between the effective nucleon-nucleon
forces $f_{eff}[\rho(x)]$ suggested by Migdal  \cite{Migdal}
and  the nucleon density of the projectile and target nuclei,
$\rho^{(0)}_1$ (\ref{rhoproj}) and $\rho^{(0)}_2$ (\ref{rhotarg}), respectively:
\begin{eqnarray}
\label{foldint}
V_{nucl}(R,\alpha_1,\alpha_2)&=&\int \rho^{(0)}_1({\bf r}-{\bf R};
\alpha_1,\beta^{(1)}_2)
f_{eff}[\rho]\nonumber\\
&\times& \rho^{(0)}_2({\bf r}; \alpha_2,\beta^{(2)}_2) d^3{\bf r}\,
,
\end{eqnarray}
\begin{equation}
f_{eff}[\rho]=300~\left(f_{in}+(f_{ex}-f_{in})
\frac{\rho(0)-\rho(r)}{\rho(0)}\right).
\end{equation}
Here  $f_{in}$=0.09, $f_{ex}$=-2.59 are the constants of the
effective nucleon-nucleon interaction; $\rho=\rho^{(0)}_1 +
\rho^{(0)}_2$. The center of
the laboratory coordinate system is placed on the target mass
center and, therefore, ${\it r}_1={\it R}$ and ${\it  r}_2=0$.

The angles between the  symmetry axis of the projectile and
target nucleus and the beam direction are $\alpha_1$ and
$\alpha_2$, respectively, (Fig.\ref{twonuc}).
The spherical coordinate system $O$ with the vector $r$, angles $\theta$
 and $\phi$ is placed at the mass center of the target nucleus and the  $Oz$
axis is directed opposite to the beam. In this coordinate system,
the direction of the vector ${\bf R}$ connecting the mass centers of
the interacting nuclei has angles $\Theta$ and $\Phi$: ${\bf r}_1={\bf R}$
and ${\bf r}_2=0$. The coordinate system is chosen  in such a way
that  the planes, in which the symmetry axes of nuclei are located, cross the $Oz$ line
and form the angle $\Phi$. For head-on (or polar) collisions $\Theta=0$ and
$\Phi=\phi$.

\begin{figure}[ht!]
 \resizebox{0.45\textwidth}{!}{\includegraphics{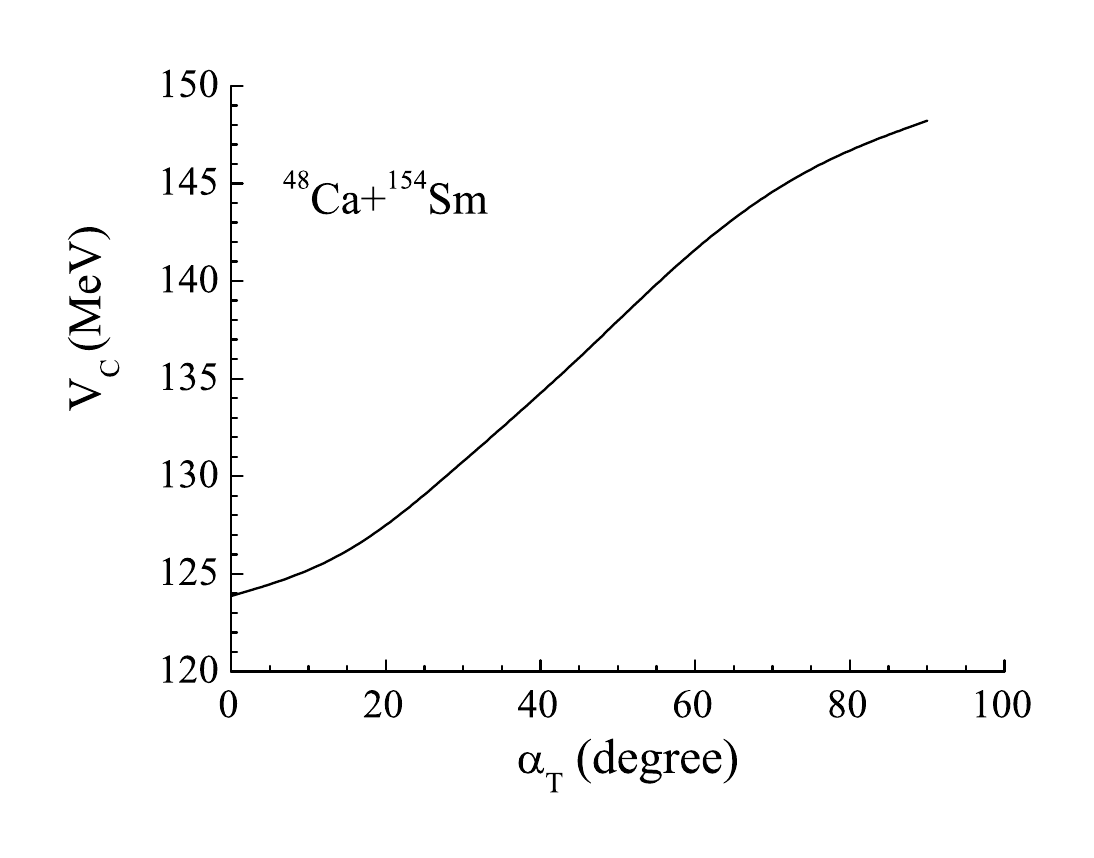}}
 \caption{Coulomb barrier V$_{\rm C}$ of the nucleus-nucleus interaction vs the orientation angle $\alpha_{\rm T}$ of the target-nucleus for the $^{48}$Ca+$^{154}$Sm reaction.\label{coulombarrier}}
\end{figure}

In this context we present in Fig \ref{coulombarrier} for the $^{48}$Ca+$^{154}$Sm reaction the dependence of the Coulomb barrier V$_{\rm C}$ on the initial orientation angle
$\alpha_{\rm T}$ of the symmetry axis of target nucleus with respect to the beam direction that it is necessary to consider in determination of nucleus-nucleus potential V
(see formula (\ref{nnV})). In this present case, when the beam-target interaction occurs with an angle $\alpha_{\rm T}=0^{\circ}$ of the target (tip collision),
the Coulomb barrier V$_{\rm C}$ is characterized by the minimum value of about 124 MeV; instead, V$_{\rm C}$ reaches the maximum value of about 148.2 MeV when the beam interacts
with target with an angle $\alpha_{\rm T}=90^{\circ}$ (equatorial collision). In this last case the Coulomb barrier value is about 20\% higher than the one
in the $\alpha_{\rm T}=0^{\circ}$ target orientation.
In our paper \cite{FazioJP77} we analyzed in detail the contributions of the capture and fusion cross sections versus the collision energy $E_{\rm c.m.}$ for various
target orientation angles $\alpha_{\rm T}$, and we presented the results of calculation in Fig.3 of the cited paper \cite{FazioJP77}.
At lower $E_{\rm c.m.}$ energies (at about $E_{\rm c.m.}<137$ MeV, only the small orientation angle of the target ($\alpha_{\rm T}\leq 45^{\circ}$) can contribute
to the capture cross section due to the low values of the Coulomb barrier for the mentioned $\alpha_{\rm T}$ angle range.
At $E_{\rm c.m.}=$148 MeV, all the $\alpha_{\rm T}$ configurations can contribute to the capture cross section with approximately the same possibilities
because the collision energy $E_{\rm c.m.}$ is sufficient to overcome the maximum value of the Coulomb barrier, depending on the $\alpha_{\rm T}$ angle orientation
at initial contact of reactants; instead, at the above-mentioned low energy range $E_{\rm c.m.}<$137 MeV the fusion cross section can only be contributed
by a small set of orientation angles of the target with $\alpha_{\rm T}\leq 45^{\circ}$ at initial beam-target interaction.
At higher $E_{\rm c.m.}$ energies (at about $E_{\rm c.m.}>$155 MeV) the contributions to the configurations with $\alpha_{\rm T}>45^{\circ}$ are larger than those with
$\alpha_{\rm T}\leq 30^{\circ}$ for both capture and fusion cross sections (see Ref. \cite{FazioJP77} for other important details).

 The shape of the dinuclear system nuclei
changes with the evolution of the  mass asymmetry degrees of
freedom: $\beta_2=\beta_2(Z,A)$ and $\beta_3=\beta_3(Z,A)$. In
order to calculate the potential energy
surface as a function of the charge number, we use the values of
$\beta_2^{(2^+)}$ from \cite{Raman}
and the values of $\beta_3^{(3^-)}$ from  \cite{Spear}.
In the $O$ system the  symmetry axis of the
target-nucleus is turned around the $\alpha_2$ angle, so its
nucleon distribution function is as follows:
\begin{eqnarray}
\label{rhotarg} &&\rho^{(0)}_2(r)=\rho_0
\biggl\{1+\exp\biggl[\frac{r-\tilde
R_2(\beta^{(2)}_2,\beta^{(2)}_3; \theta'_2)}{a}\biggr]
\biggr\}^{-1}, \\
&&\tilde R_2(\beta^{(2)}_2,\beta^{(2)}_3;
\theta'_2)=R^{(2)}_0\left(1+\beta^{(2)}_2Y_{20}
(\theta'_2)+\beta^{(2)}_3Y_{30}(\theta'_2)\right),\nonumber
\end{eqnarray}
where $\rho_0$=0.17 fm$^{-3}$, $a_0=0.54$ fm,
$$\cos\theta'_2=\cos\theta\cos(\pi-\alpha_2)+\sin\theta\sin(\pi-\alpha_2)
\cos\phi \,. \nonumber$$

The mass center of the projectile nucleus is shifted to the end of
the vector $R$ and its  symmetry axis is turned by the angle $\pi-
\alpha_1$. According to the  transformation formulae of the
parallel transfer of vectors  the variables of the transferred
system $O'$ are as follows:
\begin{eqnarray}
r'^2&=&r^2+R^2-2rR\cos(\omega_{12}),
\nonumber\hspace*{5.75cm}   \\
\cos(\omega_{12})&=&\cos\theta\cos\Theta+\sin\theta\sin\Theta
\cos(\phi-\Phi), \nonumber\\
\cos\theta'_1&=&\frac{\left(r\cos\theta-R\cos\Theta\right)}{r'}\,, \nonumber\\
\cos\phi'_1&=&(1+\tan^2\phi'_1)^{-1/2},  \nonumber\\
\tan\phi'_1&=&\frac{r\sin\phi\sin\theta-R\sin\Theta\sin\Phi}
{r\cos\phi\sin\theta-R\sin\Theta\cos\Phi}\,.  \nonumber
\end{eqnarray}
In the coordinate system $O'$, the deviation of the  symmetry axis
of projectile nuclei relative to the $O'z'$ axis is determined by
the angle
$$\cos\theta''_1=\cos\theta'_1\cos(\pi-\alpha_1)+\sin\theta'_1\cos\phi'_1.
\nonumber$$
 Now the nucleon distribution function of the projectile-nucleus looks like
this
\begin{eqnarray}
\label{rhoproj} &&\rho^{(0)}_1(r')=\rho_0
\biggl\{1+\exp\biggl[\frac{r'-\tilde
R_1(\beta^{(1)}_2,\beta^{(1)}_3; \theta'_1)}{a}
\biggr]\biggr\}^{-1}\hspace{-1.0mm},  \\
&&\tilde R_1(\beta^{(1)}_2,\beta^{(1)}_3;\theta'_1)=R^{(1)}_0
\left(1+\beta^{(1)}_2Y_{20} (\theta'_1)
+\beta^{(1)}_3Y_{30}(\theta'_1)\right). \nonumber
\end{eqnarray}

\begin{figure}[h!]
 \resizebox{0.5\textwidth}{!}{\includegraphics{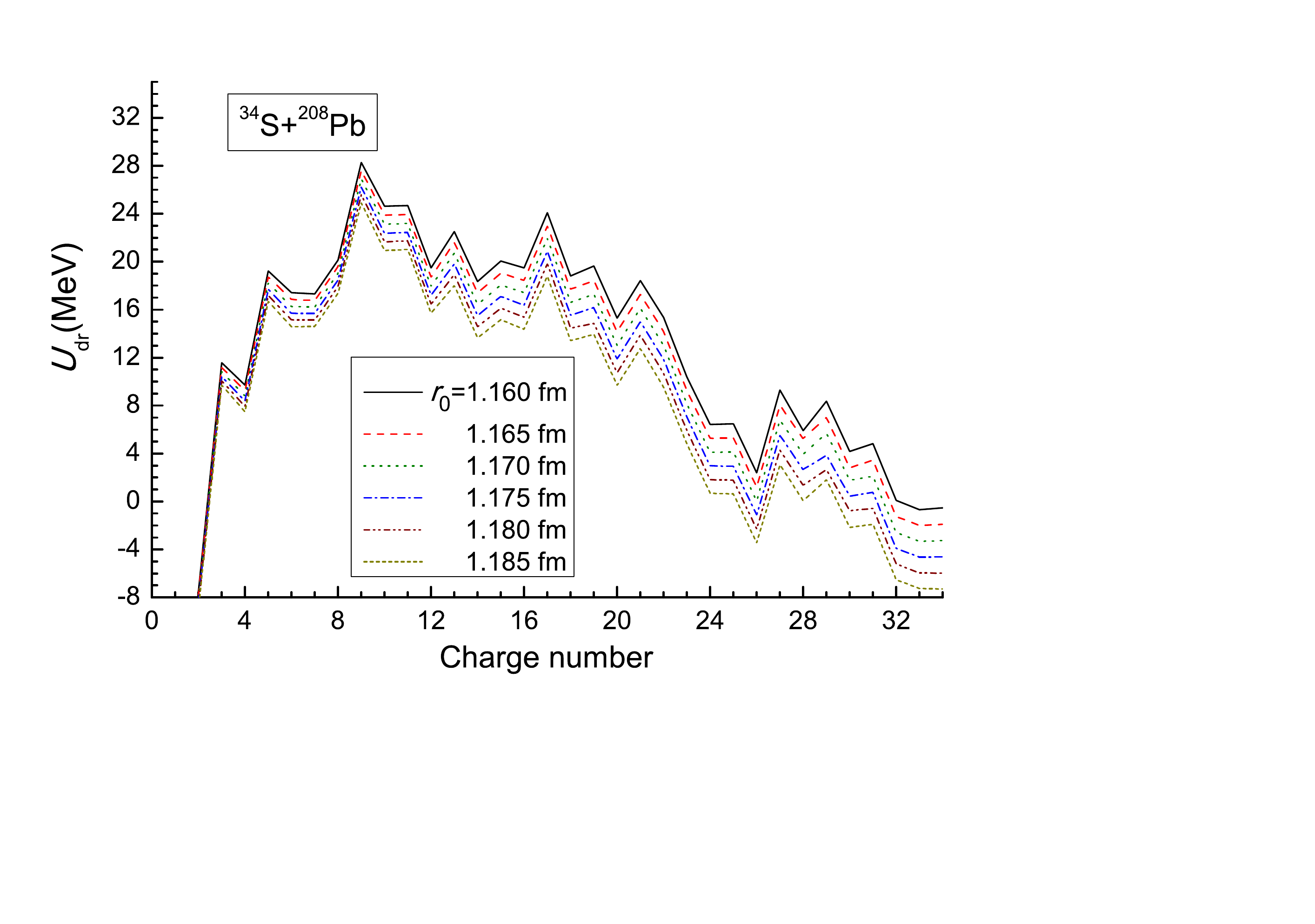}}
 \vspace*{-1.25cm}
 \caption{(Color on-line) The dependence of the driving potential calculated for the
 $^{36}$S+$^{206}$Pb reaction on the values of the radius parameter $r_0$.
\label{UdrS34Pb}}
 \end{figure}

\begin{figure}[h!]
 \resizebox{0.5\textwidth}{!}{\includegraphics{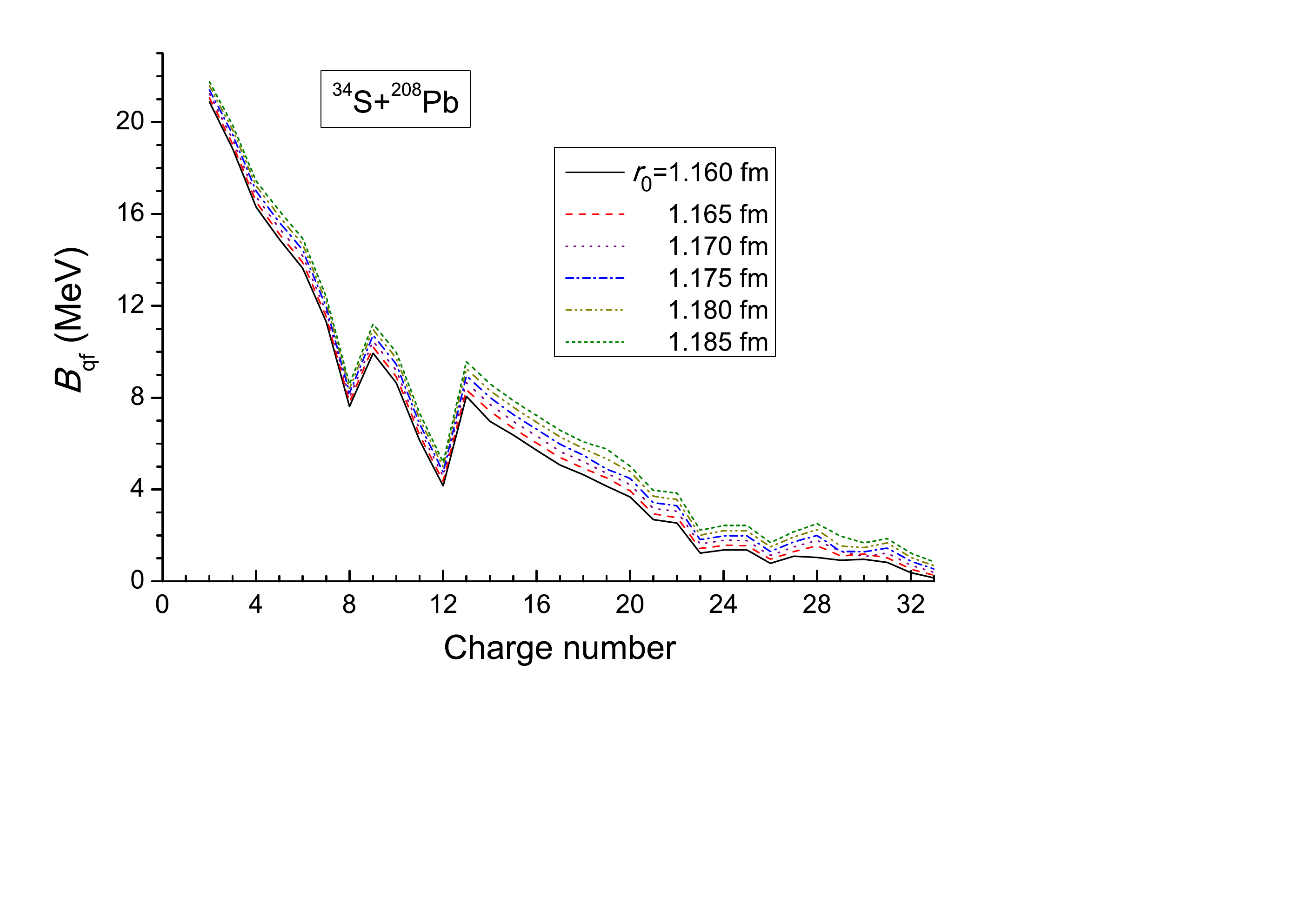}}
 \vspace*{-1.25cm}
 \caption{(Color on-line) The dependence of the quasifission barrier calculated for the
 $^{36}$S+$^{206}$Pb reaction on the values of the radius parameter $r_0$.
\label{BqfS34Pb}}
 \end{figure}

The sensitivity of the driving potential $U_{\rm dr}$ and quasifission barrier $B_{\rm qf}$  to the change of the radius parameter is presented in Fig. \ref{UdrS34Pb}
 and \ref{BqfS34Pb}, respectively.

\begin{figure}[h!]
 \resizebox{0.5\textwidth}{!}{\includegraphics{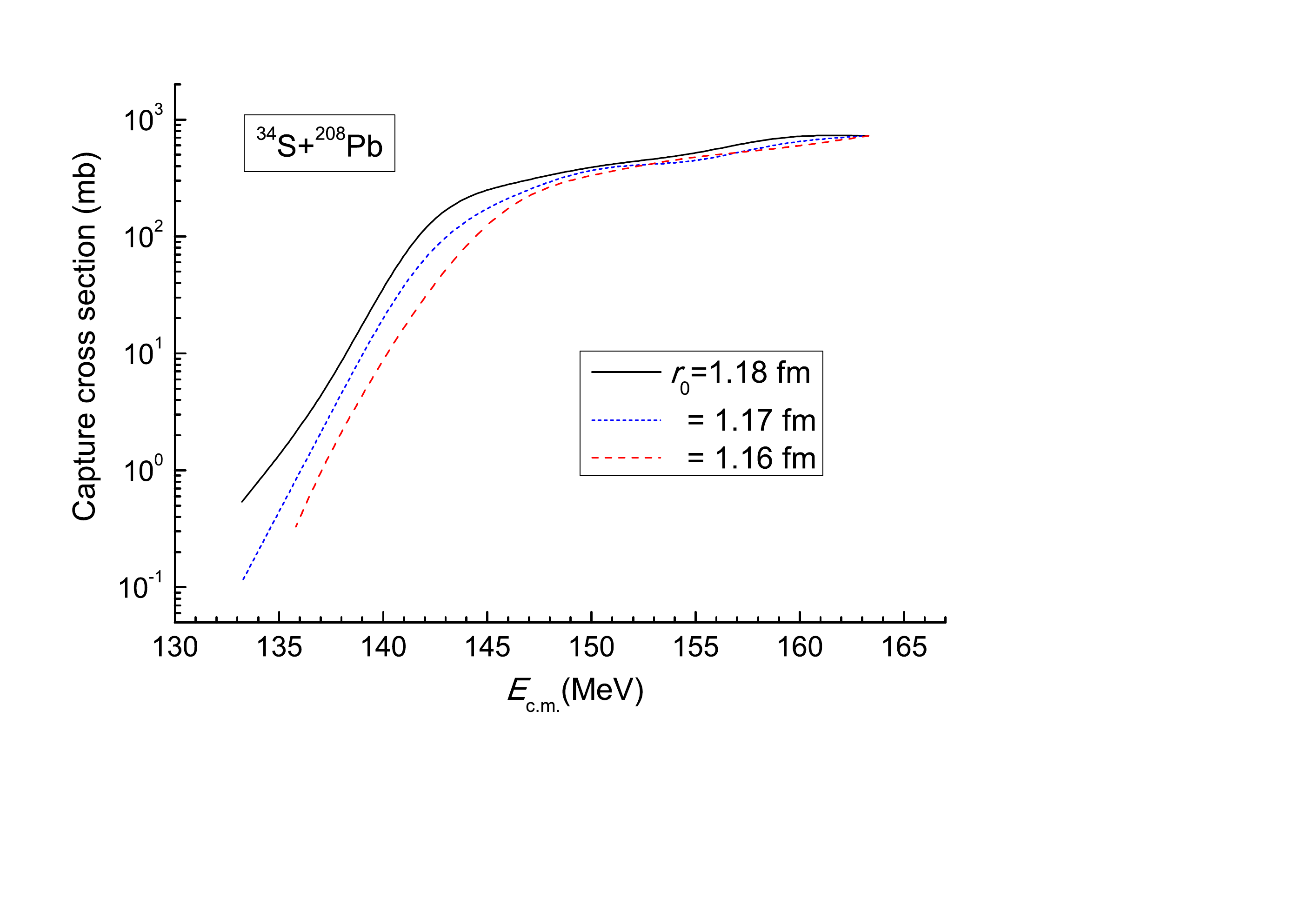}}
 \vspace*{-1.0cm}
 \caption{(Color on-line) The capture cross section $\sigma_{\rm cap}$ calculated for the
 $^{34}$S+$^{208}$Pb reaction on the values of the radius parameter $r_0$.
\label{CapS34Pb208}}
 \end{figure}

\begin{figure}[ht!]
\centering
 \resizebox{0.5\textwidth}{!}{\includegraphics{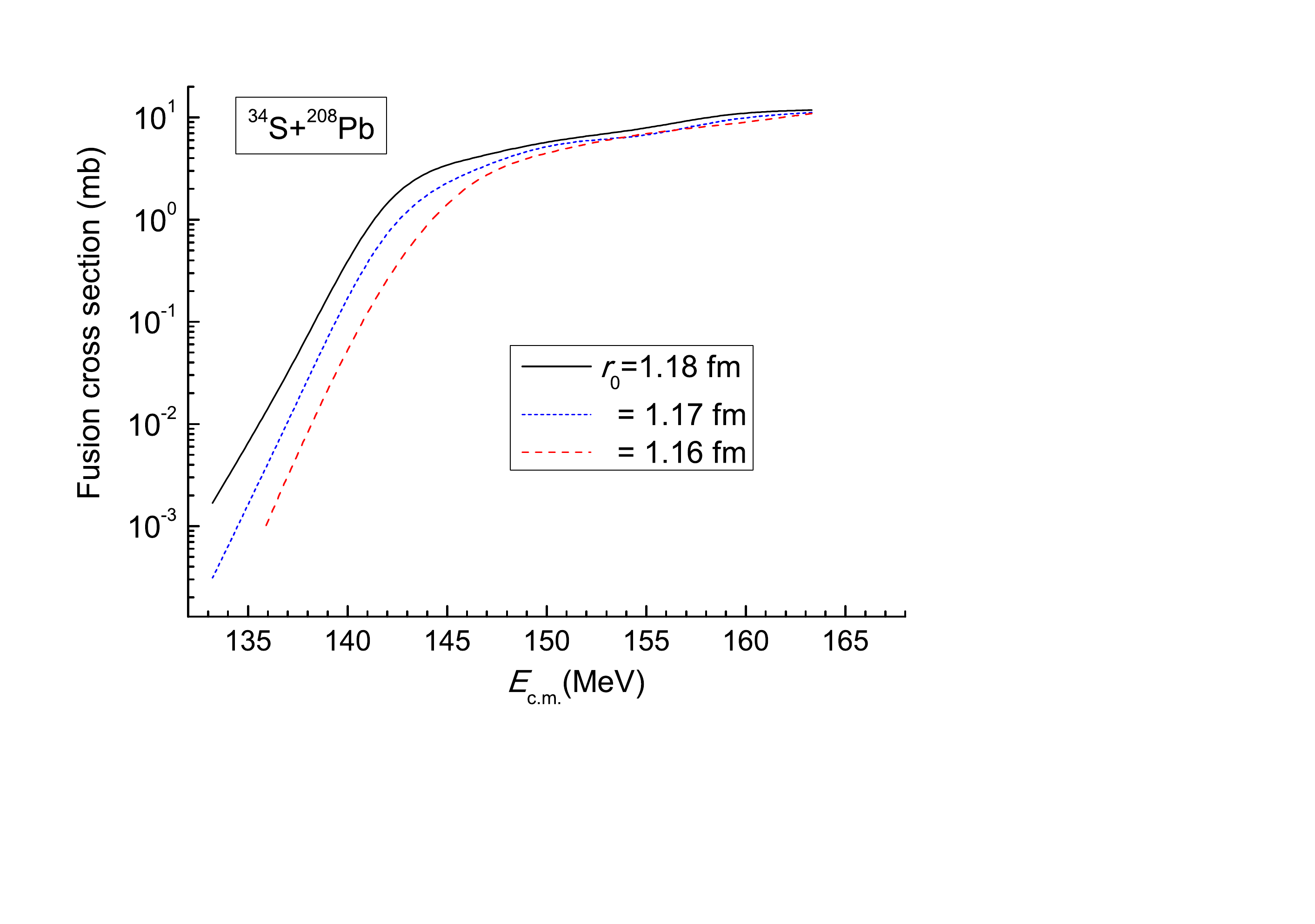}}
 \vspace*{-1.0cm}
 \caption{(Color on-line) The dependence of the fusion cross section calculated for the
 $^{34}$S+$^{208}$Pb reaction on the values of the radius parameter $r_0$.
\label{FusS34Pb208}}
 \end{figure}

\begin{figure}[ht!]
 \centering
 \resizebox{0.65\textwidth}{!}{\includegraphics{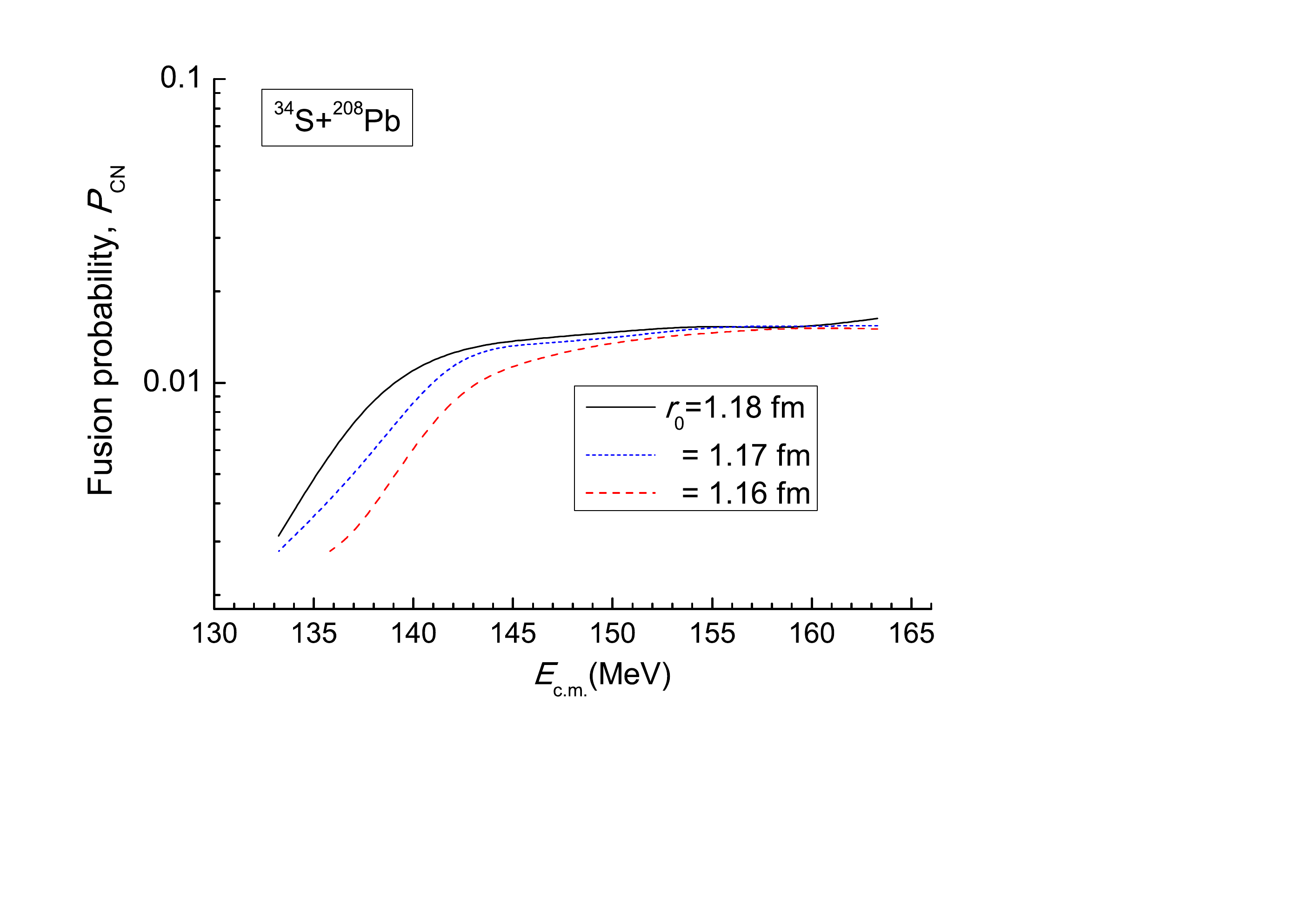}}
 \vspace{-2.0cm}
 \caption{The dependence of the fusion probability $P_{\rm CN}$ calculated for the
 $^{34}$S+$^{208}$Pb reaction on the values of the radius parameter $r_0$.
\label{PcnS34Pb208}}
 \end{figure}

The sensitivity of the capture $\sigma_{\rm cap}$ and
fusion $\sigma_{\rm fus}$ cross sections to the change of the radius
parameter is presented in Fig. \ref{CapS34Pb208}
 and \ref{FusS34Pb208}, respectively, while the sensitivity of the
 $P_{\rm CN}$ fusion probability is about 2 times at low
 beam energies $E_{\rm c.m.}$=136---140 MeV.
 Instead, at higher beam energies $E_{\rm c.m.}\geq 145$ MeV
 the $P_{\rm CN}$ values are approximately insensitive to
 the change of $r_0$ from 1.16 fm to 1.18 fm (see Fig. \ref{PcnS34Pb208}).

 As a result the position, slope and values
 of the capture and fusion excitation functions are changed significantly,
 while the $P_{\rm CN}$ fusion probability changes a little.
Moreover, as one can see that the shift of the excitation function is about
 3 MeV at lowest energies at the change of $r_0$ from 1.16 fm to 1.18 fm.
 This property of the excitation function is used in our calculation
 to reach an agreement of the capture cross section at the lowest
 energies with the experimental data.   Usually we necessity of the shift of the position of the curve of  excitation functions no more 3 MeV.

 We should note that the partial fusion cross sections are
 used in calculation of the survival probability of the excited compound
 nucleus. It is important to take into account the dependence of
 the fission barrier on the angular momentum.

\section{Procedures for determination of final products of CN}

In our model and procedures, the calculation of the effective fission barrier
$B_{\rm fis}$ is a function of the nuclear temperature $T$ and angular momentum $\ell$
as indicated in formula (\ref{fissb}) of the paper.
The damping function for the washing out of shell effect works in a very good way for general cases of heavy and superheavy compound nuclei
reached by heavy-ion reactions by comparing our theoretical results with experimental data of fission fragments and evaporation residue nuclei in a very wide set of nuclear reactions.

 \begin{figure}[h!]
 \resizebox{0.4\textwidth}{!}{\includegraphics{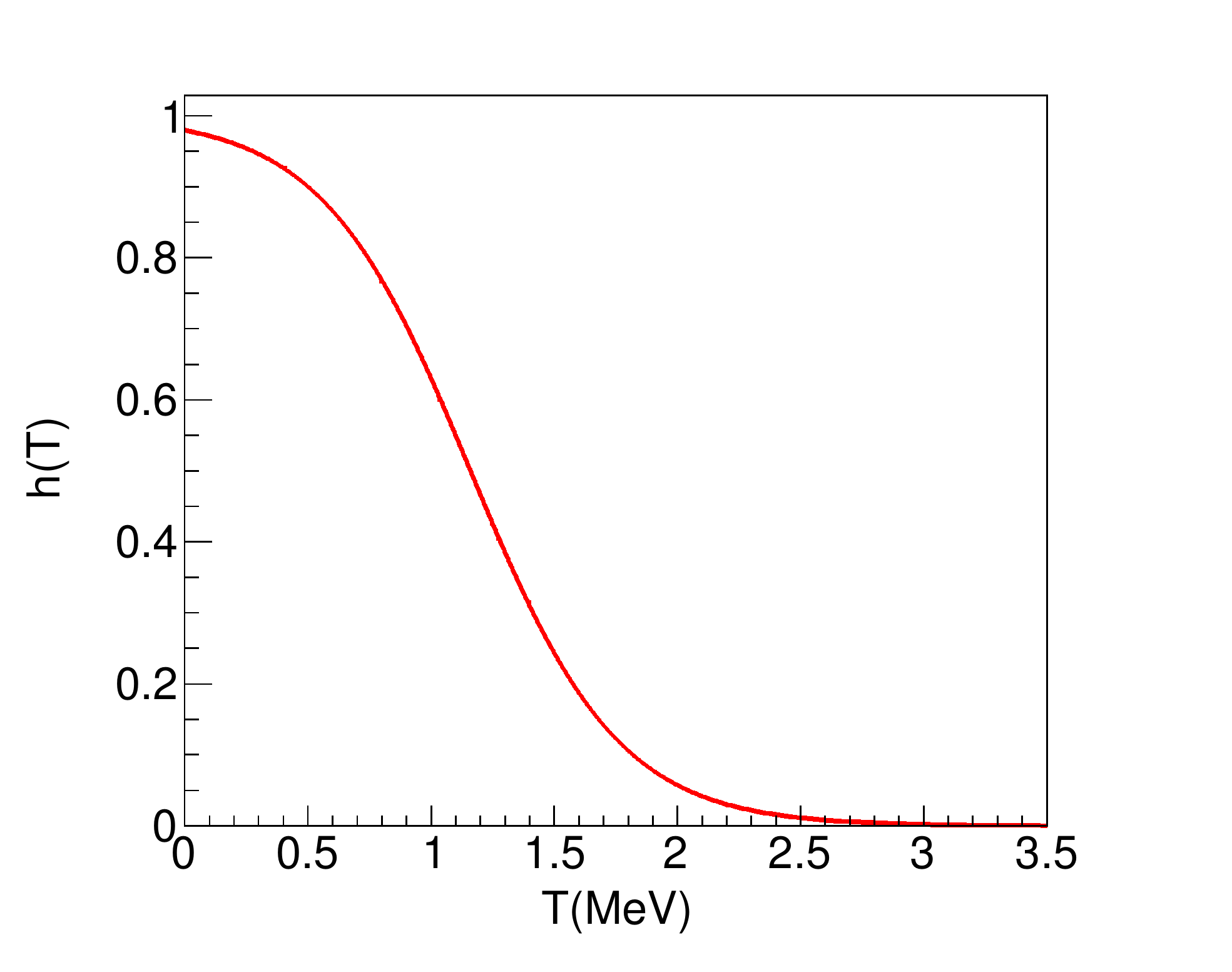}}
 \put(-60,118){\large (a)}\\
 \resizebox{0.4\textwidth}{!}{\includegraphics{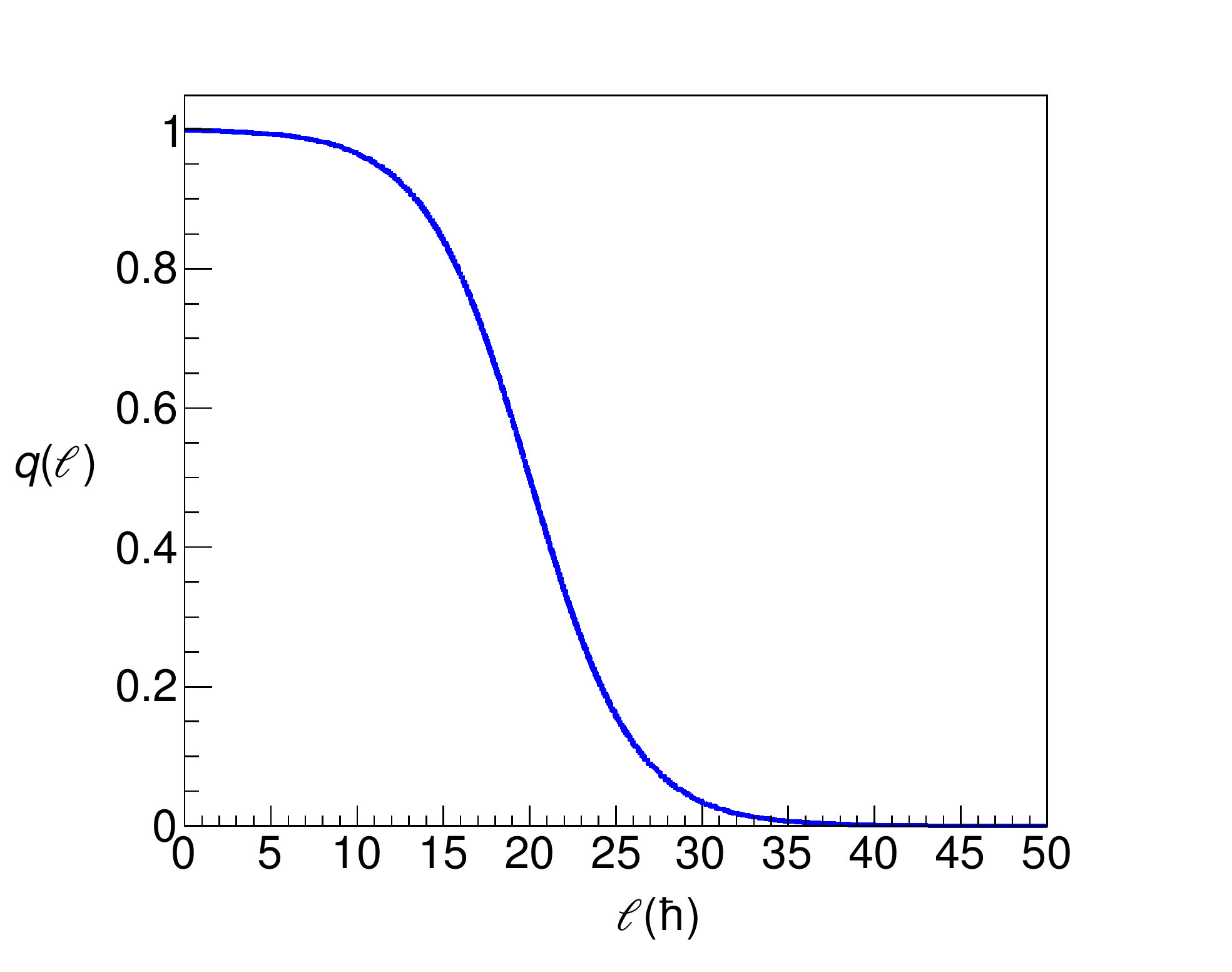}}
 \put(-60,120){\large (b)}
 \caption{(Color on-line) (a) The damping function $h(T)$ vs $T$ nuclear temperature (full line)
 for $T_{0}=1.16$ MeV and $d=0.3$ MeV.
          (b) The damping function $q(\ell)$ vs $\ell$ (dashed line) for $\ell_{1/2}=20\hbar$ and $\Delta \ell = 3\hbar$ parameters.\label{damp_shell_corr}}
 \end{figure}

 In figure \ref{damp_shell_corr} it is possible to observe the trend of the damping function as a function of $T$,
 where the nuclear temperature $T$ is connected with the excitation energy $E^{*}$ by the relation $T=\sqrt{\frac{E^*}{a}}$.
For example, in Fig. \ref{damp_shell_corr} (a) the changing of $ h(T)$ from the maximum value (close to 1) to 1/2
corresponds to the excitation energy $ E^*= a T^2 $ that, for example, in the case of a reaction
leading to the
$^{274}$Hs$^*$ CN is about 37~MeV. Moreover, $h(T)$ reaches the value 0.1 when the
nuclear temperature is about 1.83 MeV; this value corresponds to the excitation energy
$E^*$ of about 92~MeV. Therefore, the damping function $h(T)$ with the $T_0$ and $d$
values that we use leads to a very soft damping function with respect the nuclear
temperature and consequently with respect the excitation energy $E^*_{\rm CN}$ too.
The use of the parameters $d=0.3$ MeV and $T_{0}=1.16$ MeV in the damping function of the nuclear temperature is
not an arbitrary and convenient choice for some specific nuclear reactions and compound nuclei, but it is an
appropriate result obtained by investigation of a very wide set of heavy-ion reactions.
In Fig. \ref{damp_shell_corr} (b), $\ell_{1/2}=20\hbar$ and $\Delta\ell =3\hbar$ parameters reduce
the $q(\ell)$ function
from $0.9$ to $0.1$ in the (12-26) $\hbar$ interval confirming the important role of the $q(\ell)$
 damping function
in determination of the effective fission barrier $B_{\rm fis}(\ell,T)$.
It is useful to note that the values of parameters $d$, $T_{0}$, $\ell_{1/2}$ and $\Delta \ell$ used in the damping functions $h(T)$ and $q(\ell)$
are not changed in the study of heavy-ion reactions leading to heavy and superheavy nuclei as those considered in Table \ref{tab:1}.

In our code the fission and particle decay widths $\Gamma_{\rm fis}$ and $\Gamma_{\rm x}$ are calculated by the formulas

\begin{equation}
\begin{aligned}
\Gamma_{\rm fis}(E,J) = & \frac{1}{2\pi \rho (E,J)} & \int_0^{E-E_{\rm sad}(J)} \rho_{_{\rm fis}}(E-E_{\rm sad}(J)-\epsilon,J) \\
& & \times T_{\rm fis}(E-E_{\rm sad}(J)-\epsilon)d\epsilon,
\label{amp_fis}
\end{aligned}
\end{equation}

and

\begin{equation}
\begin{aligned}
\Gamma_x(E,J) = & \frac{1}{2\pi \rho (E,J)}\sum_{J'=0}^{\infty}\sum_{j=|J'-J|}^{J'+J} \\
& \times \int_0^{E-B_x} \rho_{_{\rm x}}(E-E_{\rm x}-\epsilon,J')T_{\rm x}^{\ell,j}(\epsilon)d\epsilon,
\label{amp_part}
\end{aligned}
\end{equation}

where the subscript $fis$ and $x$ refer, respectively, to the fission process and particle-$x$ emission channels (neutron, proton, $\alpha$, and $\gamma$),
and primes are used to mark an intermediate excited nucleus after particle emission and $E_{\rm sad}(J)$ is the energy of the decaying nucleus at the saddle point with angular momentum $\ell$ and total spin J.
It is known that a nucleus at the saddle point have a strong prolate deformation with the angular momentum vector perpendicular to the symmetry axis
and therefore the rotational contribution to $E_{\rm sad}$(J) is given by

\begin{equation}
\hbar^{2}J (J+1)/2(\mathcal{J}_{\perp})_{\rm sad}.
\label{E_sad}
\end{equation}

In the case of the yrast state (equilibrium state of the residual nucleus reached after particle or gamma emission),
the formation is usually slight and the shape may be prolate, oblate or even triaxial. For prolate yrast deformation,
we assume rotation around the axis perpendicular to the symmetry axis and retain expression (\ref{E_yra}) to calculate the rotational energy contribution.
In the case of the oblate deformation, however, the nucleus is assumed to rotate around its symmetry axis and the rotational energy contribution
to the potential-energy surface becomes

\begin{equation}
\hbar^{2}J (J+1)/2(\mathcal{J}_{\perp})_{\rm yr},
\label{E_yra}
\end{equation}
therefore, the effective moment of inertia $\mathcal{J}_{\rm eff}$ is defined by the relation

\begin{equation}
 \frac{1}{\mathcal{J}_{\rm eff}}=\frac{1}{\mathcal{J}_{\parallel}}-\frac{1}{\mathcal{J}_{\perp}}
\end{equation}

In addition, in formulas (\ref{amp_fis}) and (\ref{amp_part}) $\rho$, $\rho_{\rm fis}$ and $\rho_{\rm x}$ represent the collective level densities of the formed excited nucleus,
the level density of the excited nucleus at the saddle point configuration for orbital angular momentum $\ell$,
and the level density of the reached subsequent excited nucleus after particle-x emission for orbital angular momentum $\ell$, respectively.
$T_{\rm x}^{\ell ,j}$ is an optical-model transmission coefficient for particle-x with angular momentum $\ell$ coupled with particle spin to give $j$,
and the fission transmission coefficient $T_{\rm fis}$ in the Hill-Wheeler approximation is given by $T_{\rm fis}=\{1+exp[-2\pi (E^{*}-E_{\rm sad}(J)-\epsilon)/\hbar\omega]\}^{-1}$,
with $\hbar\omega=1$ MeV.
In the case of involved high excitation energies, the fission transmission coefficient is practically equal to unity,
 and therefore the particular choice of the $\hbar\omega$ value is irrelevant for the result of calculation.
The estimation of the effect of nuclear deformation at high spin values on the determination of the transmission coefficient was studied by \cite{Vig_Karw_PRC89}.
The main effect of the deformation was found to consist in the shift of the transmission coefficient threshold toward lower energies.
This shift is of the order of 1 MeV and may eventually lead to a substantial modification of the charged-particle emission close to the threshold,
but should not be relevant for the fission cross section.

It is useful to observe that in formula (\ref{amp_fis}) the calculation of $\Gamma_{\rm fis}$ fission width is characterized in the nominator
by the level density $\rho_{\rm fis}$ of the excited nucleus that reaches the saddle point state with angular momentum $\ell$
and total spin J - weighted by the transmission coefficient $T_{\rm fis}$ - and in the denominator by the level density $\rho(E,J)$ of the same excited
and rotating nucleus at the statistical equilibrated state with energy E and spin J, while in (\ref{amp_part}) the calculation of the
$\Gamma_{\rm x}$ width for particle-x emission - weighted by the $T_{\rm x}$ transmission coefficient - is characterized in the nominator
by the level density $\rho_{\rm x}(E-B_{\rm x}-\epsilon, J')$ of the intermediate excited nucleus reached after emission of one particle-x
(weighted by the $T_{\rm x }$ transmission coefficient) and in the denominator by the level density $\rho(E,J)$ of the decaying excited nucleus.

In formulas (\ref{amp_fis}) and (\ref{amp_part}) the Coulomb barriers for emission of light particles are obtained by \cite{CPC874},
the transmission coefficients are obtained by the routine SCAT2 \cite{SCAT2},
and binding energies for light particles are calculated using masses recommended by \cite{AudiNPA95} whenever available,
otherwise theoretical predictions of \cite{NM87545} are used.
The parity selection rules are also considered in calculation.

\begin{figure}
\centering
\includegraphics[scale=0.4]{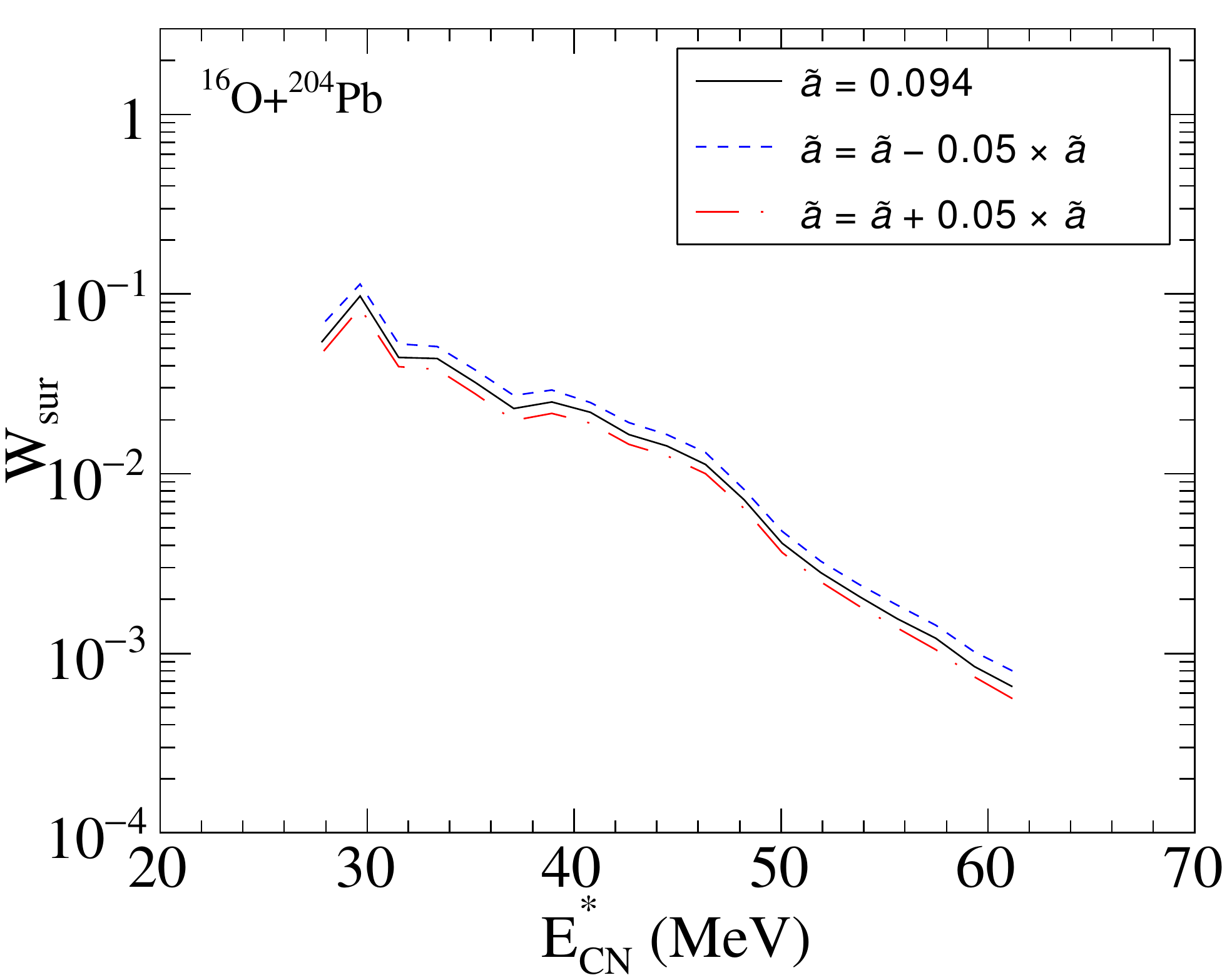}
\put(-170,50){(a)}\\
\includegraphics[scale=0.4]{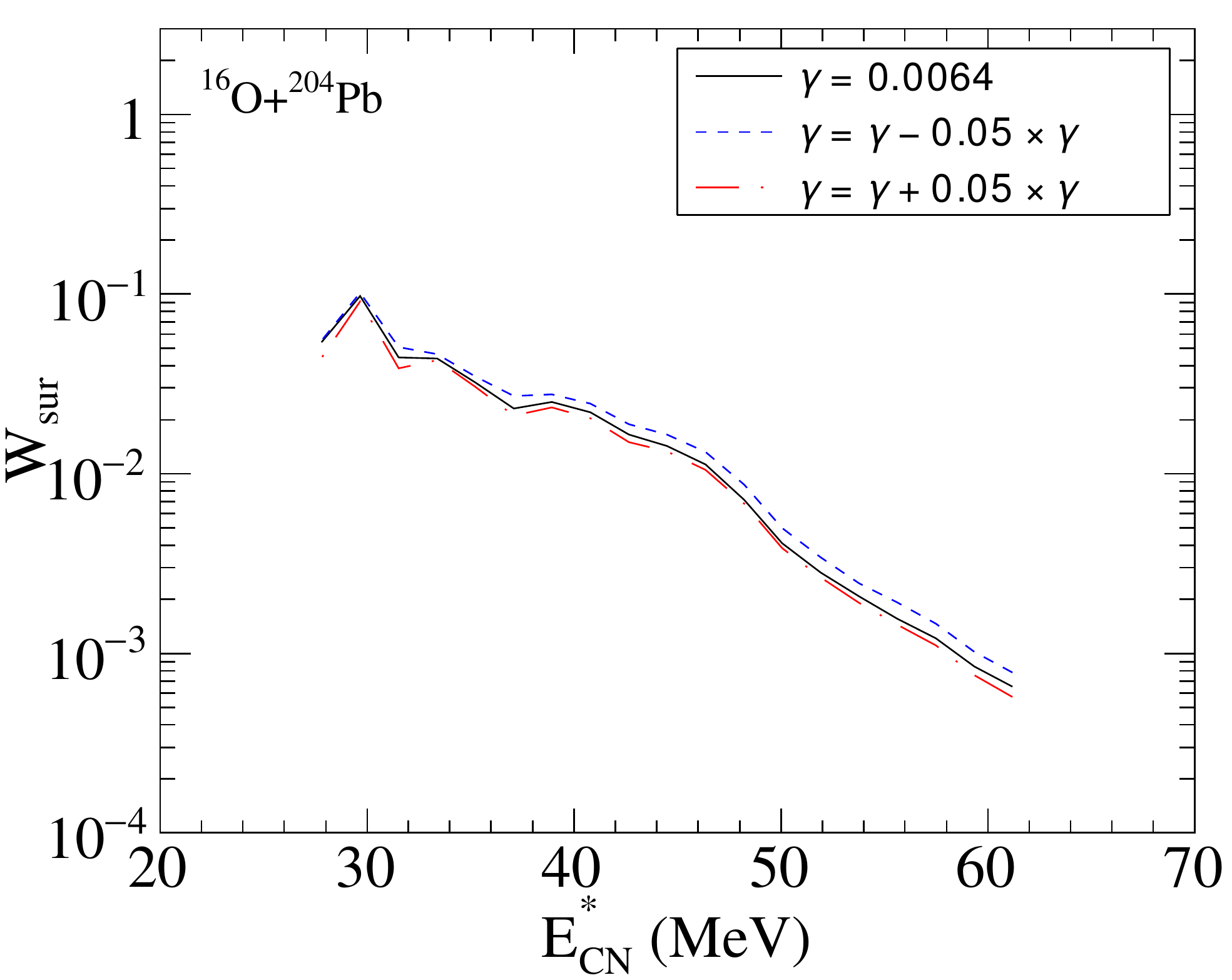}
\put(-170,50){(b)}
\caption{Survival probability $W_{\rm sur}$ vs $E^*_{\rm CN}$
for the $^{16}$O+$^{204}$Pb reaction leading to the heavy
$^{220}$Th CN.  Panel (a) represents
$W_{\rm sur}$ computed when \~{a} parameter was
changed by $\pm$5\% (dashed line for -5\%
and dash-dotted line for +5\%; full line for
the standard parameter \~{a} = 0.094$\times$A). Panel (b)
represents $W_{\rm sur}$ computed when $\gamma$
parameter was changed by $\pm$5\% (dashed
line for -5\% and dash-dotted line for +5\%; full line
for the standard parameter $\gamma$ = 0.0064
MeV$^{-1}$).\label{wsuratilde}}
\end{figure}

Figure \ref{wsuratilde} shows the sensitivity of the method in calculation of survival probability $W_{\rm sur}(E^*)$  for the $^{16}$O+$^{204}$Pb reaction
leading to the heavy $^{220}$Th CN when a relevant change of $\pm$~5\% of asymptotic level density parameter \~{a} in formula (\ref{level_density_a}) is considered (see panel (a)),
and also the sensitivity of  $W_{\rm sur}(E^*_{\rm CN})$ when a relevant change of $\pm$~5\% of the $\gamma$ parameter in the exponent term of formula (\ref{level_density_a})
is considered (see panel (b)) accounting for the rate of wash out of the shell effects with the excitation energy.
For example, at $E^*_{\rm CN}=38$~MeV of excitation energy, $W_{\rm sur}(E^*_{\rm CN})$ changes by a factor 1.2 with the relevant change of 5\% in the  
\~{a} value from 20.68 to 21.71 MeV$^{-1}$ or from 20.68  to 19.65~MeV$^{-1}$ (see panel (a)); 
the change of $W_{\rm sur}(E^*_{\rm CN})$  with the changing of $\pm$5\% is by a factor 1.1.

\begin{figure}
\centering
\includegraphics[scale=0.4]{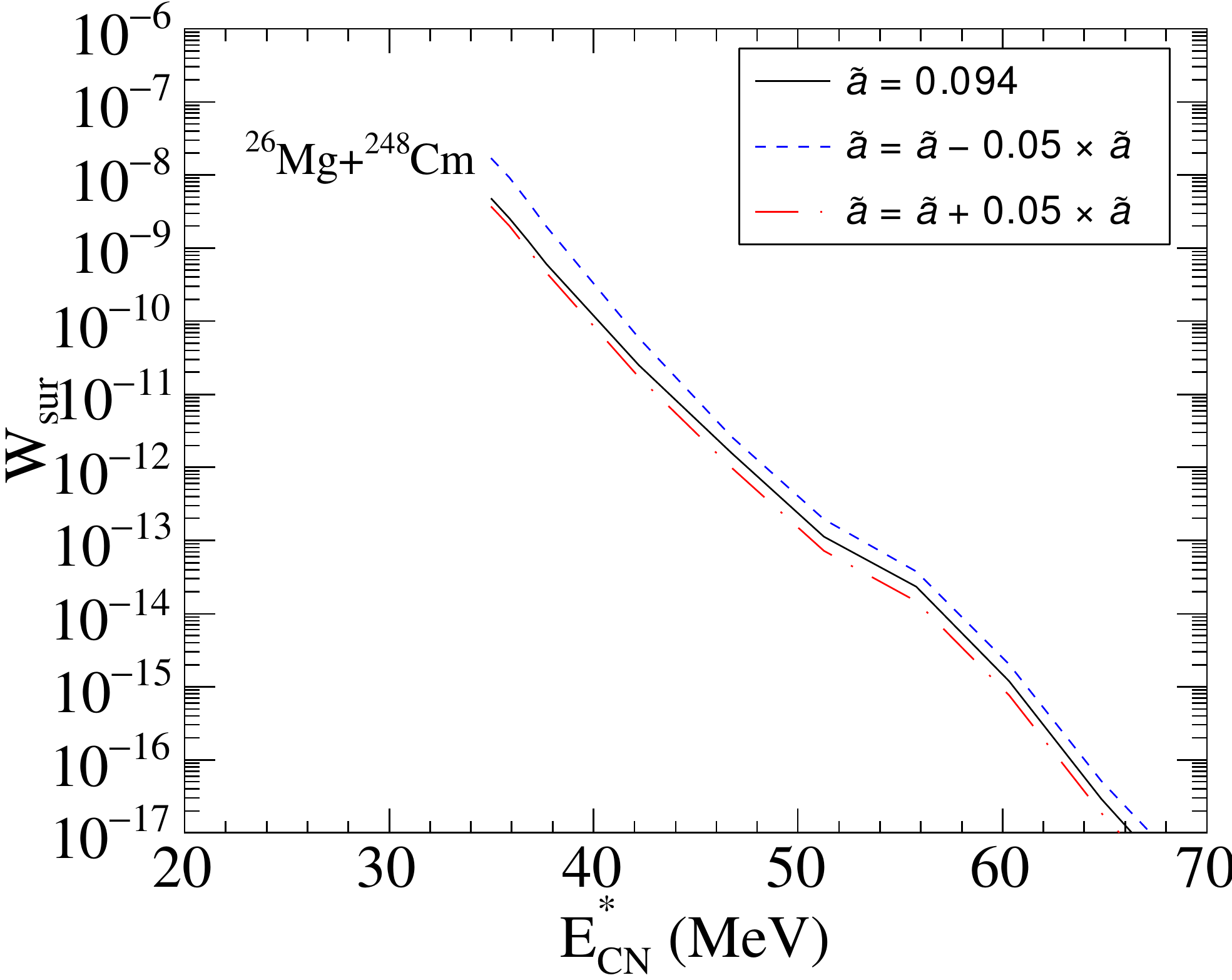}
\put(-170,50){(a)}\\
\includegraphics[scale=0.4]{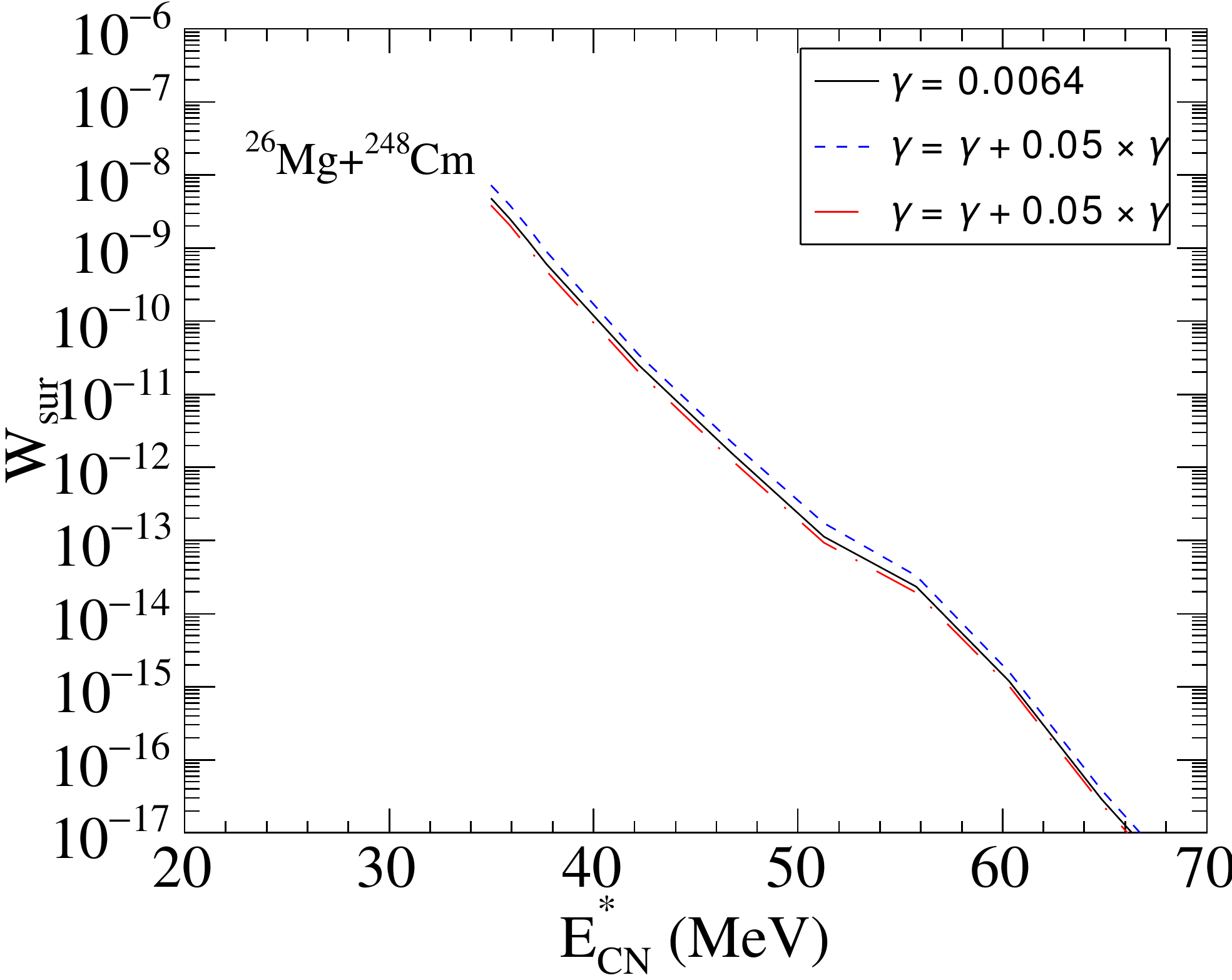}
\put(-170,50){(b)}\\
\caption{As Fig. \ref{wsuratilde}, but for the $^{26}$Mg+$^{248}$Cm reaction leading to the superheavy
$^{274}$Hs CN. \label{wsurgamma}}
\end{figure}

Analogously, the panel (a) of Fig. \ref{wsurgamma} shows an accentuate sensitivity in calculation of $W_{\rm sur}(E^*_{\rm CN})$ for the $^{26}$Mg+$^{248}$Cm reaction
leading to the superheavy $^{274}$Hs CN when the same change of $\pm$5\% of the \~{a} parameter is considered; in this case of superheavy nucleus with Z=108 and A=274,
at $E^*_{\rm CN}=44$~MeV of excitation energy, the observed $W_{\rm sur}(E^*_{\rm CN})$ values change by a factor of about 2.0 or 1.3
when \~{a} changes from 25.76 to 24.47~ MeV$^{-1}$, or from 25.76 to 27.05~ MeV$^{-1}$, respectively.
The panel (b) of Fig.\ref{wsurgamma} shows, instead, that a change of $\pm$5\% in the $\gamma$ parameter value produces a variation of $W_{\rm sur}(E^*_{\rm CN})$
at $E^*_{\rm CN}=44$~MeV by a factor of about 1.5-1.3 for the $^{274}$108 superheavy nucleus, respectively.
The present discussion is only made to show the sensitivity of the method for the calculation of $W_{\rm sur}(E^*_{\rm CN})$ for various considered compound nuclei
and nuclear reactions, but we always use in our calculation the standard parameters present in \~{a}, damping function of shell effects with $E^*_{\rm CN}$
in the determination of the level density parameter a, damping functions of the shell effects with $E^*_{\rm CN}$ and angular momentum $\ell$, on the fission barrier, respectively.
Never we use free parameters in order to study any mass asymmetric and almost symmetric reactions leading to heavy and superheavy nuclei.

By regarding $K_{\rm rot}^{\rm adiab}(E^{*})$ and $K_{\rm vibr}^{\rm adiab}(E^{*})$ collective enhancement coefficients present in
formula (\ref{rho_adiabatic}) for all studied reactions are used.
The vibrational enhancement coefficient is determined by the relation \cite{NPA1989, Ignatyuk29}
\begin{equation}
 K_{\rm vibr}^{\rm adiab}(E^{*})=exp\left[1.69\left(\frac{3m_{0}A}{4\pi\sigma_{\rm l.d.}}\frac{C_{l.d.}}{C}\right)^{\frac{2}{3}}T^{\frac{4}{3}}\right]
\label{k_vibr}
\end{equation}

where $\sigma_{\rm l.d.}=1.2$ MeV fm$^{-2}$ is the surface tension value in the liquid-drop model, $m_{0}$ is the nucleon mass, A is the mass number of the formed excited nucleus,
and the $C/C_{\rm l.d.}$ ratio is related to the different characteristics between the restoring force coefficient of the excited nucleus
and the corresponding rigidity coefficient of the liquid-drop model.
In our calculation $C=C_{l.d.}$ (liquid-drop value of rigidity coefficient) was used for all reactions forming CN.
The rotational enhancement coefficient is determined by one of the following relations

\begin{equation}
K_{\rm rot}^{\rm adiab}(E^{*})=\begin{cases}

  1, \text{for spherical nuclei,}\\
  \sigma_{\perp}^{2}, \text{for axially and mirror-symmetric nuclei,}\\
  2\sigma_{\perp}^{2}, \text{for axially-symmetric and}\\
  \, \, \, \, \text{mirror-asymmetric nuclei,}\\
  \sqrt{\frac{\pi}{2}} 2\sigma_{\perp}^{2} \sigma_{\parallel}, \text{for ellipsoidal ($D_{2}$) symmetry of nuclei,}\\
  \sqrt{8\pi} 2\sigma_{\perp}^{2} \sigma_{\parallel}, \text{for nuclei possessing no symmetry.}
 \label{K_rot}
\end{cases}
\end{equation}


In formula \ref{K_rot}, the spin-dependent parameters $\sigma_{\perp}=(\mathcal{J}_{\perp}T/\hbar^{2})^{\frac{1}{2}}$
and $\sigma_{\parallel}=(\mathcal{J}_{\parallel}T/\hbar^{2})^{\frac{1}{2}}$ are related to the perpendicular $\mathcal{J}_{\perp}$ and parallel $\mathcal{J}_{\parallel}$
moments of inertia of the deformed nucleus \cite{Ignatyuk29}, respectively

\begin{equation}
 \mathcal{J}_{\perp}=\frac{2}{5}m_{0}r_{0}^{2}A^{\frac{5}{3}}\left(1+\frac{1}{3}\beta\right)
\end{equation}

\begin{equation}
 \mathcal{J}_{\parallel}=\frac{6}{\pi^{2}}<m^{2}>a\left(1-\frac{2}{3}\beta\right)
\end{equation}
where $\beta$ is the deformation parameter of the nucleus and it represents the parameter of the internal nuclear quadrupole moment.

\textit{In order to calculate the collective level density in the non-adiabatic approach $\rho_{\rm coll}^{\rm non-adiab}(E^{*},J)$ - necessary
when the CN is formed at higher $E^{*}_{\rm CN}$ excitation energies  -}
we used a damped collective enhancement function $q(E^{*},\beta)$ with the aim to account the coupling of the collective to intrinsic degrees of freedom due to the nuclear viscosity
because while it is acceptable to treat collective modes within the adiabatic approximation at low energies, it is rather unlikely that at high energies the adiabatic assumption still holds,
due to the coupling of the elementary modes to the collective ones. So we introduce a certain general function for damping collective effects in the level density, depending strongly on
the excitation energy and deformation of the nucleus.
This is expressed in a decrease of the collective enhancement coefficient $K_{\rm coll}(E^{*})$ when the excitation energy increases;
therefore, the following expression was assumed \cite{Ignatyuk30,NPA1983}

\begin{equation}
\begin{split}
K_{\rm coll}^{\rm non-adiab}(E^{*})=\left\lbrace[K_{\rm rot}^{\rm adiab}(E^{*})-1]q(E^{*},\beta)+1\right\rbrace \\
 \times\left\lbrace[K_{\rm vibr}^{\rm adiab}(E^{*})-1]q(E^{*},\beta)+1\right\rbrace
\label{K_coll}
\end{split}
\end{equation}

where

\begin{equation}
 q(E^{*},\beta)=exp[-E^{*}/E_{1}(\beta)] .
\label{q_E_star}
\end{equation}

The expression of damping function is used in the fission and neutron (or other light particles) emission channels in the same way \cite{Rastop}.
In formula (\ref{q_E_star}) the following expression for $E_{\rm 1}(\beta)$ was assumed

\begin{equation}
 E_{1}(\beta) \cong 170\times A^{\frac{1}{3}}\beta^{2}
 \label{E_beta}
\end{equation}
in order to reach the better agreement between calculated values of fission cross sections and experimental ones for a wide set of nuclear reactions
leading to compound nuclei lighter than lead, preactinide and actinide compound nuclei, and also for nuclei with Z$>$100.
At the same time, the comparison between theoretical estimation of evaporation residue cross sections and experimental determinations have contributed
to the choice of the damping function q(E$^{*}$,$\beta$) to the $K_{\rm coll}$ collective enhancement coefficient given in formula (\ref{q_E_star})
together with the expression (\ref{E_beta}).The consequence of the quadratic dependence of $E_{\rm 1}(\beta)$ on $\beta$ (see relation (\ref{E_beta}))
is that the damping at the saddle point ($\beta \simeq 0.6 - 0.8$) is negligible in the high $E^{*}$ excitation energy region,
while the deviations of q($E^{*}$,$\beta$) from unity are already considerable for the neutron channel ($\beta \simeq 0.2 - 0.3$) at low $E^{*}$ excitation energy values.

\begin{figure}[ht!]
\centering
\includegraphics[scale=0.4]{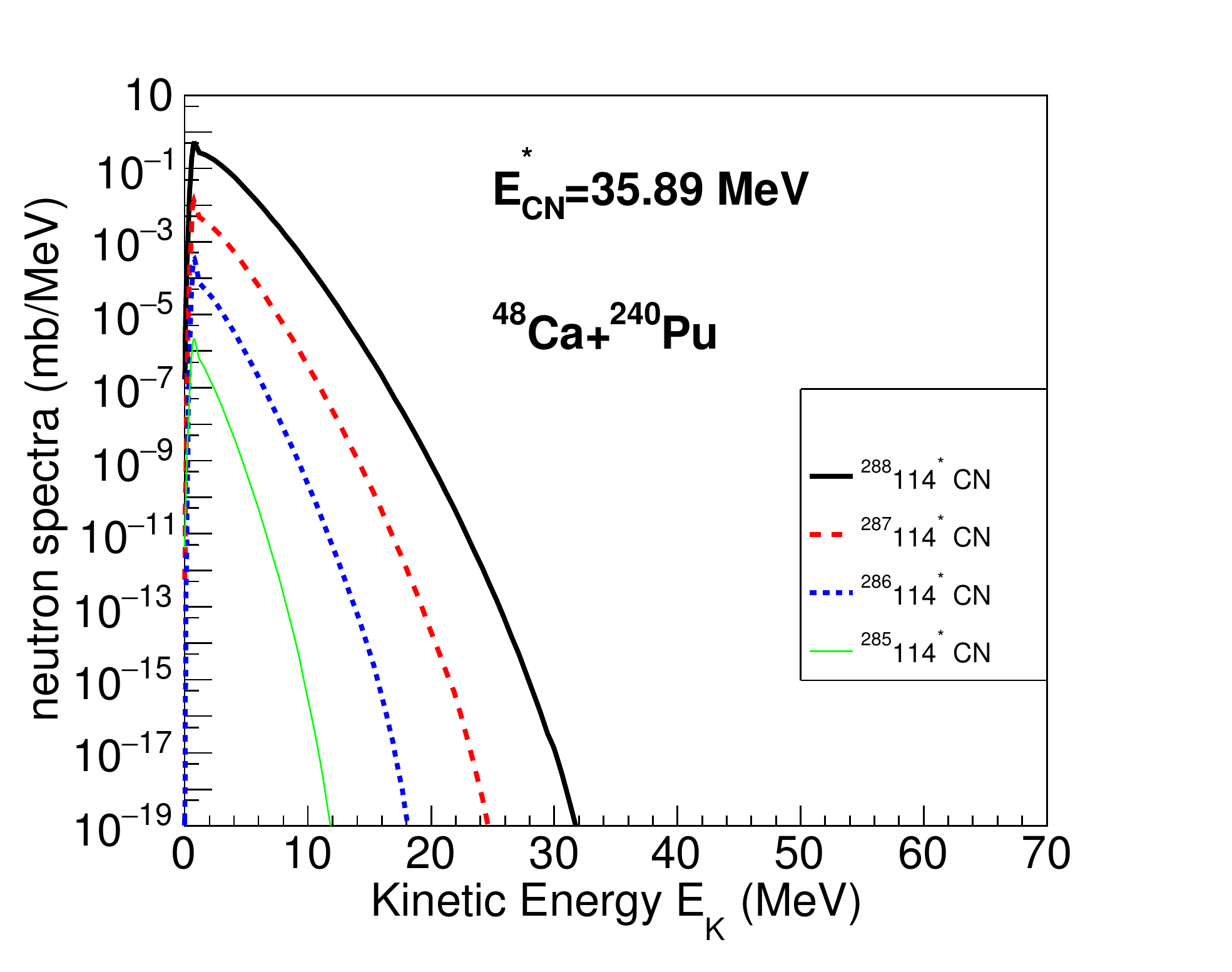}
\caption{ (Color on-line) Energy spectra of neutrons emitted at various steps of the deexcitation cascade starting from the $^{288}$114 CN
at $E^{*}_{\rm CN}=35.89$ MeV of excitation energy by the $^{48}$Ca+$^{240}$Pu reaction: thick full line for the first neutron emitted from $^{288}$114 CN$^{*}$,
dashed lines for the first neutron emitted from $^{287}$114$^{*}$, dotted line for the first neutron emitted from $^{286}$114$^{*}$
and thin full line for the first neutron emitted from $^{285}$114$^{*}$. \label{neutron_spectra}}
\end{figure}

\begin{figure}[ht!]
\centering
\includegraphics[scale=0.4]{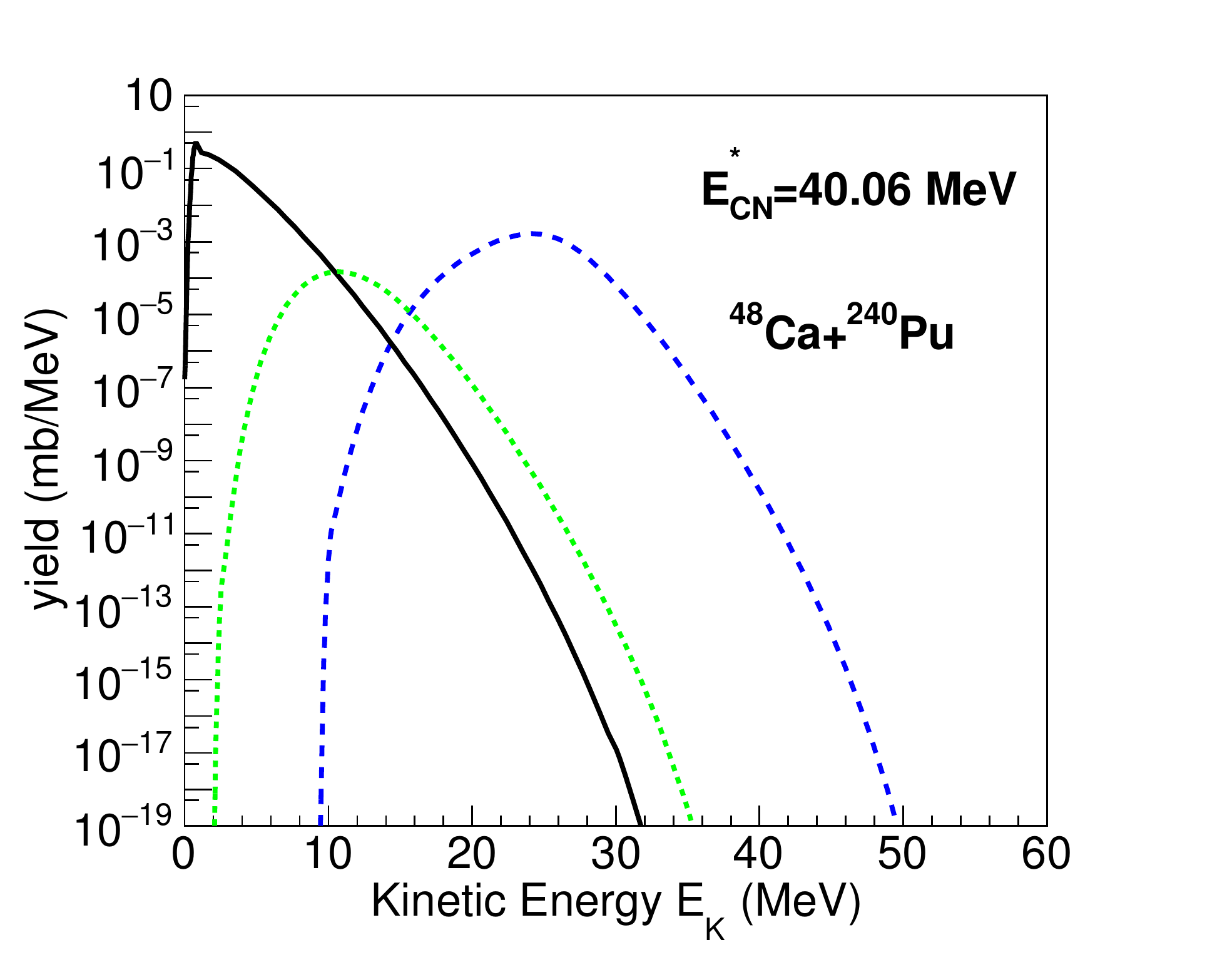}
\caption{Energy spectra of n, p, $\alpha$ emitted from the $^{288}$114 CN$^{*}$ at $E^{*}_{\rm CN}=40.06$ MeV of excitation energy
by the ${48}$Ca+$^{240}$Pu reaction: full line for the first emitted neutron, dotted line for the first emitted proton,
dashed line for the first emitted $\alpha$-particle.\label{comparison}}
\end{figure}

As an example of sensitivity of our refined model and procedures, we present in Fig.\ref{neutron_spectra} the calculated neutron energy spectrum of some emitted neutrons at various steps,
starting from the $^{288}$114 CN at $E_{\rm CN}=35.89$ MeV of excitation energy formed in the $^{48}$Ca+$^{240}$Pu reaction along the deexcitation cascade.
In Fig. \ref{comparison} we present the calculated energy spectra of neutron, proton, and $\alpha$-particle emitted
from the mentioned $^{288}$114 CN ($E_{\rm CN}=40.06$ MeV) at the first step of the cascade.

Formula (\ref{fissb}) of our manuscript describes the effective fission barrier obtained as the sum of the macroscopic fission barrier $B_{\rm fis}^{\rm m}(\ell)$
depending only on the angular momentum $\ell$ and the microscopic correction $\delta W$ due to the shell effects.
In our calculation $h(T)$ and $q(\ell)$ are the damping functions of the nuclear shell correction $\delta W$ by the increase of the excitation energy $E^*$ and angular momentum $\ell$, respectively,
and then the determination of the effective fission barrier $B_{\rm fis}(T,\ell)$ for each excited nucleus formed at various steps along the deexcitation cascade of the compound nucleus (CN) is a function of $T$ and $\ell$.
The parameters used to determine the level density, the effective fission barrier and the damping function have been extensively validated during the long-term investigation
over hundreds considered nuclear reactions (from strongly mass asymmetric reactions to almost symmetric ones) leading to heavy and superheavy nuclei.


\end{document}